\documentclass[aps,pra,noshowpacs,twocolumn,floatfix]{revtex4-2}
\usepackage{dsfont}
\usepackage{graphicx}
\usepackage{epstopdf} 
\usepackage{amsmath}
\usepackage{amsfonts}
\usepackage{mathtools}
\usepackage{braket}
\usepackage{cases}
\usepackage{ulem}
\usepackage{dsfont}
\usepackage{color}
\usepackage{soul}
\usepackage{nccmath}
\usepackage{bm}
\usepackage[utf8]{inputenc}
\usepackage{hyperref}

\graphicspath{ {./GraphicsPaper/} }

\begin{document}

\title{Caustics in quantum many-body dynamics}

\date{\today}
\author{W. Kirkby}
\affiliation{Department of Physics and Astronomy, McMaster University, 
1280 Main St. W., Hamilton, Ontario, Canada, L8S 4M1}
\author{Y. Yee}
\affiliation{Department of Physics and Astronomy, McMaster University, 
1280 Main St. W., Hamilton, Ontario, Canada, L8S 4M1}
\author{K. Shi}
\affiliation{Department of Physics and Astronomy, McMaster University, 
1280 Main St. W., Hamilton, Ontario, Canada, L8S 4M1}
\author{D. H. J. O'Dell}
\affiliation{Department of Physics and Astronomy, McMaster University, 
1280 Main St. W., Hamilton, Ontario, Canada, L8S 4M1}

\begin{abstract} 
We describe a new class of nonequilibrium quantum many-body phenomena in the form of networks of caustics that dominate the many-body wavefunction in the semiclassical regime following a sudden quench. It includes the light cone-like propagation of correlations as a particular case. Caustics are singularities formed by the birth and death of waves and form a hierarchy of universal patterns whose natural mathematical description is via catastrophe theory.  Examples in classical waves range from rainbows and gravitational lensing in optics to tidal bores and rogue waves in hydrodynamics.  Quantum many-body caustics are discretized by second-quantization (``quantum catastrophes'')  and live in Fock space which can potentially have many dimensions. We illustrate these ideas using the Bose Hubbard dimer and trimer models which are simple enough that the caustic structure can be elucidated from first principles and yet run the full range from integrable to nonintegrable dynamics.  The dimer gives rise to discretized versions of fold and cusp catastrophes whereas the trimer allows for higher catastrophes including the codimension-3 hyperbolic and elliptic umbilics which are organized by, and projections of, an 8-dimensional corank-2 catastrophe known as $X_9$. These results describe a hitherto unrecognized form of universality in quantum dynamics organized by singularities that manifest as strong fluctuations in mode population probabilities.
\end{abstract}

\maketitle

\section{Introduction}
Despite playing a fundamental role in many-body dynamics, the first observations of light cone-like spreading of correlations were only made recently using ultracold atomic gases in optical lattices \cite{Cheneau2012,Fukuhara2013,Langen2013} and trapped ions \cite{Richerme14,Jurcevic14}. These systems offer long relaxation times, the ability to vary external potentials and interparticle interactions, and spatially resolved imaging at the level of single sites/ions. The experiments proceed by creating a highly nonequilibrium state through a sudden quench, e.g.\ by rapidly changing the lattice depth, and then monitoring the time evolution of  site occupations.  This success has been followed-up with observations of the many-body localization transition \cite{Schreiber2015,Choi2016,Smith2016,Lukin2019} to a non-thermalizing dynamical phase of matter \cite{Nandkishore2015,Abanin2019} related to localization in Fock space \cite{Altshuler1997}. Another highly controllable system which allows individual site addressing and imaging is provided by arrays of Rydberg atoms; starting from high-energy states experiments have revealed the surprising existence of long-lived periodic revivals  \cite{Schauss2012,Labuhn2016,Bernien2017}, dubbed `quantum many-body scars' \cite{Turner2018,Turner2018b,Moudgalya2018,Khemani2019,Ho2019,Choi2019}. These discoveries have foundational implications for our understanding of how isolated quantum systems reach thermal equilibrium \cite{Deutsch1991,Srednicki1994,Rigol2008,Eisert2015} and whether nonequilibrium dynamics can display universality akin to that seen at equilibrium phase transitions \cite{Polkovnikov2011,nicklas15,Erne2018,Heyl2018,Link2020}. There are also technological implications because quantum information processors are themselves out-of-equilibrium many-particle systems \cite{Smith2019}.

In this paper we introduce the idea of quantum many-body caustics. Like the above-mentioned phenomena, caustics occur in out-of-equilibrium quantum many-body wavefunctions but unlike scars, which arise from individual eigenstates, these come from the interference of multiple eigenstates. Caustics are the result of wave bifurcations, which are violent events where waves are born or die. This results in a locally large amplitude that diverges in the classical (mean-field) limit and caustics can thus dominate wavefields. Remarkably, certain shapes of caustic are \textit{structurally stable} against perturbations and hence occur generically. These form a hierarchy described by catastrophe theory \cite{thom75,arnold75,Zeeman77}. They also obey scaling laws in which each member of the hierarchy has its own set of scaling exponents comprised of Arnold and Berry indices \cite{berry81}. This universality, like that in equilibrium phase transitions, ultimately derives from the presence of singularities.  In previous work we have considered caustics in integrable systems such as the Bose-Hubbard (BH) dimer (bosonic Josephson junction)  \cite{odell12,Mumford2017,Goldberg2019}  and the transverse field Ising model with infinite-range \cite{Mumford2019} and short-range \cite{Kirkby2019} interactions, respectively. In particular, the latter paper showed that light-cone wave fronts on a spin chain are in fact caustics arising from the coalescence of two waves (fold catastrophe). This allowed us to predict new properties of cones including a nontrivial scaling with respect to the spin coupling strength and the existence of a hierarchy of new structures such as double cones when the spin-spin coupling symmetry is broken, e.g.\ in the anisotropic XY model where three waves coalesce (cusp catastrophe).  

The present work has two goals: firstly to explore higher catastrophes beyond the fold and cusp in Fock space, and secondly to see if caustics occur in nonintegrable systems and hence are a generic feature of many-body wavefunctions. For this purpose we choose the BH trimer which is simple enough to permit exact numerical solutions for moderate particle numbers, and even analytic calculations in the case of a $\delta$-kicked Hamiltonian, such that the caustics can be easily identified, and yet is nonintegrable (classically chaotic) with direct connections to current experiments in optical lattices and spin-1 Bose-Einstein condensates (BECs)  \cite{Evrard2021}.  The power of the catastrophe theory approach, which derives from its origins in topology, lies in its ability to make robust qualitative predictions for the hierarchy of allowed caustics and their morphologies. In this paper we verify these mathematically rigorous predictions with detailed numerical and analytical calculations.

\begin{figure}\centering
	\includegraphics[width=0.6\columnwidth]{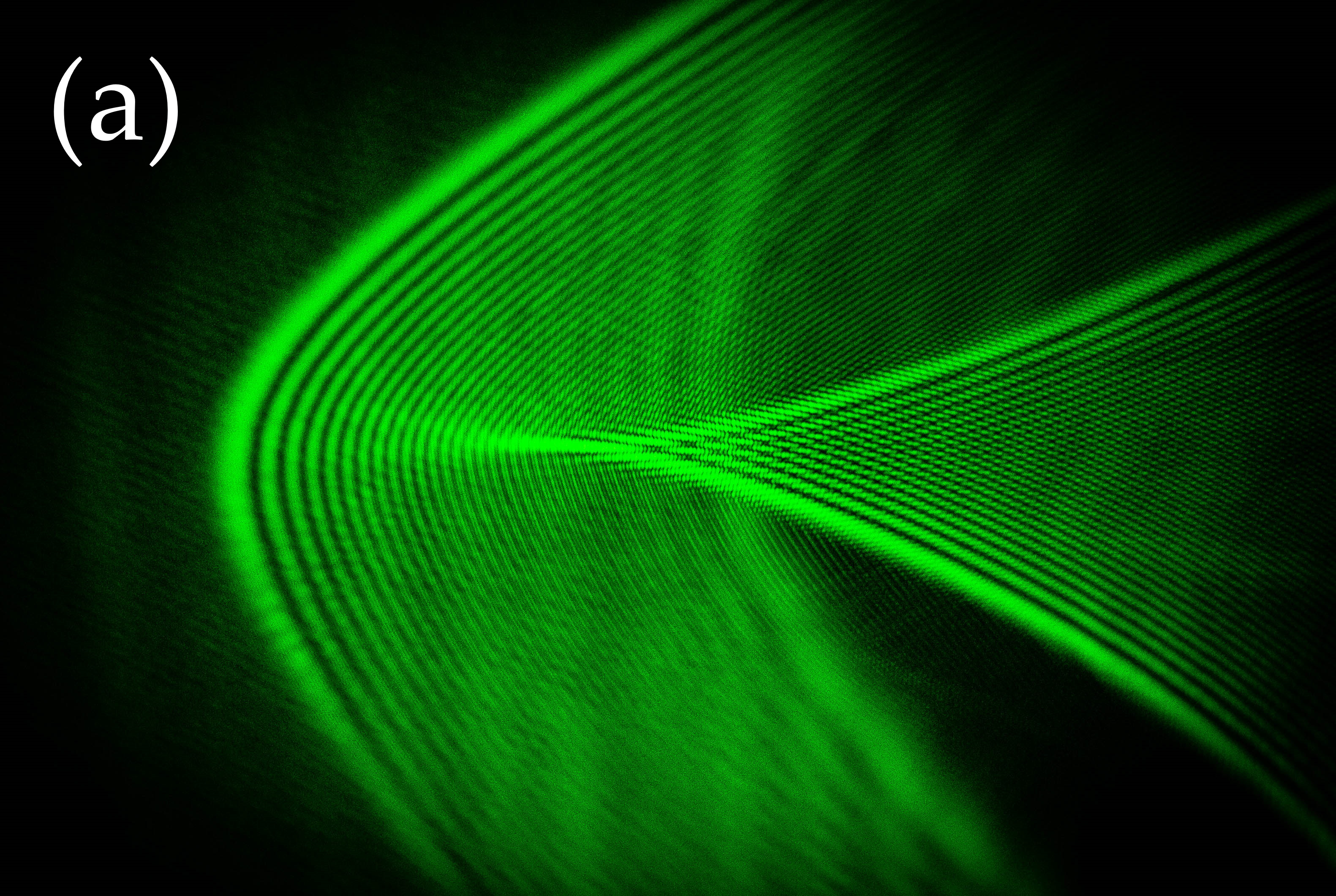}
	\includegraphics[width=0.6\columnwidth]{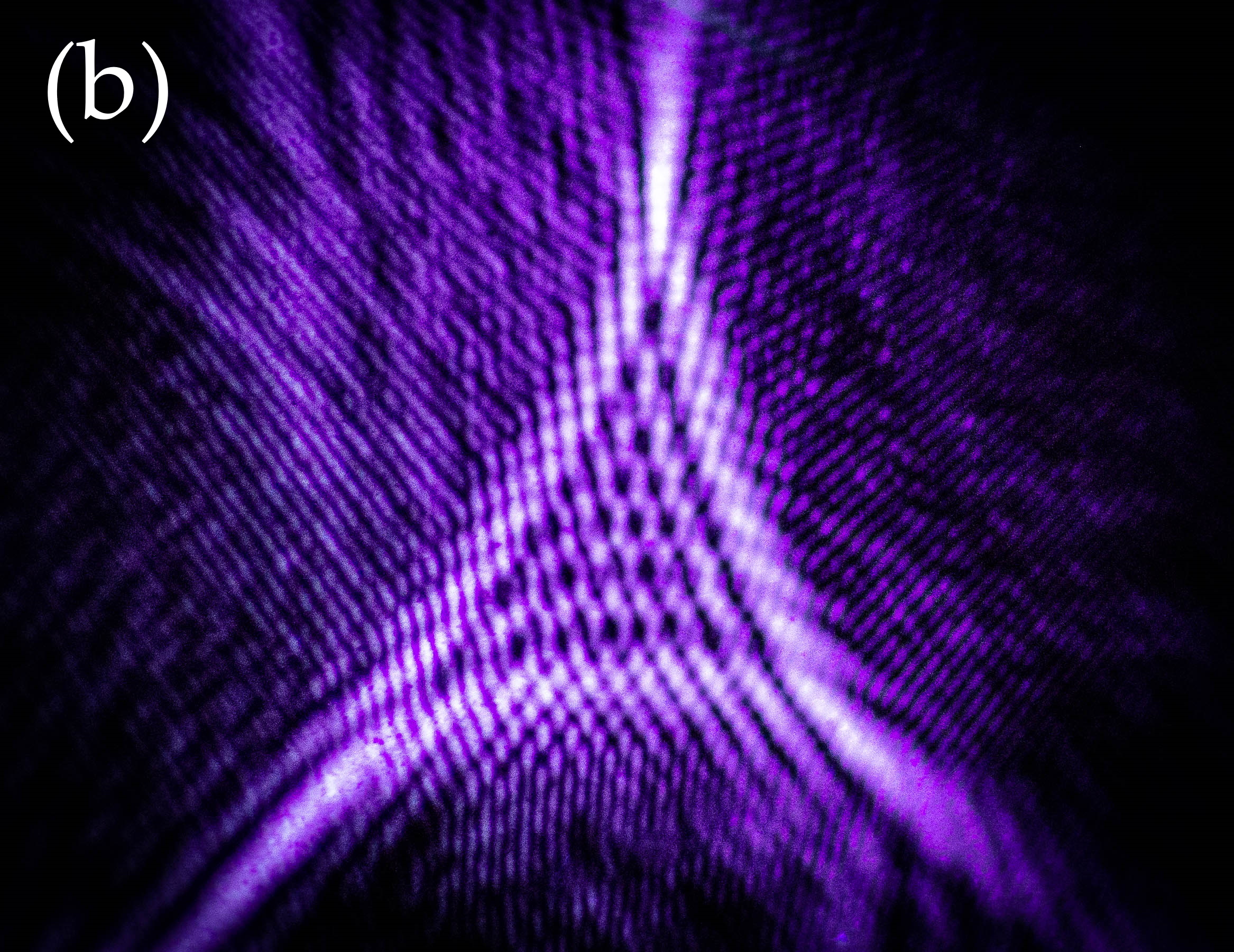}
	\includegraphics[width=0.6\columnwidth]{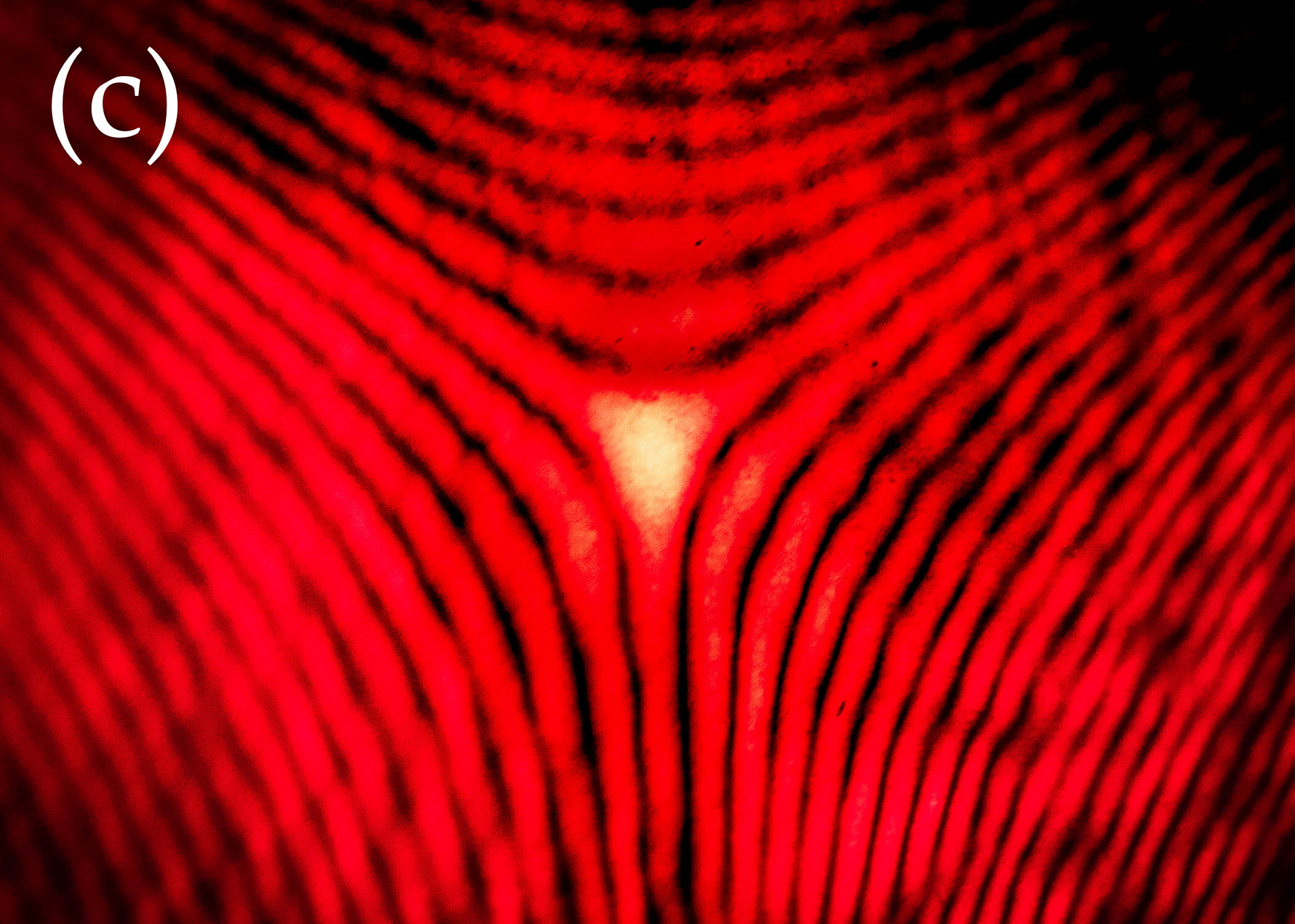}
	\caption{\label{fig:causticgallery} Real optical caustics made by shining a laser pointer through a water droplet of radius $\sim 1$mm and photographed on a screen at a distance of several meters. The droplet has a triangular perimeter imposed by placing it in a triangle cut out of tape stuck on a microscope slide. This mimics the triangular Fock space found in the BH trimer model and leads to the same families of caustics. \textbf{Panel (a):} Hyperbolic umbilic,  \textbf{Panel (b):}  Elliptic umbilic, \textbf{Panel (c):} Elliptic umbilic near its most singular point. What we see in these photos are slices through three dimensional catastrophes that also contain lower catastrophes: the hyperbolic umbilic contains a single cusp (which is the only structurally stable catastrophe in 2D and is dressed by the Pearcey function wave catastrophe) and the elliptic umbilic contains three cusps. Taking a 1D slice across a cusp gives the simplest catastrophe of all, the fold catastrophe (whose wave pattern is the Airy function), and in fact folds and cusps are the basic elements in `light cones' in Ising and XY models, see Fig.\ 2 in reference \cite{Kirkby2019}. These images and the methods used to make them were inspired by the experiments reported in reference \cite{BerryElliptic1978}. The different colors arise from using three different color lasers.}
\end{figure}

Some images of caustics are shown in Fig.\ \ref{fig:causticgallery}. They were made by shining laser pointers through water droplets and photographing the resulting pattern on a screen. These striking morphologies occur without special tuning: they are structurally stable and hence occur in `typical' or generic situations \cite{BerryElliptic1978,BerryUpstill1980}. For example, an isolated point focus arising from a perfect lens is not structurally stable in two or more dimensions and instead evolves into an extended caustic in the presence of aberrations. Note that higher catastrophes contain the lower ones.  Indeed, the structurally stable catastrophe in 2D is the cusp and in both panels (a) and (b) we can see, respectively, one and three cusp shapes embedded in these slices through what are actually patterns in 3D space. The wave catastrophe dressing the cusp is known as the Pearcey function \cite{pearcey46}, and is defined through a so-called `diffraction integral' (which can be viewed as an elementary form of path integral, see Eq.\ \ref{eq:wavecatastrophe}]) \cite{handbook}. The three-fold forked pattern seen in panel (c) of Fig.\ \ref{fig:causticgallery} corresponds to a section of the elliptic umbilic catastrophe which can be described by a pair of separable Airy functions \cite{BerryElliptic1978}. Zooming out to larger scales the fringes bunch up so that the caustics appear as singular intersecting lines with diverging intensity, but at the wavelength scale we see that they are softened by interference. At the finest scales (not shown) the interference pattern contains a network of vortices  \cite{berry81} that we predict are also present in light cones and many-body caustics more generally.

Everyday optical examples of caustics include rainbows, bright lines on the bottom of swimming pools, and twinkling starlight  \cite{berry77}. Less everyday examples include Cherenkov radiation \cite{Ginzburg2005,Gladush2008} and gravitational lensing \cite{Nye1999}.   Caustics occur in hydrodynamics as ship wakes (as first understood by Kelvin \cite{Kelvin1905}), and also as tsunamis \cite{Berry2005,Degueldre2016} and tidal bores \cite{berry18}, and have been identified as one of the causes of freak waves and extreme events \cite{white98,hohmann10,Onoratoab13,Adcock14,Heller21}. This has inspired recent studies comparing freak waves in linear and nonlinear optical systems \cite{Solli2007,Arecchi2011,Akhmediev2013,Marsal2014,Mathis2015,Pierangeli15,Mattheakis2016,Safari17,Zannottibook}. Cosmology is another field where caustics appear because smooth distributions of matter evolving under gravity will generically develop caustics (singularities in the density distribution) and this has been proposed as an explanation for the large scale structure of the universe \cite{Arnold1982,Feldbrugge2018}.

Caustics also occur in quantum waves. Historically, rainbows have been studied in nuclear scattering \cite{Silveira1973}, and more recently have been observed in electron microscopy \cite{Petersen2013}, atom optics  \cite{Rooijakkers2003,Simula2013,Rosenblum2014,Mossman2021}, and in the experiment described in reference \cite{Huckans09} a cusp caustic was recorded in the time-dependent atom density distribution of a dilute BEC moving in a 1D optical lattice. In these examples the matter waves are adequately described by the single-particle Schr\"{o}dinger wave equation, but caustics are not limited to this scenario and also arise in solutions of the Gross-Pitaevskii equation (nonlinear Schr\"{o}dinger equation) which can equally describe self-interacting BECs \cite{Plestid2018} and nonlinear optics in a fibre \cite{Solli2007}. Nevertheless, these cases still correspond mathematically to the ``classical'' wave scenario, i.e.\ an actual or effective single-particle wave, whether linear or nonlinear.  The many-body caustics we study in this paper are a new kind of object: they live in Fock space which is fundamentally discrete. This second quantization is crucial for regulating the singularities present in the classical (mean-field) limit \cite{odell12}, and thus they are in the same spirit as quantized phase singularities in quantum optics  \cite{Leonhardt02,Berry04}.
  
The BH dimer and trimer models we focus on here give rise to dynamical caustics that live in (1+1)- and (2+1)-dimensional spaces: 1- and 2-dimensional Fock space plus time, respectively (assuming total number conservation). Catastrophe theory predicts, and we shall indeed find, that the dimer displays discretized fold and cusp catastrophes which are the simplest two in Thom's hierarchy, while the trimer hosts the codimension-3 catastrophes: the hyperbolic elliptic and umbilic catastrophes. Our results are the first steps in elaborating the hierarchy of many-body caustics and, by fully incorporating quantum fluctuations, they go beyond previous applications of catastrophe theory to many particle systems such as equilibrium thermodynamic   \cite{Schulman73,stewart2012} and quantum \cite{gilmore77,gilmore78,gilmore77b,Gilmore1981} phase transitions  in  the Lipkin-Meshkov-Glick and Dicke models, which were limited to mean-field theory.

The plan for the rest of this paper is as follows:  in Section \ref{sec:Model} we present the BH dimer and trimer models 
 and in Section \ref{sec:Cats} explain the relevant parts of catastrophe theory, including the associated interference patterns. In Sections \ref{sec:BHdimer} and \ref{sec:BHtrimer} we present a gallery of images of caustics in the dimer and trimer found using exact numerical solutions of the quantum equations of motion. However, although we can numerically compute the wavefunction for $N \sim 150$ particles, we are unable to obtain analytic mappings onto the canonical catastrophe wavefunctions for the trimer because it is non-integrable. In order to provide some analytic examples, in Section \ref{sec:kicked} we instead study $\delta$-kicked dynamics where the interactions are flashed on and off such that the mapping can be achieved analytically (interactions can be engineered in cold atom experiments using Feshbach resonances, see e.g.\ \cite{Abeelen99}). In Section \ref{sec:Stability} we go beyond the quantum phase model (rigid pendulum model), which assumes all three modes are significantly occupied, and find corrections that break circular symmetry in Fock space in favor of triangular symmetry and identify the particular sub-family of the $X_{9}$ catastrophe at work. In order to do this we introduce a path integral representation for the wavefunction. In Section \ref{sec:interactions} we compare repulsive and attractive interactions and discuss the crucial role interactions play in the formation of caustics in BH dynamics.  Finally, in Section \ref{sec:Discussion} we give our conclusions. There are also three appendices which contain details of some of the calculations, including a derivation of the path integral. As the caustics described in this paper live in Fock space, their main experimental signature would be singularity dominated fluctuations in mode populations, e.g.\ populations of sites in an optical lattice or populations of spin states in a spinor gas.  Experimental considerations are discussed mainly in Sections \ref{sec:Model} and \ref{sec:Discussion}.

\section{\label{sec:Model} Two- and Three-Mode Bose-Hubbard models: experiment and theory}

We choose the BH model to illustrate the basic ideas of many-body caustics because it is a
key model in statistical physics \cite{Fisher89,Sachdevbook} that describes interacting bosons hopping on a lattice, and has been realized in celebrated experiments using ultracold atoms \cite{Jaksch98,Greiner2002}.  Due to the ability of these experiments to create sudden quenches, the dynamical states of the BH model have received ongoing theoretical  \cite{Altman2002,Isella2005,Schutzhold2006,Kollath2007,Rigol2007,Rigol2008,Dziarmaga2012,Barmettler2012,Daley2012,Lacki2013, Kordas2015,Vicentini2018,Cosme2018,Fitzpatrick2018,Nagao19,Mokhtari2021} and experimental \cite{Fallani2004,Tuchman2006,Will2010,Trotzky2012,Cheneau2012,Fukuhara2013,Meinert2013,Braun2015,Cosme2018b,Boulier2019} attention including: studies of the timescales for many-body quantum revivals and the establishment of coherence, light-cone-like propagation of correlations, effective Hamiltonians in periodically driven ``Floquet'' systems, and relaxation to equilibrium, to name just a few. The BH dimer and trimer models are particular cases that consider two and three lattice sites (modes), respectively. The dimer describes bosonic Josephson junctions \cite{Milburn97,smerzi97,Vardi2001,Paraoanu01,Pitaevskii01,Graefe2007,Chuchem2010,Veksler15} that have been realized experimentally with BECs trapped in double-well potentials \cite{albiez05,Levy07,Leblanc2011,Folling2007,Ryu2013,trenkwalder16}, and also with spinor BECs  exhibiting the internal version of the Josephson effect  \cite{zibold10}. Additionally, the same Hamiltonian  describes trapped ions with two internal states and long-range interactions  \cite{Das06,Britton12,Jurcevic14,Richerme14}.

The BH trimer model describes BECs in triple-well potentials as well as spin-1 BECs \cite{Law1998,Zhang05,StamperKurn2013} where the atoms share a common external state (as in a tight trap).  Spin-1 BECs have been realized in experiments on  $^{23}$Na \cite{Stenger98,Black2007,Farolfi2019,Evrard2021} and $^{87}$Rb  \cite{Sadler2006,Gerving2012,Linnemann2016,Lange2018,Kunkel2018,Kunkel2019} where the three internal states are provided by the Zeeman sublevels of the F=1 hyperfine manifold. Due to conservation of the angular momentum during collisions, these models do not naturally realize the full trimer model, but this can enforced by applying an integrability breaking RF field that drives transitions between the $\vert m \rangle$ and $\vert m \pm 1 \rangle$ states \cite{Evrard2021,Rautenberg2020}. Like the dimer, these systems can display macroscopic quantum self-trapping \cite{Franzosi2003,Liu2007,Buonsante2010}. However, unlike the dimer  the trimer is nonintegrable and its classical dynamics exhibits chaos \cite{Franzosi2003,GarciaMarch2018,Thommen2003,Kolovsky2007,Hiller2009,Viscondi2011,Rautenberg2020,Wittmann2021} giving behaviour qualitatively closer to the many-site model. The trimer also accomodates next-to-nearest-neighbor interactions which are important when the atoms have dipole-dipole interactions \cite{Muller2011,Peter2012,DellAnna2013,Baier2016}.

\subsection{\label{subsec:TwoModeModel} Two-Mode Equations of Motion}
We first consider the BH dimer, which will form up to (1+1)-dimensional caustics in the dynamics \cite{odell12,Mumford2017}. The Hamiltonian is \cite{Leggett01,gati07}
\begin{equation}
	\hat{H}^{\text{dimer}}=-J\left(\hat{a}_l^\dagger\hat{a}_r+\hat{a}_r^\dagger\hat{a}_l\right)+U\left(\hat{a}_l^\dagger\hat{a}_l-\hat{a}_r^\dagger\hat{a}_r\right)^2\;,
\end{equation}
where $\hat{a}_{l/r}^{(\dagger)}$ annihilates (creates) a particle in the left/right well, $J$ is the hopping energy and $U$ is the on-site interaction energy between particles. The operators obey the usual bosonic commutation relations $[\hat{a}_{i},\hat{a}_{j}^{\dagger}]=\delta_{ij}$, where $i$ and $j$ correspond to either $l$ or $r$.

In this paper we study caustics that form in Fock space. The Fock states $\vert n \rangle$ are eigenstates of the half-number-difference operator $\hat{n}\equiv(\hat{a}_l^\dagger \hat{a}_l-\hat{a}_r^\dagger\hat{a}_r)/2$. A general quantum state can be expanded as
\begin{equation}\label{eq:fockexpn}
	\ket{\Psi(t)}=\sum\limits_nc_n(t)\ket{n}.
\end{equation}
Inserting Eq.\ \eqref{eq:fockexpn} into the time-dependent Schr\"{o}dinger equation, $\mathrm{i}\hbar\partial_t\ket{\Psi(t)}=\hat{H}\ket{\Psi(t)}$, we obtain a set of $N+1$ coupled differential equations for the Fock-space amplitudes $c_{n}(t)$ which we refer to as the generalized Raman-Nath (RN) equations (a similar set of differential difference equations were derived by Raman and Nath in the context of dynamical diffraction  \cite{RamanNath,Berry1966,odell01}),
\begin{align}
	\mathrm{i}\hbar \dot{c}_n(t)=&\;4Un^2c_{n}(t)-Jc_{n-1}(t)\sqrt{\frac{N^2}{4}+\frac{N}{2}-n^2+n}\nonumber\\&-Jc_{n+1}(t)\sqrt{\frac{N^2}{4}+\frac{N}{2}-n^2-n} \label{eq:TwoModeRN}
\end{align}
where the dot represents a time derivative.

The mean-field limit is given by the Heisenberg substitution rules, replacing operators with complex amplitudes \cite{Thommen2003,Mossmann2006}
\begin{equation}
	\hat{a}_{r/l}\to\sqrt{N_{r/l}}\mathrm{e}^{\mathrm{i}\theta_{r/l}}\;,
\end{equation}
and leads to the Hamiltonian
\begin{equation}\label{eq:HMF2Mode}
	H_{\text{MF}}^\text{dimer}=4Un^2-J\sqrt{N^2-4n^2}\cos\phi\;,
\end{equation}
where $\phi=\theta_{r}-\theta_{l}$ is the phase difference between the two modes and is the conjugate variable to $n$. $H_{\text{MF}}^\text{dimer}$ describes a classical nonrigid pendulum where $n$ is angular momentum and $\phi$ angular position. The variable length of the nonrigid pendulum is accounted for by the square root factor \cite{smerzi97}.  In fact, because they have simultaneously well-defined position and momentum as a function of time, the mean-field solutions are analogous to geometric rays. Hamilton's equations of motion give Josephson's equations for two coupled superfluids \cite{Raghavan1999}
\begin{align}
	\dot{n}=&\;-\frac{1}{\hbar}\frac{\partial}{\partial \phi} H_{\text{MF}}^\text{dimer} =-\frac{J}{\hbar} \sqrt{N^2-4n^2}\sin\phi \label{eq:hamilton1} \\
	\dot{\phi}=&\;\frac{1}{\hbar}\frac{\partial}{\partial n} H_{\text{MF}}^\text{dimer} = 8 \frac{U}{\hbar}n + \frac{J}{\hbar} \frac{8n}{\sqrt{N^2-4n^2}} \cos\phi\;. \label{eq:hamilton2}
\end{align}

Attempts to semiclassically quantize the mean-field problem are complicated by the appearance of both number and phase variables in the potential energy term, meaning that the Hamiltonian is not separated into the sum of a `kinetic' term proportional to $n^2$ and a `potential' term $V(\phi)$,  but can be pushed through with some care \cite{Graefe2007,Anglin2001}. However, providing the population difference is always small in comparison to the total particle number ($n \ll N$) the square root term in $H_{\text{MF}}^\text{dimer}$ can be set to unity reducing it to that of a standard rigid pendulum. This is the relevant Hamiltonian in 
atomic BECs in optical lattices when there are many atoms per site \cite{Cataliotti2001,Orzel2001,Hadzibabic2004,Schori2004,Xu2006,Schweikhard2007}, and also in
superconducting Josephson junctions where it is known as the \textit{quantum phase model} (QPM) \cite{Otterlo1995,Fazio01}. In this paper we shall sometimes make use of the QPM for simplicity but  will also consider corrections to it (we will see in Section \ref{sec:Stability} that this can make a difference to the caustics that occur).

\subsection{\label{subsec:ThreeModeModel} Three-mode Equations of Motion}
The Hamiltonian for the BH trimer can be written as
\begin{align}\label{eq:QuantumThreeMode}
	\hat{H}=&\;-K_{L}(\hat{a}_1^\dagger\hat{a}_2+\hat{a}_2^\dagger\hat{a}_1)-K_{R}(\hat{a}_2^\dagger\hat{a}_3+\hat{a}_3^\dagger\hat{a}_2)\\
	&\nonumber-K_X(\hat{a}_3^\dagger\hat{a}_1+\hat{a}_1^\dagger\hat{a}_3)+\frac{U}{2}\sum_{i=1}^3\hat{n}_i(\hat{n}_i-1)+\sum_{i=1}^3\epsilon_i\hat{n}_i \ .
\end{align}
The parameters $K_L,K_R,K_X$ correspond to the hopping energies between wells 1 and 2, 2 and 3, and 1 and 3, respectively.  The $\epsilon_i$ allow for different well depths.

In its linear configuration ($K_X=0$), the BH trimer system has been used as a spatial model for rapid adiabatic passage by controlling the well depths $\epsilon_{i}(t)$ as functions of time  \cite{Greentree2004,Cole2008,Rab2008,Opatrny2009,Bradly2012,Olsen2014}. Indeed, adding a magnetically-induced tilt to the lattice allows additional rich behaviour including control of correlations \cite{Guo2014,Dutta2019}. The linear configuration has also been studied from the point of view of an ultracold atom transistor-like device \cite{Stickney2007,Wilsmann2018,Zhang2019,Caliga2016}. In its fully-connected triangular configuration ($K_X \neq 0$), the trimer system provides a minimal model for superfluid circuits and discrete vortices \cite{Arwas2014,Arwas2015,Gallemi2015,Lee2006}.  Both chain and triangle have been discussed in the context of quantum steering \cite{Olsen2015a,Olsen2015b,Kalaga2016,Kalaga2017}. To the best of our knowledge the BH trimer  has not been studied experimentally using triple-well BECs, although detailed proposals with tuneable hopping and interaction parameters via Feshbach resonances exist   \cite{Olsen2018}. As mentioned above, spin-1 BECs  provide another physical system where the three mode BH model can provide the natural theoretical description  \cite{Rautenberg2020}.

When $U=0$, the mean-field  BH trimer model exhibits regular dynamics, while for nonzero interactions it exhibits chaos indicating nonintegrability. Close to the ground state chaotic trajectories are mixed with islands of regular dynamics \cite{Franzosi2003,Mossmann2006}. Energy level statistics in the quantum version tell a similar story: in general they obey neither the Poisson nor Wigner distributions but are better described by a Berry-Robnic distribution \cite{Kolovsky2020} which is a signature of a classical limit  containing both regular and chaotic dynamics \cite{berry84}. The significance of this for the results we present below is that we find caustics in a nonintegrable model even though caustics are usually associated with integrable behavior \cite{Berry_leshouches2}. We therefore conjecture that the caustics are at least stable against weak integrability breaking terms as described by the famous Kolmogorov-Arnold-Moser theorem \cite{Arnoldbook}.

\begin{figure}[t]\centering
	\includegraphics[width=0.6\columnwidth]{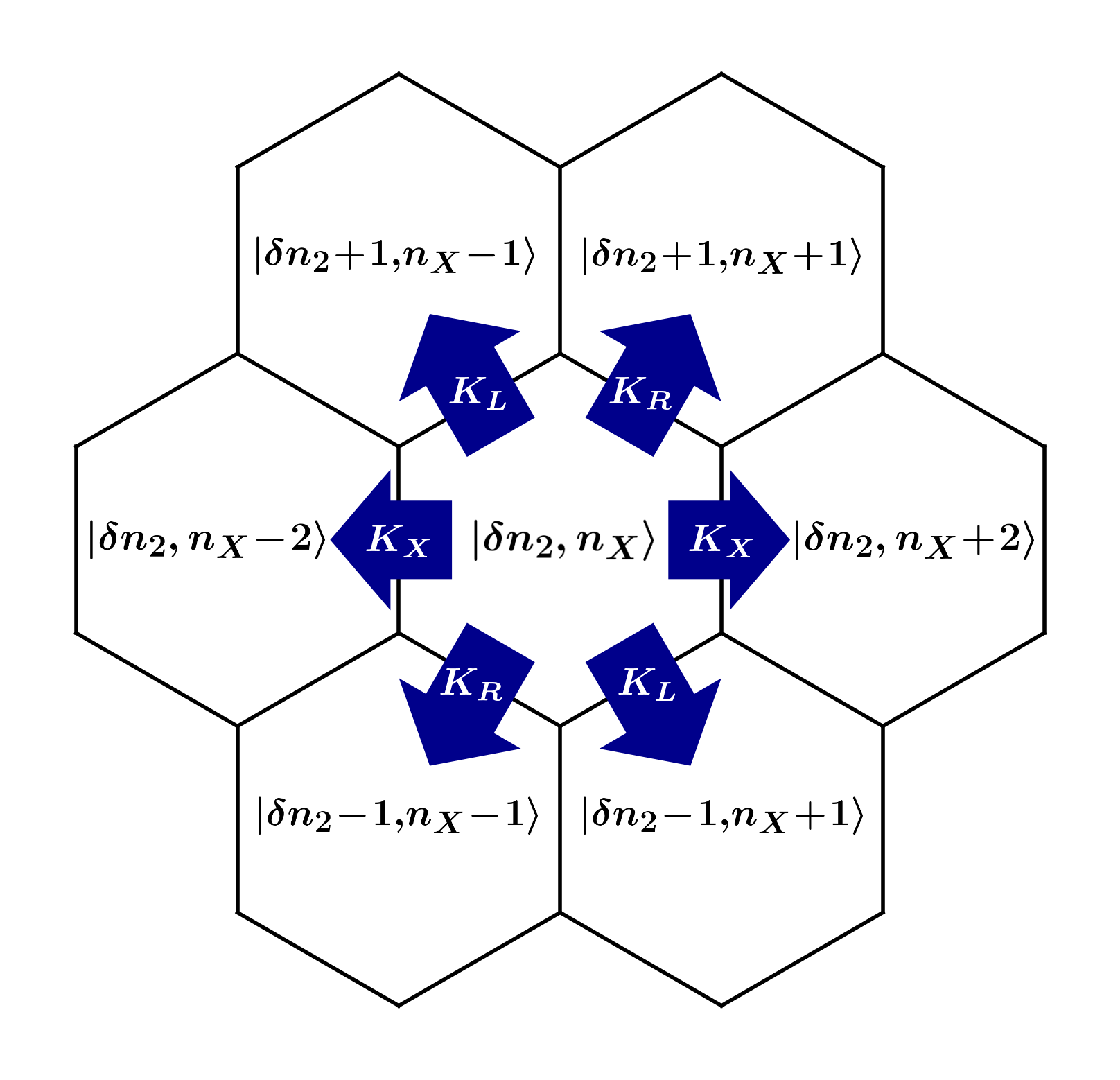}
	\caption{\label{fig:Introplot} A small region of Fock space for the BH trimer. It can be tiled by hexagonal cells in the coordinates $\{\delta n_2,n_X\}$ we use throughout this paper. Hopping terms shift $\delta n_2$ by only 1 unit, while $n_X\equiv n_1-n_3$ can change by 1 or 2 units. If the reader zooms in on the BH trimer Fock space figures in this paper they will see this underlying hexagonal lattice. This discretization is a true feature of second quantization not present in classical wave catastrophes (it is
not a false effect due to pixilation of the images).}
\end{figure}

The quantum many-body state can be expanded as 
\begin{align}\label{eq:FockRepresentation}
\ket{\Psi(t)}=&\;\sum_{n_1n_2n_3}M_{n_1n_2n_3}(t)\ket{n_1,n_2,n_3}\\
=&\;\sum_{n_1n_X}M_{\delta n_2n_X}(t)\ket{\delta n_2,n_X}    \label{eq:3modeFock}
\end{align}
where  we have introduced new Fock space coordinates $\delta n_2\equiv n_2-N/3$ and $n_X\equiv n_1-n_3$, and have assumed the total particle number is conserved so that one of the sums is eliminated (the variables $\delta n_2$ and $n_X$ are similar to those used by Arwas \textit{et al} \cite{Arwas2014}).  The allowed Fock space is then triangular in shape and contains $(N+2)(N+1)/2$ states. The schematic in Fig.\ \ref{fig:Introplot} depicts a small region of it. Since each Fock state is coupled to six others by all the possible hopping terms, Fock space can be tiled by a hexagonal pattern as shown.

Inserting Eq.\ (\ref{eq:3modeFock}) into Schr\"{o}dinger's equation gives the following generalized Raman-Nath equations for the Fock space amplitudes $M_{ij}(t)$ (we have put all $\epsilon_i=0$)
	{\small\begin{align}  
	\mathrm{i}\hbar & \dot{M}_{\delta n_2,n_X}(t) = \nonumber \\ & -{\displaystyle\frac{K_L}{\sqrt{2}}\sqrt{\left(\tfrac{N}{3}+\delta n_2+1\right)\left(\tfrac{2N}{3}-\delta n_2+n_X\right)}}M_{\delta n_2+1,n_X-1} \nonumber 
\\ & -\frac{K_L}{\sqrt{2}}\sqrt{\left(\tfrac{N}{3}+\delta n_2\right)\left(\tfrac{2N}{3}-\delta n_2+n_X+2\right)}M_{\delta n_2-1,n_X+1} \nonumber \\
	&-\frac{K_R}{\sqrt{2}}\sqrt{\left(\tfrac{N}{3}+\delta n_2\right)\left(\tfrac{2N}{3}-\delta n_2-n_X+2\right)}M_{\delta n_2-1,n_X-1} \nonumber 
 \\ & -\frac{K_R}{\sqrt{2}}\sqrt{\left(\tfrac{N}{3}+\delta n_2+1\right)\left(\tfrac{2N}{3}-\delta n_2-n_X\right)}M_{\delta n_2+1,n_X+1} \nonumber  \\
	&-\frac{K_X}{2}\sqrt{\left(\tfrac{2N}{3}-\delta n_2+n_X+2\right)\left(\tfrac{2N}{3}-\delta n_2-n_X\right)}M_{\delta n_2,n_X+2} \nonumber \\ & -\frac{K_X}{2}\sqrt{\left(\tfrac{2N}{3}-\delta n_2-n_X+2\right)\left(\tfrac{2N}{3}-\delta n_2+n_X\right)}M_{n_2,n_X-2}\nonumber\\
	&+\frac{U}{4}\biggl[3\delta n_2^2+n_X^2\biggr]M_{\delta n_2,n_X}. \label{eq:RamanNath}
		\end{align}}

The mean-field approximation is obtained as above by replacing operators with complex amplitudes, $\hat{a}_i\to \sqrt{n_i}\mathrm{e}^{\mathrm{i}\phi_i}$. The resulting Hamiltonian is,
\begin{align}
	H_{\text{MF}}=&\;-2K_{L}\sqrt{n_1n_2}\cos(\varphi_2-\varphi_1)\nonumber\\&-2K_{R}\sqrt{n_2n_3}\cos(\varphi_3-\varphi_2)\label{eq:HMF}\\&-2K_{X}\sqrt{n_3n_1}\cos(\varphi_1-\varphi_3)\nonumber\\&+\frac{U}{2}\sum_{i=1}^3n_i(n_i-1)+\sum_{i=1}^3\epsilon_in_i\nonumber.
\end{align}
Like in the quantum case, we can change our coordinates to $\delta n_2$ and $n_X$ and eliminate the third member due to number conservation. However, in the mean-field problem we also require the phase variables conjugate to the number variables and these are  $\phi_X\equiv\frac{1}{2}(\varphi_1-\varphi_3)$ and $\phi_{C}\equiv\frac{1}{2}(2\varphi_2-\varphi_1-\varphi_3)$, respectively. The third phase variable $\Theta \equiv \varphi_1+\varphi_2+\varphi_3$ is irrelevant to the mean-field dynamics studied here, and in the quantum wavefunction becomes a global phase. The resulting mean-field equations of motion in these variables come out to be:
	{\small
	\begin{align} 
		\dot{n}_X  = & -K_L\sqrt{2\left(\tfrac{N}{3}+\delta n_2\right)\left(\tfrac{2N}{3} -\delta n_2+n_X\right)}\sin\left(\phi_X-\phi_C\right) \nonumber \\ & -K_R\sqrt{2\left(\tfrac{N}{3}+\delta n_2\right)\left(\tfrac{2N}{3}-\delta n_2-n_X\right)}\sin\left(\phi_X+\phi_C\right) \nonumber \\
		& -2K_X\sqrt{\left(\tfrac{2N}{3}-\delta n_2\right)^2-n_X^2}\cos\left(2\phi_X\right)  \label{eq:classicalEOM1}  \\
		\dot{\delta n}_2  = & K_L\sqrt{2\left(\tfrac{N}{3}+\delta n_2\right)\left(\tfrac{2N}{3}-\delta n_2+n_X\right)}\sin\left(\phi_X-\phi_C\right) \nonumber \\ & -K_R\sqrt{2\left(\tfrac{N}{3}+\delta n_2\right)\left(\tfrac{2N}{3}-\delta n_2-n_X\right)}\sin\left(\phi_X+\phi_C\right)  \\
		\dot{\phi}_X  = & \frac{U}{2}n_X-K_L\frac{\left(\frac{N}{3}+\delta n_2\right)\cos(\phi_X-\phi_C)}{\sqrt{2\left(\frac{N}{3}+\delta n_2\right)\left(\frac{2N}{3}-\delta n_2+n_X\right)}} \nonumber \\ &+K_R\frac{\left(\frac{N}{3}+\delta n_2\right)\cos(\phi_X+\phi_C)}{\sqrt{2\left(\frac{N}{3}+\delta n_2\right)\left(\frac{2N}{3}-\delta n_2-n_X\right)}} \nonumber \\ & +K_X\frac{n_X\cos(2\phi_X)}{\sqrt{\left(\frac{2N}{3}-\delta n_2\right)^2-n_X^2}} \\
		\dot{\phi}_C=& \frac{3U}{2}\delta n_2-K_L\frac{\left(\frac{N}{3}-2\delta n_2+n_X\right)\cos(\phi_X-\phi_C)}{\sqrt{2\left(\frac{N}{3}+\delta n_2\right)\left(\frac{2N}{3}-\delta n_2+n_X\right)}} \nonumber \\ &-K_R\frac{\left(\frac{N}{3}-2\delta n_2-n_X\right)\cos(\phi_X+\phi_C)}{\sqrt{2\left(\frac{N}{3}+\delta n_2\right)\left(\frac{2N}{3}-\delta n_2-n_X\right)}} \nonumber \\ & +K_X\frac{\left(\frac{2N}{3}-\delta n_2\right)\cos(2\phi_X)}{\sqrt{\left(\frac{2N}{3}-\delta n_2\right)^2-n_X^2}} \, . \label{eq:classicalEOM4}
	\end{align}}
While the dimer case gave mean-field equations describing a non-rigid pendulum, the trimer case can be interpreted as describing three coupled anharmonic oscillators \cite{Mossmann2006}. Eqns.\ (\ref{eq:classicalEOM1})-(\ref{eq:classicalEOM4}) must in general be solved numerically, but a problem can potentially arise for trajectories that touch the boundaries of  Fock space where the square root factors in the denominators vanish. Physically, the boundaries correspond to situations where one of the modes is empty. We find empirically that this becomes less of a problem as $N$ is increased and almost never occurs in the semiclassical regime we consider in this paper where $N \sim 150$ because trajectories spend most of their time in the central region of Fock space.

\begin{table}[t]
	\begin{tabular}{|c|c|c|c|c|}\hline
		\textbf{Catastrophe} & \textbf{Symbol} & $n$ & $Q$ & $\Phi_{Q}(\mathbf{s};\mathbf{C})$\\\hline
		Fold & $A_2$ & 1 & 1 & $s^3+Cs$\\\hline
		Cusp & $A_3$ & 1 & 2 & $s^4 + C_2s^2 + C_1s$\\\hline
		Swallowtail & $A_4$ & 1 & 3 & $s^5+C_3s^3+C_2s^2+C_1s$\\\hline
		Butterfly & $A_5$ & 1 & 4 & $s^6+C_4s^4+C_3s^3+C_2s^2+C_1s$\\\hline
		$\begin{array}{c} \mathrm{Hyperbolic} \\ \mathrm{Umbilic} \end{array}$ & $D_4^+$ & 2 & 3 & $s_1^3+s_2^3+C_3s_1s_2+C_2s_2+C_1s_1$\\\hline
		$\begin{array}{c} \mathrm{Elliptic} \\ \mathrm{Umbilic} \end{array}$  & $D_4^-$ & 2 & 3 & $\begin{array}{c} 3s_1^2s_2-s_2^3+C_3(s_1^2+s_2^2) \\ +C_2s_2+C_1s_1 \end{array}$\\\hline
		$\begin{array}{c} \mathrm{Parabolic} \\ \mathrm{Umbilic} \end{array}$ & $D_5$ & 2 & 4 & $\begin{array}{c} s_2^4+s_1^2s_2+C_4s_2^2 +C_3s_1^2 \\ +C_2s_2+C_1s_1 \end{array}$\\\hline
	\end{tabular}
	\caption{Thom's seven elementary catastrophes, their symbols, and generating functions $\Phi_{Q}(\mathbf{s};\mathbf{C})$, organized by corank $n$, and codimension $Q$ \cite{BerryUpstill1980}.}
	\label{tab:catastrophetable}
\end{table}

\section{\label{sec:Cats} Wave Catastrophes}
Catastrophe theory describes the structurally stable singularities of gradient maps. This includes all theories that can be posed in terms of a minimum principle, e.g.\ the principle of stationary action, and hence applies to both classical and quantum mechanics. 
The seven elementary catastrophes introduced by Ren\'{e} Thom  \cite{thom75}  are listed in Table \ref{tab:catastrophetable}, and some examples of higher catastrophes are listed in Table \ref{tab:catastrophetable2}. They are organized by corank (number of ``state'' variables $\mathbf{s}=\{s_{1},s_{2},\ldots \}$ that label paths) and codimension (number of control parameters  $\mathbf{C}=\{C_{1},C_{2},\ldots \}$ which in our case is the dimension of Fock space plus time and any other parameters in the Hamiltonian). The key objects are the generating functions $\Phi_Q (\mathbf{s};\mathbf{C})$, and in physical applications they give the \textit{local action} close to the caustic. Stationary points $\partial_{s} \Phi_Q =0$ specify classical paths or rays, which in many-particle problems correspond to mean-
field solutions. Caustics occur where the action is stationary to higher order, i.e.\ $\partial^{2}_{s} \Phi_Q =0$ (in two or more dimensions this condition becomes the vanishing of the Hessian matrix), and thus are regions where
classical paths either coalesce or are born (bifurcations). The main point is that this can only happen in certain ways if the bifurcation is to be structurally stable against perturbations.

For example, in the case of light-like cones the action is $\Phi(k;x,t)=kx - \epsilon_{k}t/\hbar$, where $\epsilon_{k}$ is the dispersion relation for quasiparticles of wavenumber $k$  \cite{Kirkby2019}. The Lieb-Robinson bound, which gives the maximum speed of quasiparticles and hence defines the cone is \cite{Stephan2011,Calabrese2012}
\begin{equation}
v_{\text{\tiny{LR}}}=\max_k \left\vert \frac{\mathrm{d} \epsilon_k}{\mathrm{d} k} \right\vert
\end{equation}
which is exactly equivalent to the two conditions  $\partial_{s} \Phi_Q =0$ and $\partial^{2}_{s} \Phi_Q =0$ defining caustics.

 Each catastrophe has a `germ', which is the part of $\Phi_Q$ that remains when it is evaluated at the origin of control space $\mathbf{C}=0$. The germ characterizes the order of the singularity. The other terms show how the catastrophe `unfolds' as one moves away from the origin of control space. The corank 1 catastrophes, $A_{Q+1}$ are called the \textit{cuspoids}, and extend beyond the butterfly to the \textit{wigwam} and \textit{star} catastrophes (not listed). The remaining catastrophe types, $D_{Q+1}$, $E_{Q+1}$, and above are typically called \textit{umbilics}, referring to the classification of cubic forms near an umbilic point (a point on a surface with locally spherical curvature) which become the germs for these catastrophes.

\begin{table}[t]
	\begin{tabular}{|c|c|c|}\hline
		\textbf{Symbol} & $n$  & $\Phi_{Q}(\mathbf{s};\mathbf{C})$ \\\hline
		$A_{Q+1}$ & 1 & $s^{Q+2}+\sum_{i=1}^{Q-1} C_i s^i$\\\hline
		$D_{Q+1}^{\pm}$ & 2 & $s_1^Q\pm s_1s_2^2+C_{Q+1}s_2^2+\sum_{i=2}^{Q-1}C_is_1^i+C_{1}s_2$\\\hline
		$E_6$ & 2 & $s_1^3+s_2^4+C_5s_1s_2^2+C_4s_2^2+C_3s_1s_2+C_2s_2+C_1s_1$\\\hline
		$E_7$ & 2 & $\begin{array}{c}
		s_1^3+s_1s_2^3+C_6s_2^4+C_5s_2^3+C_4s_2^2\\+C_3s_1s_2+C_2s_1+C_1s_2
		\end{array}$\\\hline
		$E_8$ & 2 & $\begin{array}{c}
			s_1^3+s_2^5+C_7s_1s_2^3+C_6s_1s_2^2+C_5s_2^3\\+C_4s_1s_2+C_3s_2^2+C_2s_1+C_1s_2
		\end{array}$\\\hline
		$X_9^{\pm}$ & 2 & $\begin{array}{c}
			s_2^4+Ks_1^2s_2^2\pm s_1^4+C_7s_2^2s_1+C_6s_2s_1^2\\+C_5(s_2^2+s_1^2)
			+C_4(s_2^2-s_1^2)\\+C_3s_2s_1+C_2s_2+C_1s_1
		\end{array}$\\\hline
	\end{tabular}
	\caption{Catastrophe organization beyond Thom's seven elementary list \cite{Nye1999}, with general control space dimension $Q$. Many of the higher catastrophes do not have names, and are referenced by their group-theoretic symbol, yet are often split into cuspoids $A_{Q+1}$, and umbilics $\{D_{Q+1},E_{Q+1},X_9\}$.}
	\label{tab:catastrophetable2}
\end{table}

Catastrophes obey projection identities: higher catastrophes contain lower ones, e.g.\ the swallowtail contains two cusps and three fold lines when projected into two dimensions. It is not, however, guaranteed that catastrophes of high order (i.e. higher codimension and/or corank) contain all catastrophes of lower order. In section \ref{sec:kicked}, we will briefly discuss distinctions between families of the high-order catastrophe $X_9$, which have different projection identities ($X_9$ is the catastrophe that organizes all the structures we see in the BH trimer dynamics). With special tuning one could engineer focusing events with any shape, but catastrophe theory instead describes structurally stable caustics that result from natural  focusing, and so are more likely to appear generically.

The geometric ray theory (mean-field theory in Fock space) gives the basic shape of the caustic, but these ray sums give divergent amplitudes.  To remove these one should include interference and we enter the realm of wave catastrophes (diffraction integrals) \cite{BerryUpstill1980,Nye1999}. Each ray catastrophe is dressed by a characteristic wave interference pattern described by a diffraction integral,
\begin{equation}
	\Psi(\textbf{C})\propto\lambda^{n/2}\int...\int\mathrm{d}\textbf{s}\;\mathrm{e}^{\mathrm{i}\lambda\Phi_Q(\textbf{s};\textbf{C})}\; .
	\label{eq:wavecatastrophe}
\end{equation} 
This wavefunction lives in the space of control parameters (Fock space + time) and resembles a path integral where the generating function plays the role of the action and we integrate over state variables $\mathbf{s}$ which label paths. The parameter $\lambda$ acts as the inverse of Planck's constant, and in the BH model is proportional to the total number of particles $N$. The precise connection to path integrals will be explained in Section \ref{sec:Stability} and Appendix \ref{Appdx:PathIntegral}. It is interesting to note that while wave theory removes geometric singularities it also introduces new ones, namely phase singularities where the phase takes all values and hence is undefined  \cite{Nye2009}. These are more commonly known as dislocations in optics and vortices in condensed matter systems. A genuinely new feature of caustics in quantum many-body wavefunctions in comparison to classical waves is that phase singularities are removed by second-quantization because it discretizes the vortices \cite{Mumford2019}. We shall not dwell on this `fine structure' of caustics here, and focus instead on their gross features.

\begin{figure}\centering
	\includegraphics[width=\columnwidth]{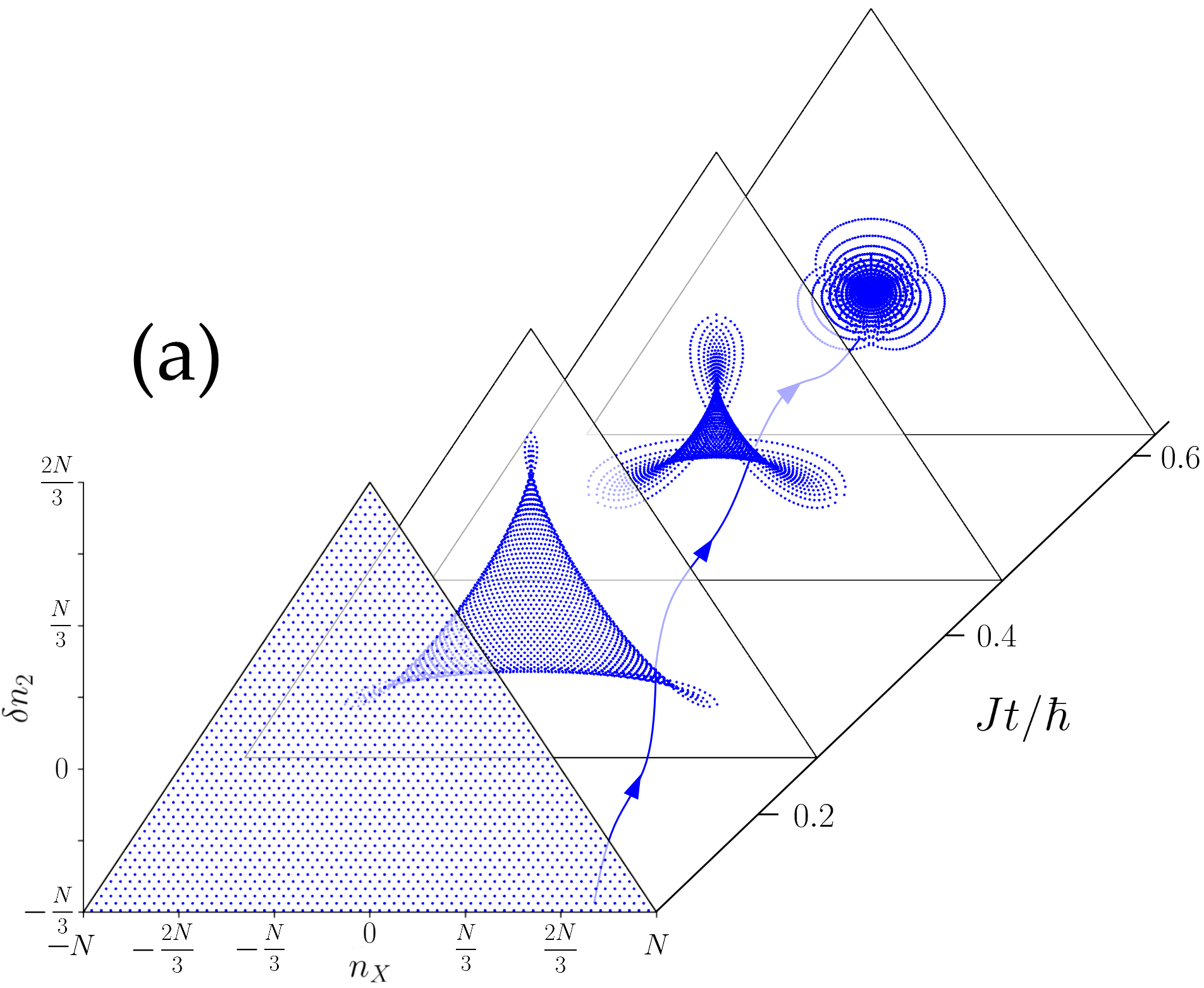}
	\includegraphics[width=\columnwidth]{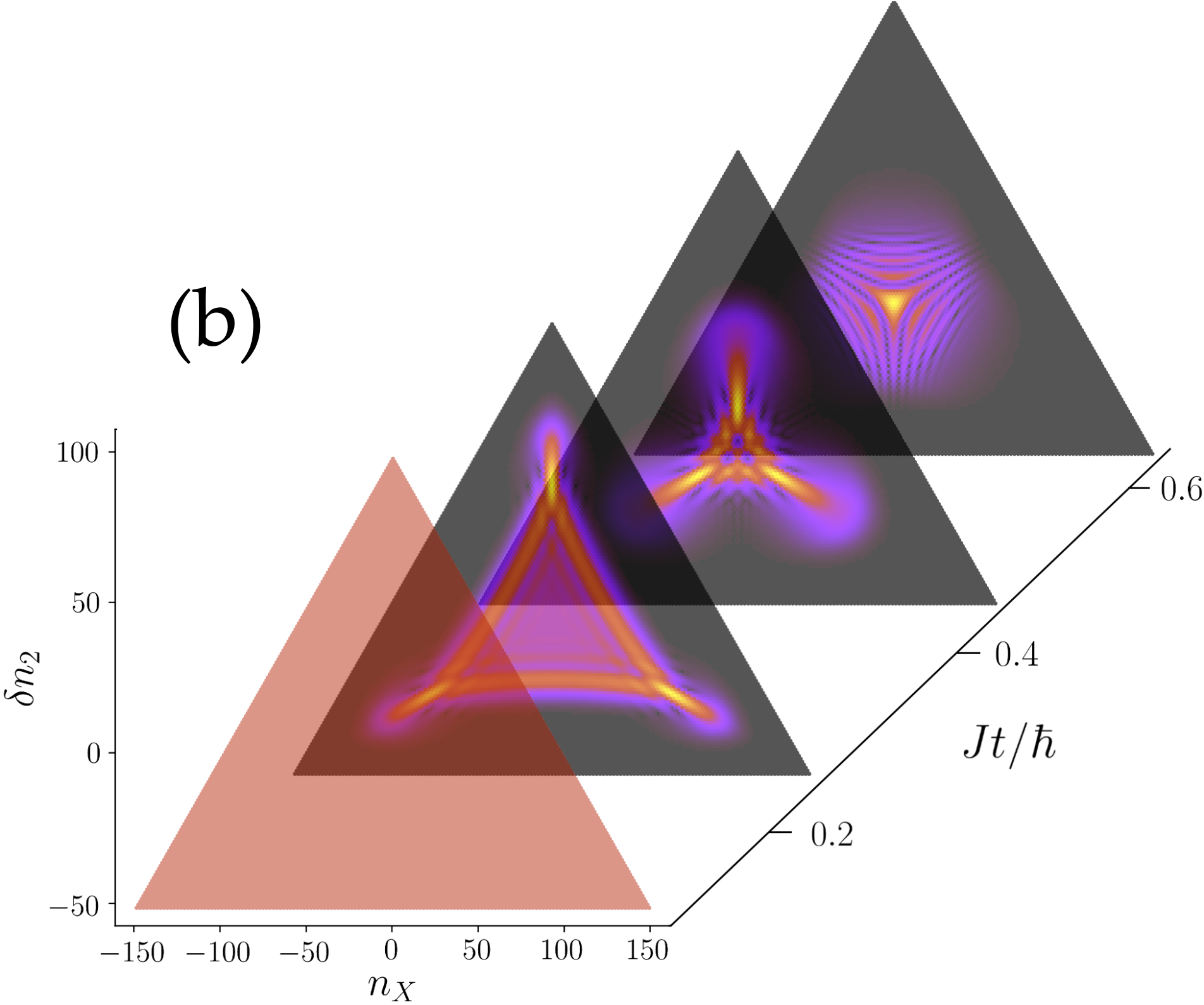}
	\caption{\label{fig:3DPlot} Time slices comparing mean-field and quantum dynamics in the triangular Fock space found in the BH trimer under the constraint of total number conservation. \textbf{Panel (a):} An ensemble of initial points are evolved in time using the classical equations of motion Eqns.\ (\ref{eq:classicalEOM1})--(\ref{eq:classicalEOM4}) and form a space-time version of the elliptic umbilic catastrophe, with one particular trajectory picked out for illustration. In the TWA the initial conditions are drawn from a quantum probability distribution and in this simple case the initial state is a phase state meaning that the relative phases are sharply defined (here taken to be $\phi_{X}=\phi_{C}=0$) but with the consequence that their conjugate number differences take all possible values with equal probability, thereby uniformly populating the allowed triangular region of Fock space. \textbf{Panel (b):} The quantum dynamics are obtained by solving Eq.\ (\ref{eq:RamanNath}). The initial phase state appears as a discretized plane wave in Fock space (although here N = 150 and the discreteness of Fock space is hard to see unless the reader zooms in). The correspondence between the quantum and classical dynamics is clear: the quantum wavefunction is brightest where the density of trajectories is highest. In both images, $K_L=K_R=K_X\equiv J$, $U=-0.005J$, and $\epsilon_{i}=0$.}
\end{figure}

In the semiclassical regime  $N \gg 1$ the discretization is hardly visible and we tend to a continuous theory. However,  there are two possible limits that distinguish many-body versus one-body interference \cite{Mumford2019,Anglin2001}
\begin{enumerate}
\item A wave-like theory where commutators between operators like $[\hat{a},\hat{a}^{\dag}]=1$, or approximate macroscopic versions such as $[\hat{\phi},\hat{n}]\approx i$, are maintained (\textit{interference fringes in Fock space preserved})
\item A geometric ray-like theory where  commutators vanish. This is the Gross-Pitaevskii equation limit (\textit{interference fringes in Fock space removed})
\end{enumerate}
In Fig.\ \ref{fig:3DPlot} we compare the mean-field (ray) and fully quantum (2nd quantized) theories for a caustic in the 3-mode BH model. Although present, the discretization of Fock space is hard to see (unless zoomed in) and the quantum waves appear smooth at large scales. The interference fringes in Fig.\ \ref{fig:3DPlot}(b)  are true many-body fringes not present in the one-body (ray) theory shown in Fig.\ \ref{fig:3DPlot}(a). The ray theory we apply is the truncated Wigner approximation (TWA) where an \textit{ensemble} of classical rays are propagated  using the classical equations (\ref{eq:classicalEOM1})--(\ref{eq:classicalEOM4})  with initial conditions sampled from a quantum quasiprobability distribution (the Wigner function)  \cite{Sinatra01,Sinatra02,Polkovnikov2010,Javanainen2013}. Summing these rays gives the mean-field approximation to the quantum dynamics. The initial state in Fig.\ \ref{fig:3DPlot} is a phase state, which is a state with narrow relative phase distributions but which consequently has a flat probability distribution in Fock space because number and phase are conjugate variables. In the mean-field case shown in panel (a), we see that the first time slice contains a representative set of points approximating an equal superposition of Fock states. Panel (b) plots the absolute values of the quantum amplitudes of the Fock states found by solving the Raman-Nath equations Eq.\ (\ref{eq:RamanNath}). In both cases the dynamics leads to focusing and clearly forms an elliptic umbilic caustic which can be compared to those shown in Fig.\ \ref{fig:causticgallery}. An equal superposition of Fock states is a plane wave-like state analogous to the initial state often considered in optics, but here it is the BH dynamics that acts as an imperfect lens which focuses the wave in Fock space. The key feature of this initial state is that it is broad in Fock space, and structural stability means that the caustics it generates will not be qualitatively different from those generated by other broad states such as the gaussian-shaped ground state in the case with strongly coupled sites such that the hopping dominates interactions (this case will be discussed in Section \ref{sec:triangulardimer}).

In the next two sections we investigate the hierarchy of wave catastrophes that appear in BH dimer and trimer dynamics, building up to the high-order catastrophe $X_9$ which ultimately organizes the lower catastrophes we see.  It should be borne in mind that  because the trimer is not integrable our ability to analytically describe the appearance of each catastrophe is limited and for this reason we go to $\delta$-kicked dynamics in Section \ref{sec:kicked}. Furthermore, unlike free-space optics, classical rays in Fock space do not travel in straight lines in the BH model even for the integrable case ($U=0$).

\begin{figure}\centering
	\includegraphics[width=0.49\columnwidth]{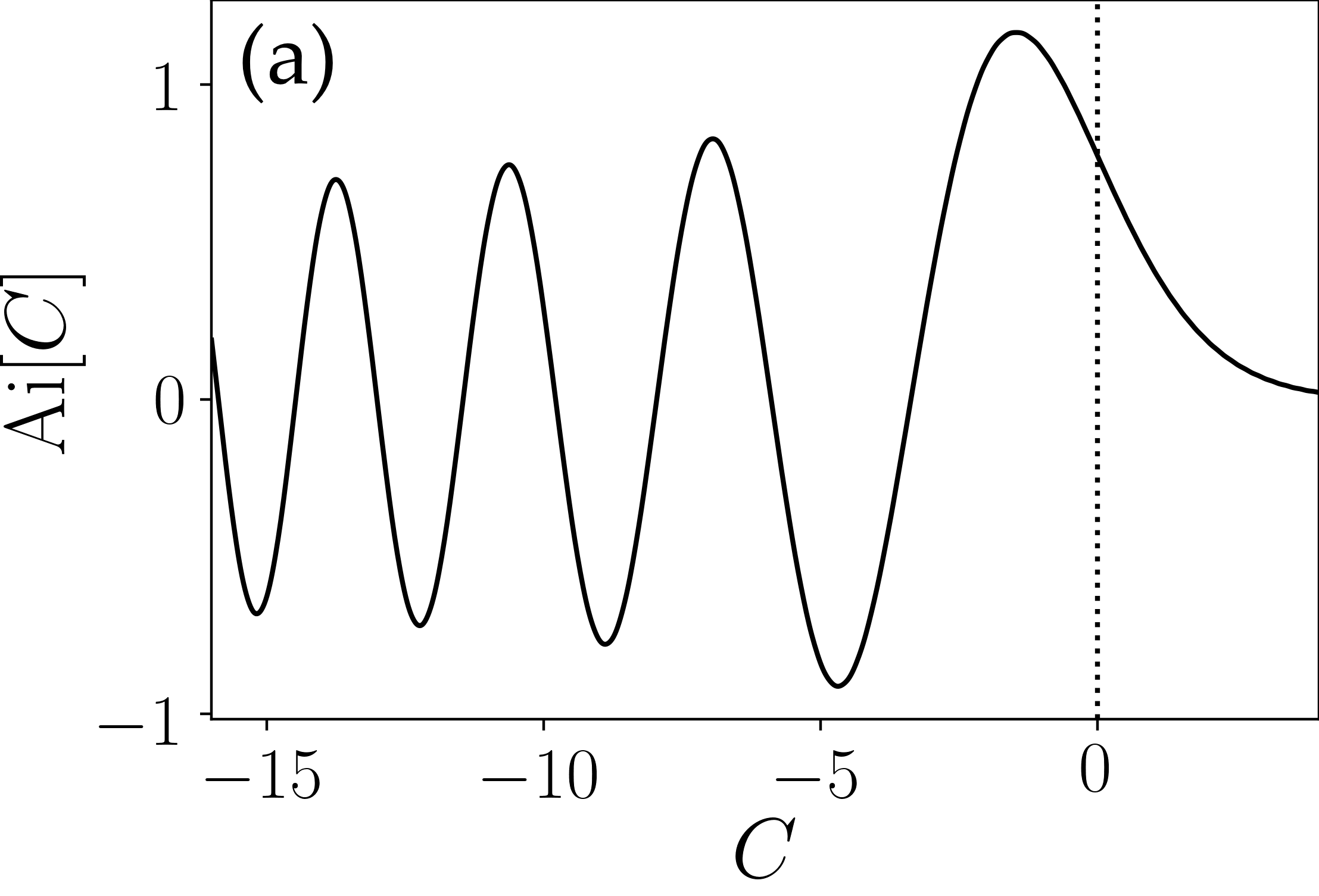}
	\includegraphics[width=0.49\columnwidth]{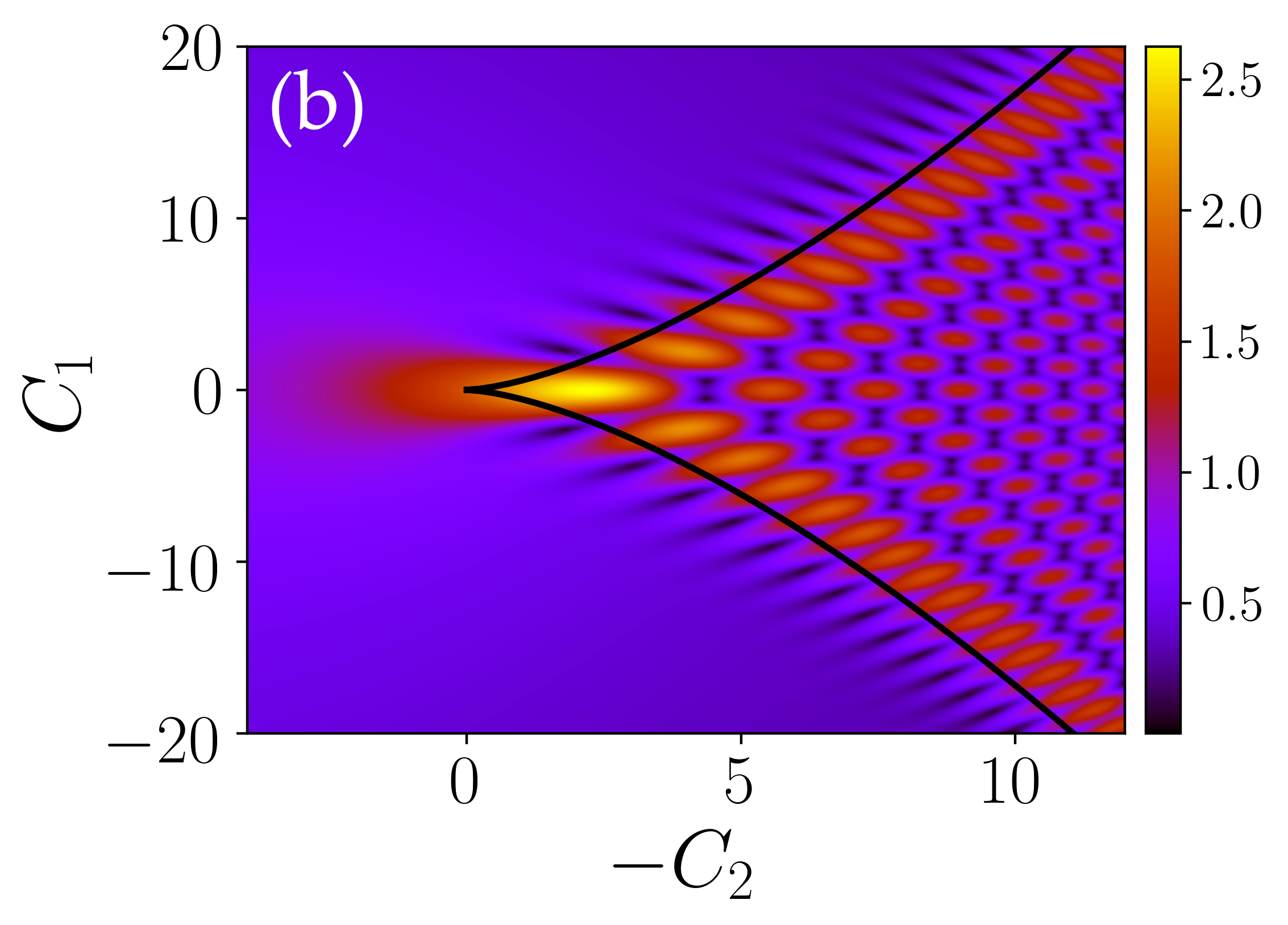}
	\caption{\label{fig:AiryPearcey} The wave catastrophes associated with the fold and the cusp. \textbf{Panel (a):} The fold catastrophe is decorated by an Airy function, Eq.\ \eqref{eq:Airy}. The location of the classical caustic is indicated by a vertical dotted line. In the classical (ray) theory the intensity is divergent at this point and falls off as $1/\sqrt{-C}$ on the bright side. By contrast, the Airy function is finite at the caustic and two-wave interference gives rise to oscillations; these two waves coalesce at $C=0$ and become evanescent on the dark side.    \textbf{Panel (b):} The cusp catastrophe is decorated by the Pearcey function which is a complex-valued function (here we plot the modulus) given by Eq.\ \eqref{eq:Pearcey}, with the divergent classical cusp caustic shown as a solid black curve.}
\end{figure}

\section{\label{sec:BHdimer} Caustics in the dimer}

Dynamical caustics in the BH dimer live in the (1+1)D space formed by Fock space and time. The only structurally stable catastrophes in two dimensions are fold lines which can meet at cusp points.

\subsection{\label{subsec:Fold} Fold}
The simplest catastrophe is the fold. Folds are corank-1, codimension-1 objects with a cubic generating function,
\begin{equation}
	\Phi_{1}(s;C)=s^3+Cs\;.
\end{equation}
 Folds arise where two families of rays coalesce and this can be at a point on a line, a line in a plane, a surface in 3D etc.  The corresponding wave catastrophe is 
 \begin{equation}\label{eq:Airy}
	 \frac{2\pi}{3^{1/3}}\mathrm{Ai} \left( \frac{C}{3^{1/3}} \right) =\int_{-\infty}^{\infty}\mathrm{d}s\;\mathrm{e}^{\mathrm{i}(s^3+Cs)} 
\end{equation}
and is plotted in Fig.\ \ref{fig:AiryPearcey}(a). This function is the well-known  Airy function introduced as the wave description for light at rainbows in 1838 \cite{Airy1838}. It not only removes the singularity in the ray theory but its interference fringes also explain the supernumerary arcs that are sometimes visible inside the main bow in optical rainbows.

\begin{figure}\centering
	\includegraphics[width=0.6\columnwidth]{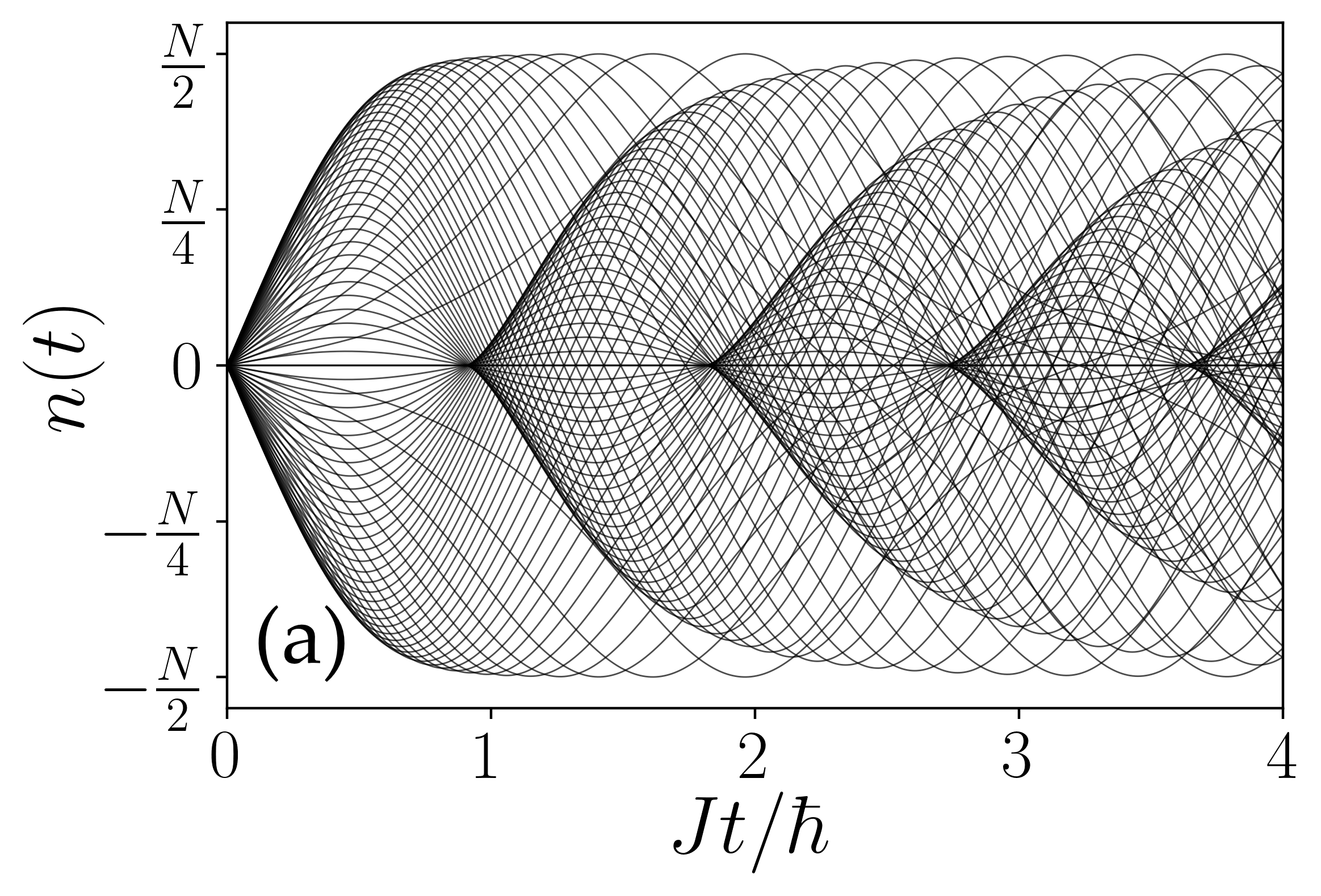}	
	\includegraphics[width=0.6\columnwidth]{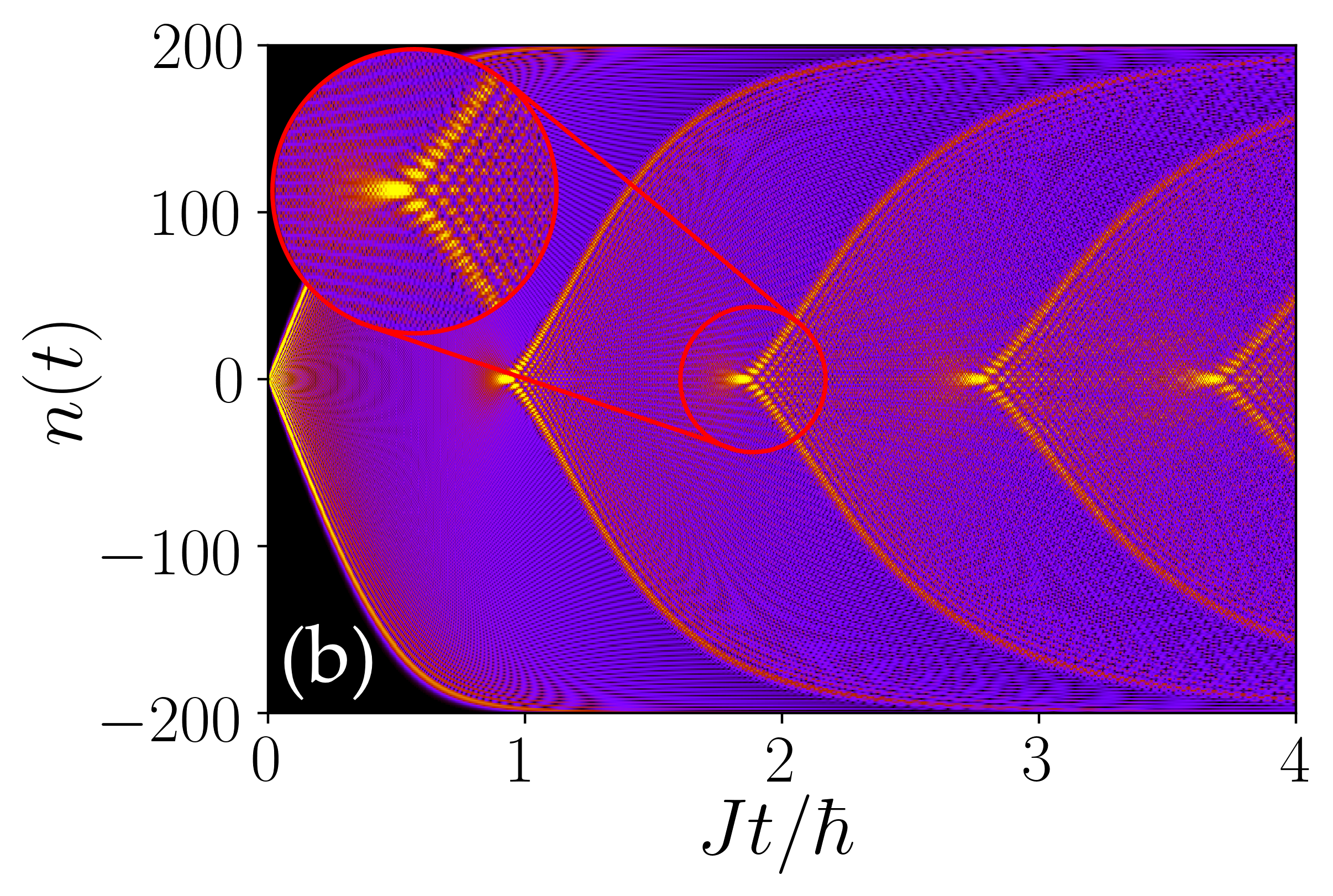}
	\caption{\label{fig:TwoMode} Recurring cusp caustics in the two-mode BH model. \textbf{Panel (a):} Each curve is a mean-field configuration obtained from Eqns.\ (\ref{eq:hamilton1})--(\ref{eq:hamilton2}) starting from $n=0$ and with an initial phase sampled from the distribution $\phi=[0...2 \pi )$ corresponding to a definite (equal) number of bosons in each well. \textbf{Panel (b):} Modulus of the quantum wavefunction calculated using the two-mode RN equations Eq.\ (\ref{eq:TwoModeRN}) with $N=400$ and where the initial state is the single Fock state $\vert n=0 \rangle$. In both cases we can identify a series of cusp caustics corresponding to partial revivals of the initial state. In the inset we see that in the immediate vicinity of each cusp the wavefunction resembles a Pearcey function  \cite{odell12,Mumford2019} (the wave function appears continuous but the discretization of Fock space can be seen if the reader zooms in). }
\end{figure}

\subsection{\label{subsec:Cusp} Cusp}

In two dimensions, generic focusing events are fold lines that  meet at cusps. These have quartic generating functions,
\begin{equation}
	\Phi_{2}(s;C_1,C_2)=s^4+C_2s^2+C_1s\;.
\end{equation}
Inside the cusp, three families of classical rays coexist, while only one family exists outside. This means that along each fold line, two sets of rays coalesce, while at the highly singular cusp point, all three families coalesce. 
The wave catastrophe associated with the cusp is a 2D wavefunction known as the Pearcey function \cite{pearcey46,handbook},
\begin{equation}\label{eq:Pearcey}
	\mathrm{Pe}[C_1,C_2]=\int_{-\infty}^{\infty}\mathrm{d}s\;\mathrm{e}^{\mathrm{i}(s^4+C_2s^2+C_1s)} \ ,
\end{equation}
and is plotted in Fig.\ \ref{fig:AiryPearcey}(b).
As can be seen, the Pearcey function consists of an interference pattern inside the cusp (due to three-wave interference), while outside it becomes exponentially suppressed. Since the cusp catastrophe consists of the meeting of two fold lines, a one-dimensional slice of the Pearcey function across one of these lines projects onto an Airy function.

We show examples of these two wave catastrophes appearing in Fock-space dynamics of the BH dimer following a quench in Fig.\ \ref{fig:TwoMode}. The initial condition is a single Fock state $\vert n=0 \rangle$ (which is the opposite case to the phase state shown in Fig.\ \ref{fig:3DPlot}). Physically, it describes the situation where two independent BECs are suddenly coupled so that at $t=0$ particles can begin to hop between them, thereby building up coherence  \cite{Zapata03}. The structural stability of catastrophes means that qualitatively similar behaviour is found for similar initial states such as narrow gaussians in Fock space that can also be centered away from $n=0$. Fig.\ \ref{fig:TwoMode}(a) shows the set of TWA trajectories propagating from $n=0$, each with a different relative phase $\phi$. The complete certainty in $n$ means complete uncertainty in $\phi$ so the phases are drawn with equal probability from the interval $\phi=[0...2 \pi)$. The imperfect focusing of trajectories leads to cusps which revive at times $Jt=m\pi/(2\sqrt{1+2NU/J})$ for $m=1,2,3,...$. Cusps only form because of the nonlinearity due to interactions: setting $U=0$ leads to isolated focal points. In the quantum theory Pearcey patterns dress each cusp. Moving away from the cusp tip, the wavefunction rapidly tends to back-to-back Airy functions describing the two fold lines emanating from the cusp \cite{odell12}.

\section{\label{sec:BHtrimer} Caustics in the trimer}

\begin{figure}\centering
	\includegraphics[width=0.6\columnwidth]{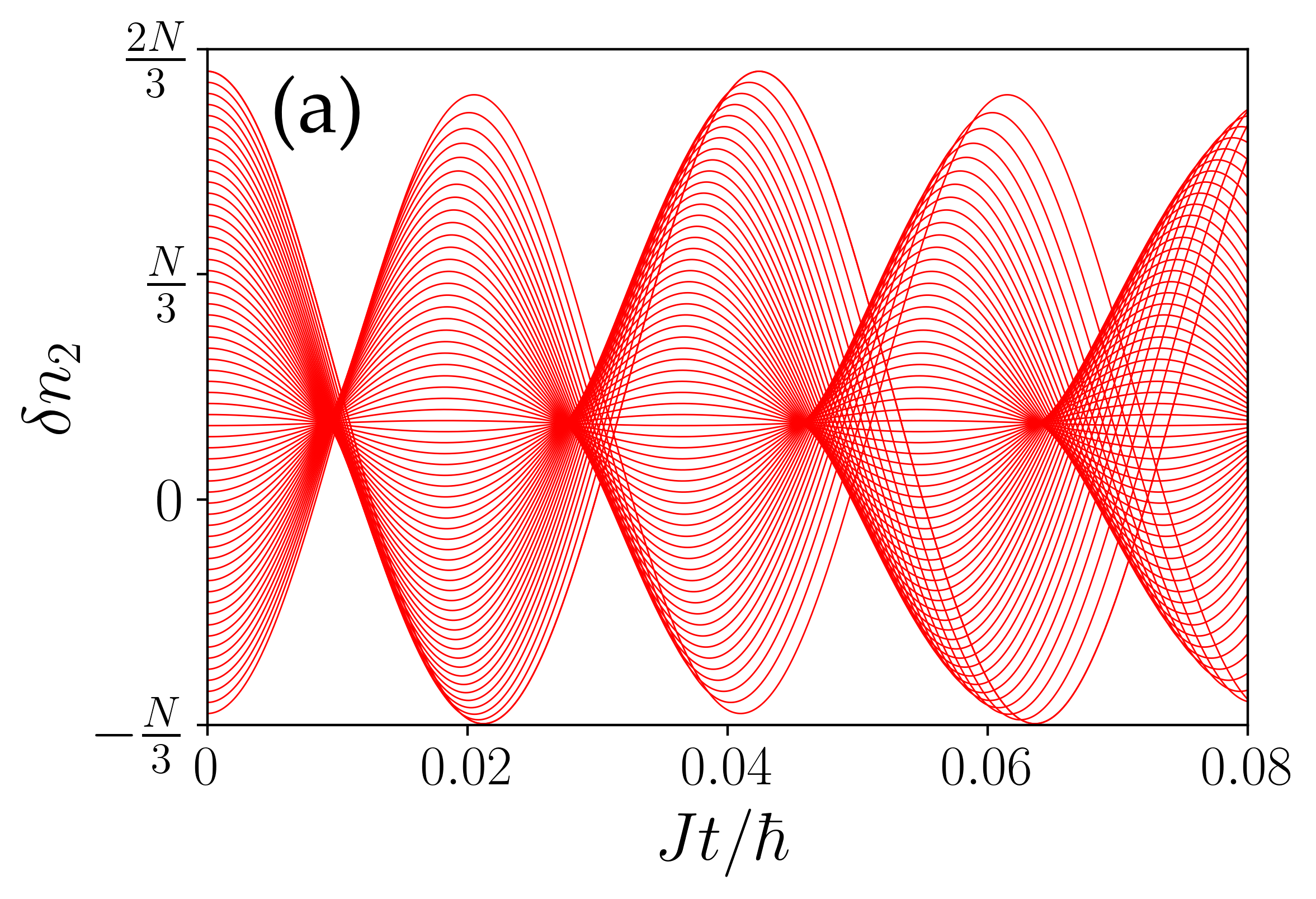}
	\includegraphics[width=0.6\columnwidth]{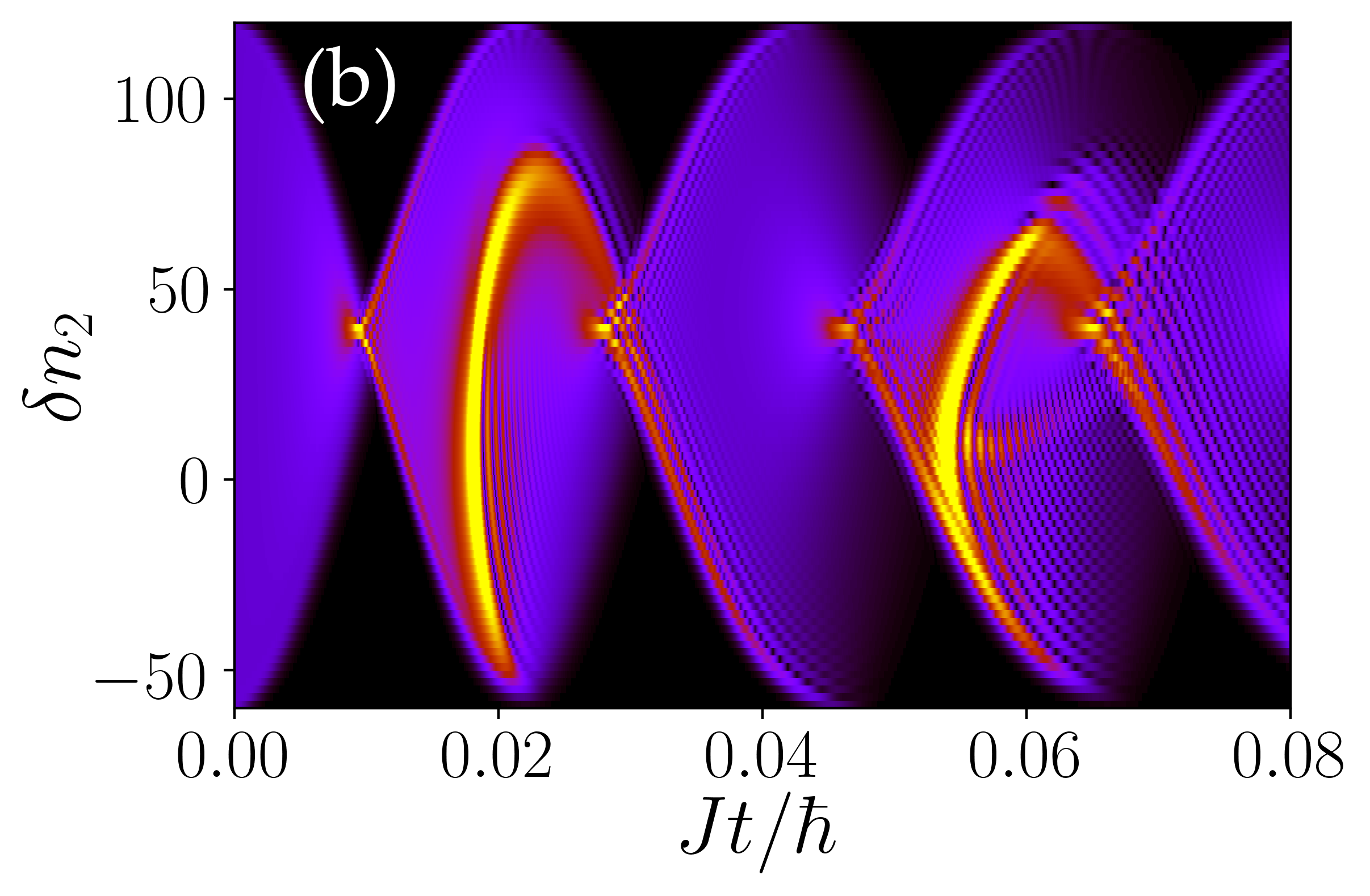}
	\caption{\label{fig:ThreeModeSlices} Recurring cusps in the BH trimer model, starting from an equal superposition of Fock states, the three-mode version of a phase state. \textbf{Panel (a):} Classical trajectories along the plane of fixed $n_X=0$. \textbf{Panel (b):} Quantum dynamics in the same plane as (a) where the classical cusps are now dressed by interference fringes, locally approximated by the Pearcey function. The bright streaks are finite-size effects due to reflections off the Fock-space boundary in the $n_X$ direction, but structural stability ensures the cusps survive. In both panels, $K_L=K_R\equiv J=100U$, $K_X=0$ (linear configuration), while for the quantum dynamics we used $N=180$.}
\end{figure}

The BH trimer model also yields cusps if we restrict attention to just one of the Fock space variables, as illustrated in Fig.\ \ref{fig:ThreeModeSlices}. For variety the initial condition is this time chosen to be an equal superposition of all Fock states. Like an individual Fock state, this does not correspond to an eigenstate of the Hamiltonian that we use to propagate the system in time. Instead it is a highly-excited state made up of a broad superposition of eigenstates, and this allows the system to explore its nonlinearity and produce caustics which revive periodically.

\begin{figure}
	\centering
	\textbf{\hspace{0.12\columnwidth}BH Trimer \hspace{0.22\columnwidth} Wave Catastrophe}\\
	\includegraphics[width=0.49\columnwidth]{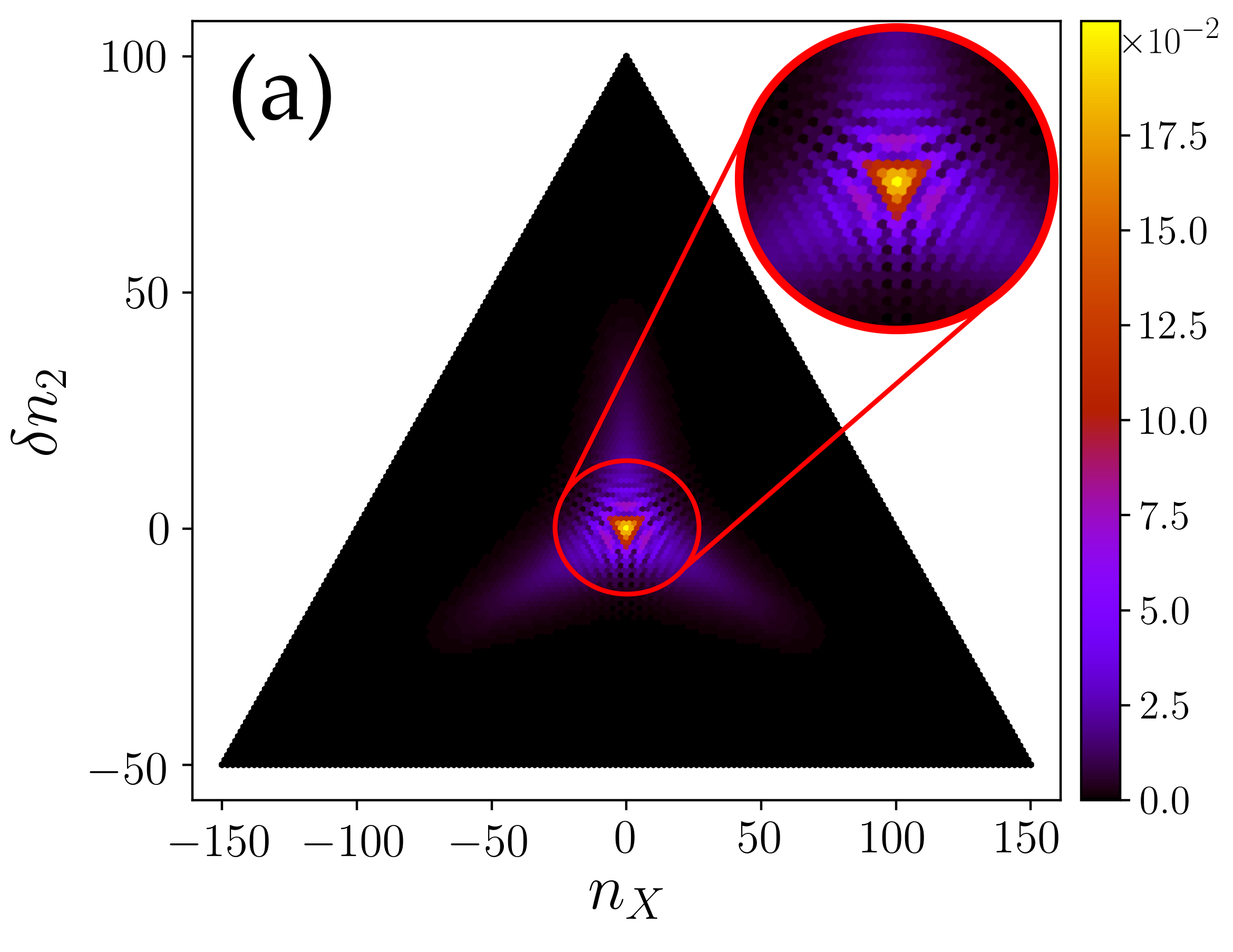}
	\includegraphics[width=0.49\columnwidth]{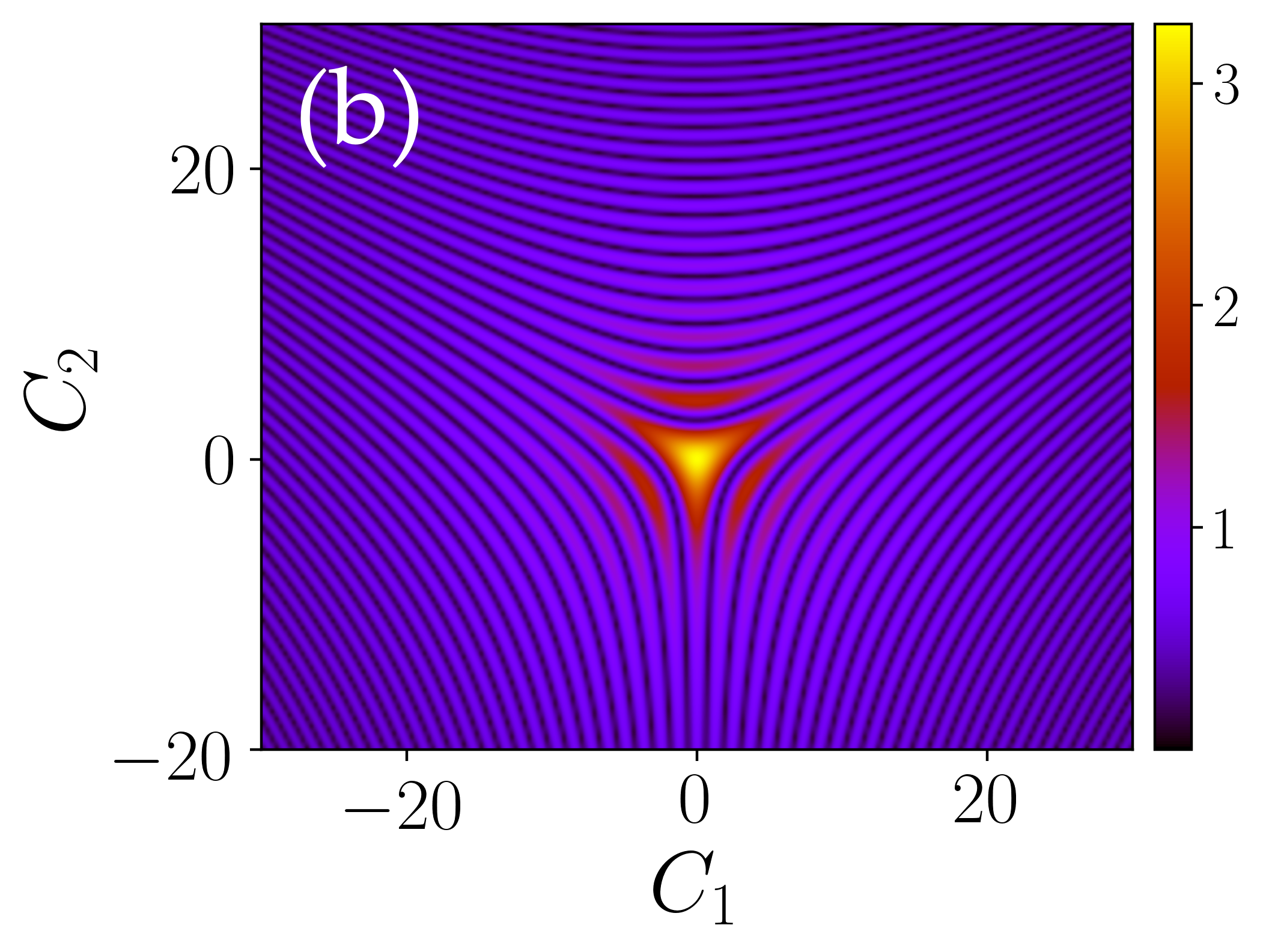}
	\includegraphics[width=0.49\columnwidth]{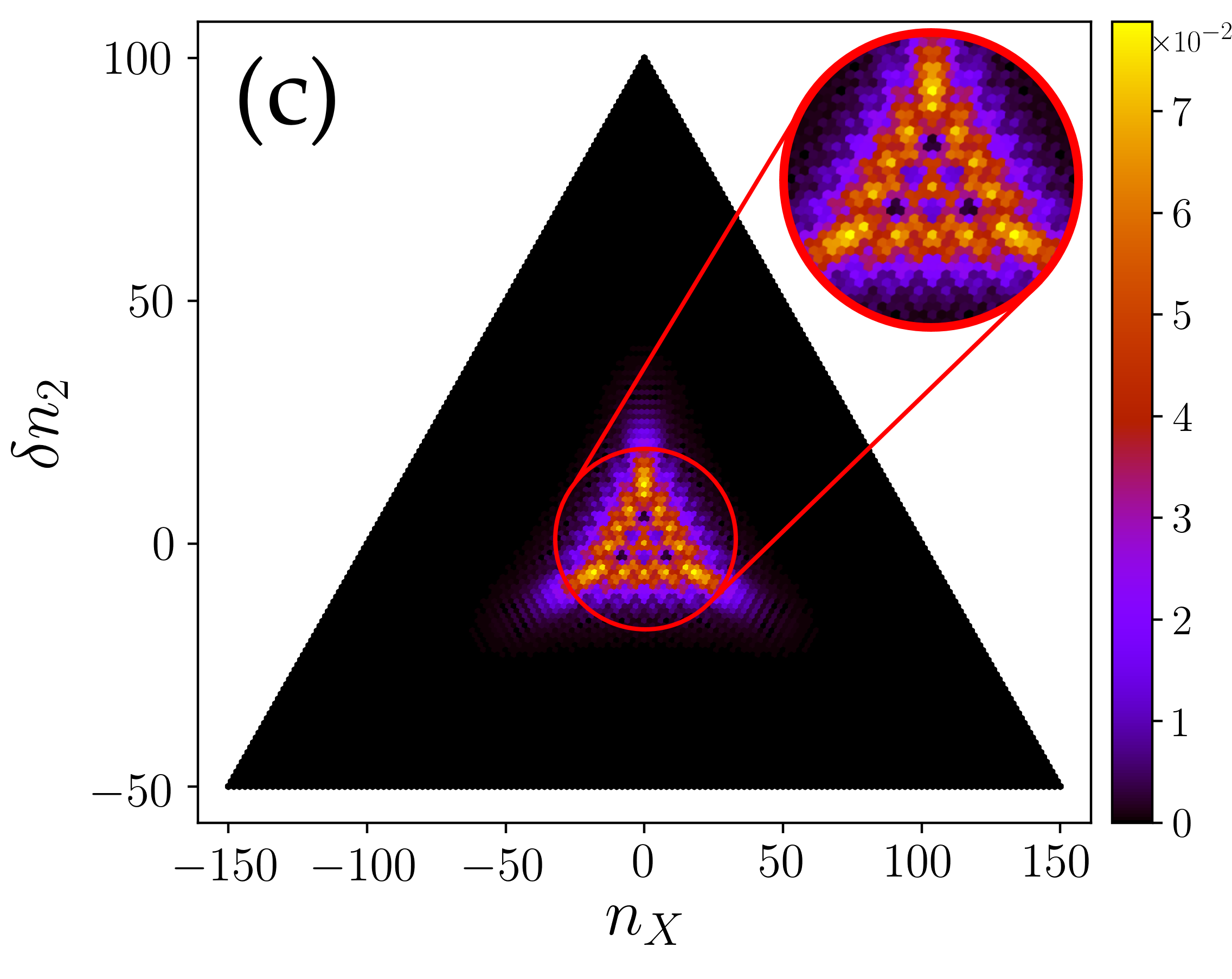}
	\includegraphics[width=0.49\columnwidth]{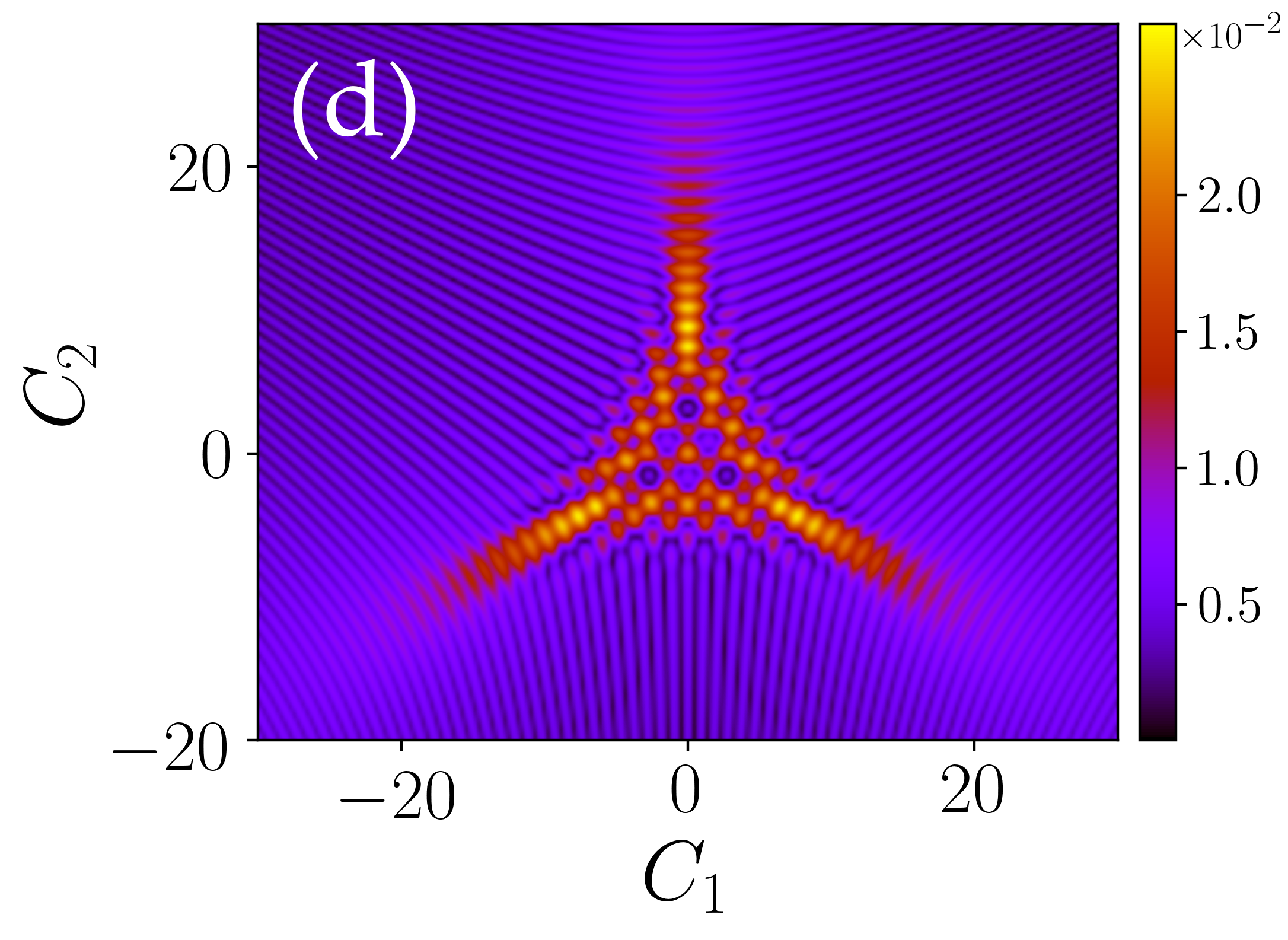}
	\includegraphics[width=0.4\columnwidth]{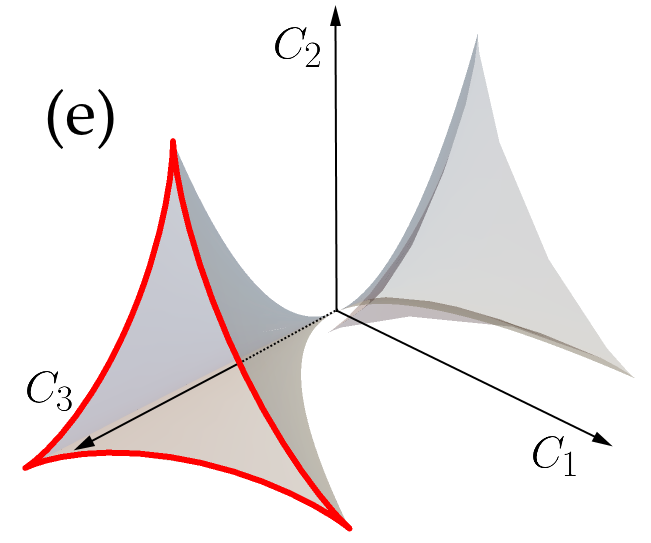}
	\caption{\label{fig:EllipticUmbilic} Comparison of BH trimer  dynamics with the canonical elliptic umbilic catastrophe. \textbf{Panel (a):} Fock-space amplitudes starting from an even superposition of Fock states [Eq.\ \eqref{eq:FockSuper}] for $K_L=K_R=K_X\equiv J$, $U=0.01 J$ and $N=150$ at $Jt/\hbar = 0.47$. \textbf{Panel (b):} Diffraction pattern amplitude in the plane around the elliptic umbilic focus. \textbf{Panel (c):} Same conditions as panel (a), but now at $Jt/\hbar = 0.54$. \textbf{Panel (d):} Diffraction pattern for a slice of the elliptic umbilic at $C_3=3.8$.
	\textbf{Panel (e):} Caustic surface of the elliptic umbilic catastrophe. The triple-cusped intersection with a $C_3=\text{const.}$ plane is highlighted in red.}
\end{figure}

\begin{figure}
	\centering
	\textbf{\hspace{0.12\columnwidth}BH Trimer \hspace{0.22\columnwidth} Wave Catastrophe}\\
	\includegraphics[width=0.49\columnwidth]{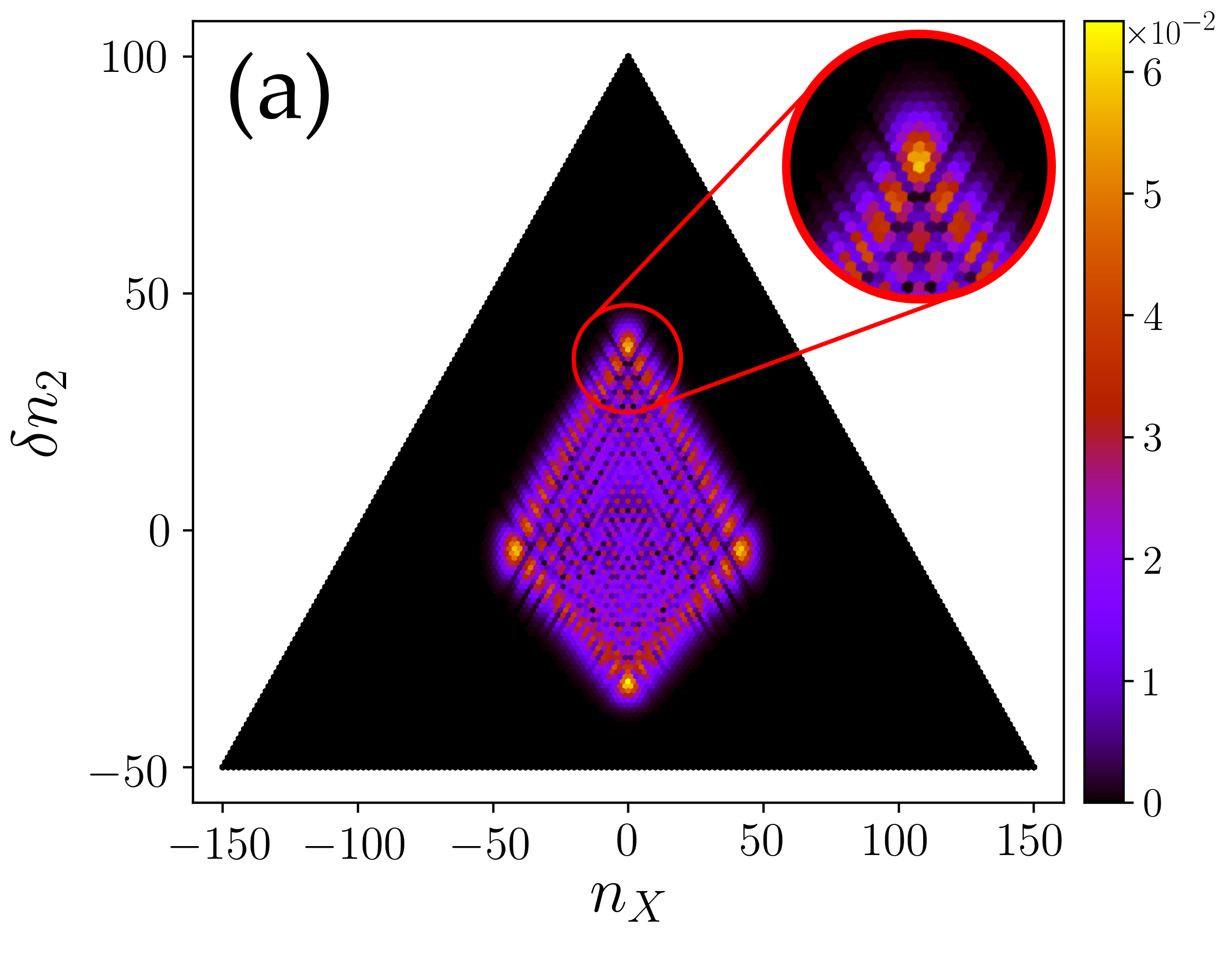}
	\includegraphics[width=0.49\columnwidth]{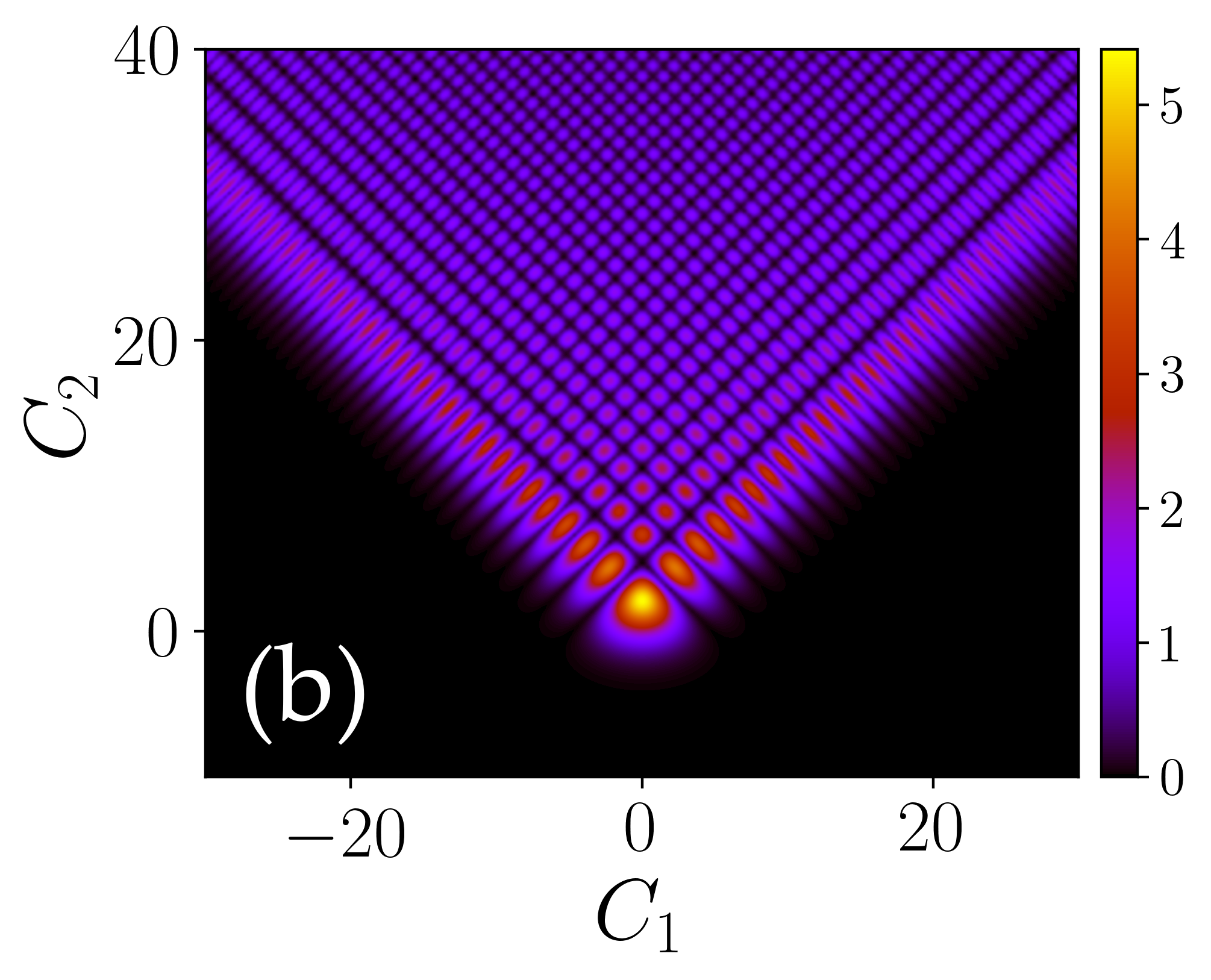}
	\includegraphics[width=0.49\columnwidth]{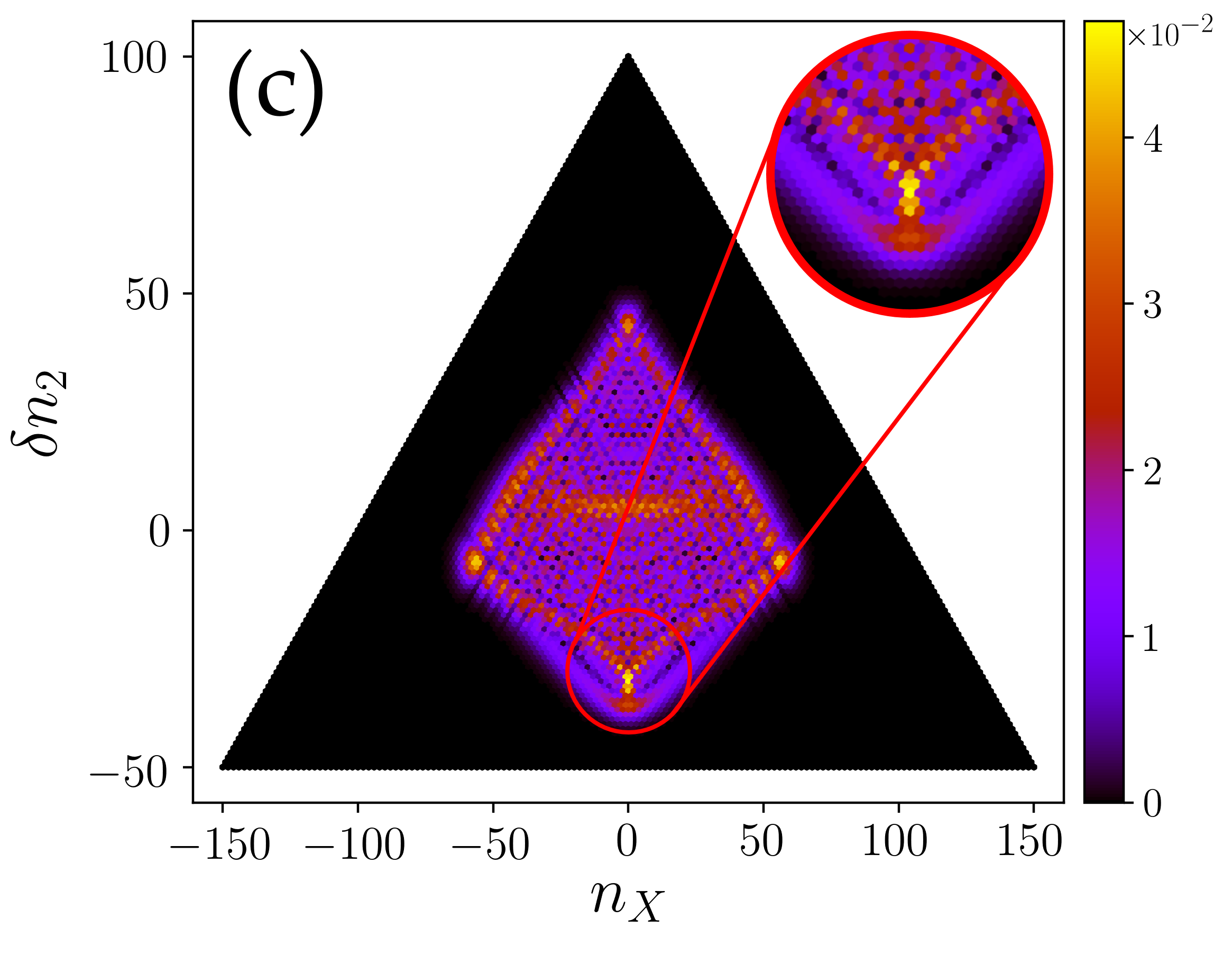}
	\includegraphics[width=0.49\columnwidth]{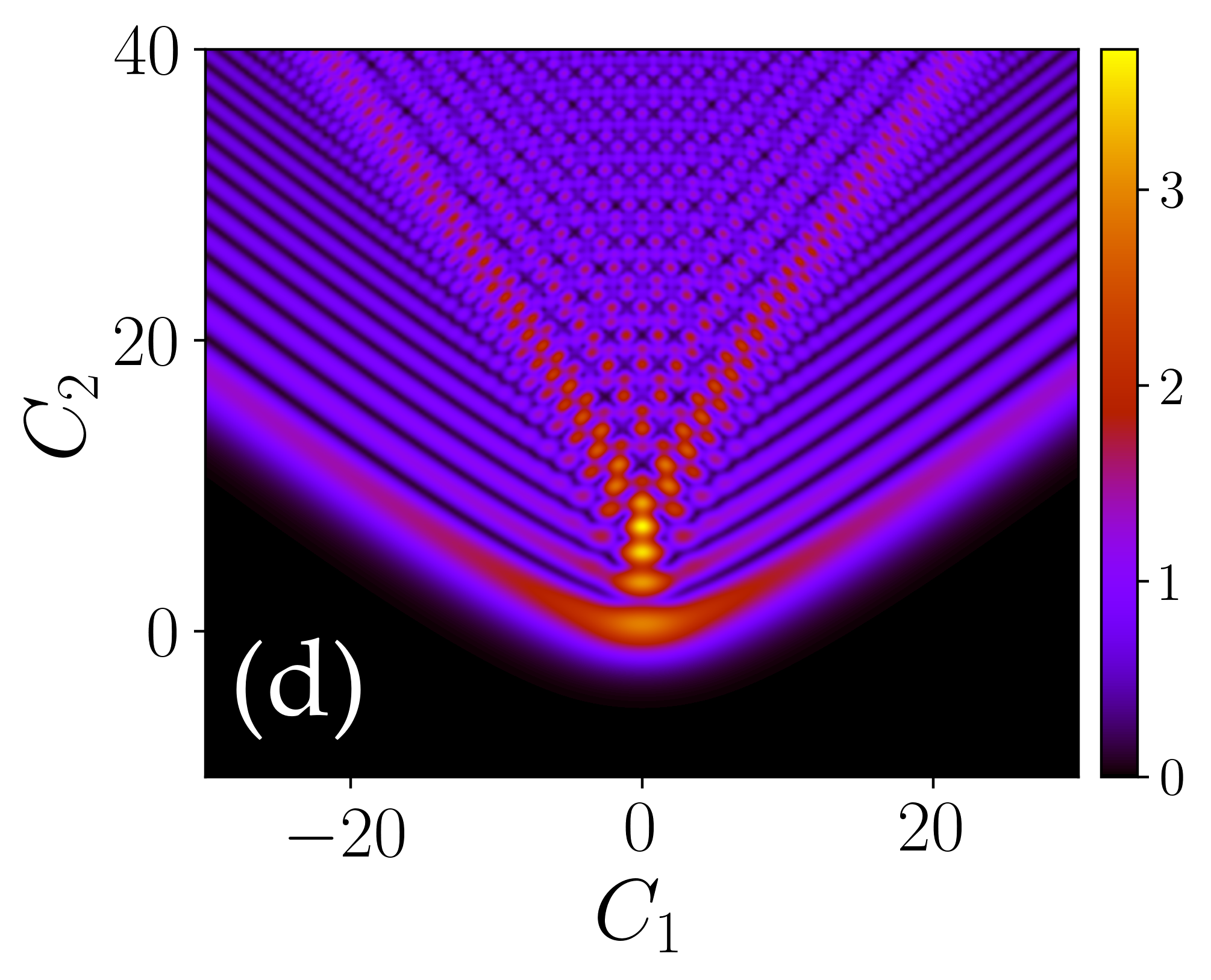}
	\includegraphics[width=0.5\columnwidth]{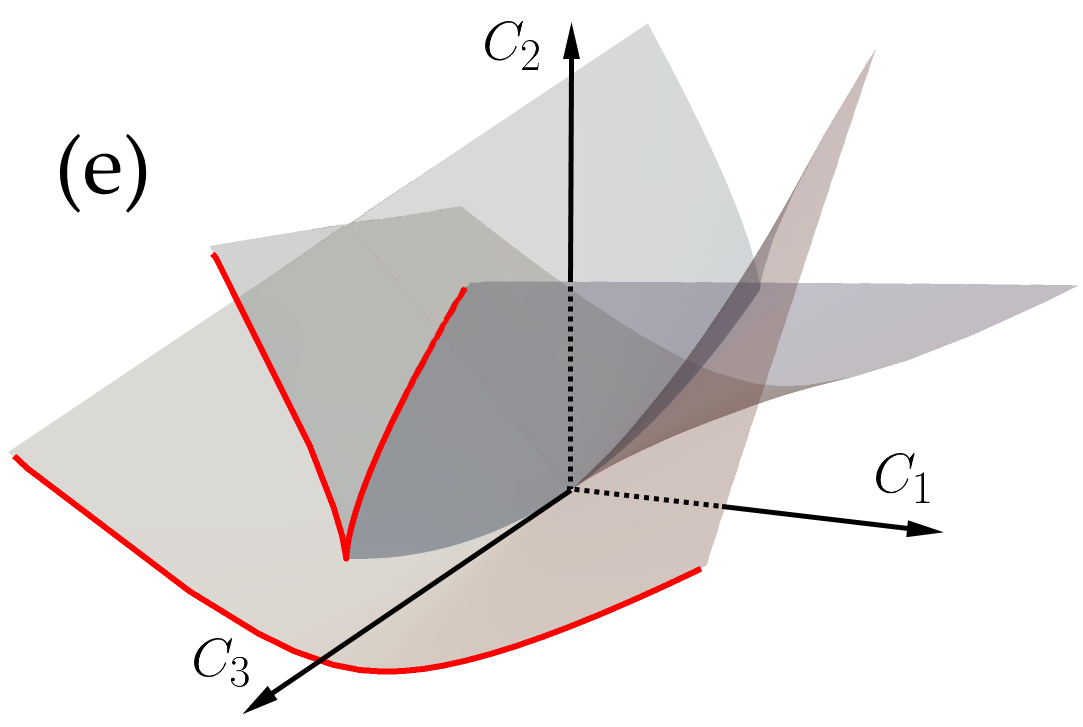}
	\caption{\label{fig:HyperbolicUmbilic} Comparison of BH trimer  dynamics with the canonical hyperbolic umbilic catastrophe. \textbf{Panel (a):} Wavefunction amplitudes for hopping strengths $K_L=K_R\equiv J=4U$, $K_X=0$, and $N=150$ at $Jt/\hbar=0.24$. The initial state was $\ket{\delta n_2,n_X}=\ket{0,0}$. \textbf{Panel (b):} Hyperbolic umbilic diffraction pattern on the plane $C_3=0$. \textbf{Panel (c):} Same conditions as panel (a) now at $Jt/\hbar=0.34$. The bottom umbilic is now completely unfolded. \textbf{Panel (d):} Diffraction pattern for the hyperbolic umbilic in the plane $C_3=3$. \textbf{Panel (e):} Caustic surface of the elliptic umbilic catastrophe. A projection onto the plane $C_3=\text{const.}$ is highlighted in red.}
\end{figure}

When we examine the full (2+1)D space available in the trimer then, as expected, we discover codimension-3 catastrophes, namely the elliptic and hyperbolic umbilics. This is shown in Figs.\ \ref{fig:EllipticUmbilic} and \ref{fig:HyperbolicUmbilic} which compare numerical solutions of the generalized Raman-Nath equations given in Eq.\ (\ref{eq:RamanNath}) with the canonical wave catastrophes. According to Table \ref{tab:catastrophetable}, there is a third codimension-3 catastrophe known as the swallowtail, but this is a corank 1 catastrophe meaning that it only has a single state variable. To realize the swallowtail catastrophe in the trimer we could freeze one of  the conjugate phases $\{\phi_{X}, \phi_{C} \}$, but we will not pursue this possibility here. 

It is important to emphasize that although we have chosen parameters where the match in Figs.\ \ref{fig:EllipticUmbilic} and \ref{fig:HyperbolicUmbilic} between the numerical results and the canonical catastrophes is quite good, we have not optimized the parameters and these catastrophes occur generically. If we could solve the trimer model analytically we could perform an exact mapping but due to its nonintegrability this is not possible. We must therefore satisfy ourselves with qualitative rather than quantitative matches which are in any case fully within the spirit of catastrophe theory which is a topological theory \cite{thom75,arnold75,Zeeman77}.  In Section \ref{sec:kicked} we consider simplified kick dynamics where we \textit{can} precisely map the wavefunctions onto canonical wave catastrophes. A brief note on the figures in this paper: we present the bifurcation sets in orientations that are convenient to visualize, hence some of the caustic surfaces are plotted using the negative axis of control parameters, specifically Figs.\ \ref{fig:AiryPearcey}, \ref{fig:K6Hyperbolic}, and \ref{fig:K2SpunCusp}.

\subsection{\label{subsec:EllipticUmbilic} Elliptic Umbilic}

The elliptic umbilic catastrophe, 
\begin{equation}
\Phi_{D_4^-}(\textbf{C})=3s_1^2s_2-s_2^3+C_3(s_1^2+s_2^2)+C_2s_2+C_1s_1\;,
\end{equation}
is one of the two catastrophes of codimension 3 with corank 2. As can be seen in Fig.\ \ref{fig:EllipticUmbilic}, the dominant feature of the elliptic umbilic is its three-fold symmetry. Taking two-dimensional slices at fixed values of $C_{3}$ (which for us is the time direction), the elliptic umbilic appears as three curved fold lines that meet at three cusps. In its full three dimensional form we see that these are really three fold surfaces that meet at cusp shaped ribs.  In the geometric theory, there are four rays at all points inside the caustic, while only two exist at any  point outside because two coalesce on the caustic. 

The corresponding diffraction catastrophe has been studied both theoretically and experimentally by Berry \textit{et al.} in Ref.\ \cite{BerryElliptic1978} by focusing light through a triangular water droplet lens (as repeated by us in Fig.\ \ref{fig:causticgallery}). A Pearcey diffraction pattern locally dresses each cusp, as apparent by considering any particular corner of Fig.\ \ref{fig:EllipticUmbilic}(d). The three-fold cusp structure expands/contracts as the control parameter $C_3$ is changed until it collapses at $C_3=0$. The most singular part of the caustic, the umbilic focus,  is at the centre of control space ($\textbf{C}=\bm{0}$).  In the wave theory the focus is dressed by a diffraction pattern composed of a three-fold symmetric fork with the brightest patch at the center and surrounded by an Airy fringe pattern, as shown in Fig.\ \ref{fig:EllipticUmbilic}(b).

\subsection{\label{subsec:HyperbolicUmbilic}Hyperbolic Umbilic}

The hyperbolic umbilic catastrophe, 
\begin{equation}
\Phi_{D_4^+}=s_1^3+s_2^3+C_3s_1s_2+C_2s_2+C_1s_1\;,
\end{equation}
is the remaining catastrophe of codimension 3 with corank 2. The hyperbolic umbilic caustic surface corresponds to an overlapping cusp and fold extended into three-dimensional space, as shown in Fig.\ \ref{fig:HyperbolicUmbilic}(e). Within the cusp, there are four classical rays at every point, two of which annihilate each other as the cusp surface is crossed. The remaining two rays annihilate as the fold is crossed. The resulting diffraction pattern projected on a plane of constant $C_3$ is a Pearcey-like function surrounded by an Airy fringe pattern as seen in Fig.\ \ref{fig:HyperbolicUmbilic}(d). At the plane $C_3=0$, the caustic is only partially unfolded, and now two sets of fold lines overlap to form a right-angle corner, dressed by a pattern described by a product of Airy functions in two dimensions, as shown in Fig.\ \ref{fig:HyperbolicUmbilic}(b).

We can observe the hyperbolic umbilic in the dynamics of the trimer system by starting from the single Fock state $\ket{\delta n_2,n_X}=\ket{0,0}$ and with $K_X=0$ (linear spatial configuration of the triple-well). The initial compact state spreads non-uniformly into a two-fold symmetric polygon with partially unfolded hyperbolic umbilic corners as shown in Fig.\ \ref{fig:HyperbolicUmbilic}(a). As the dynamics continue, the edges unfold completely, and the cusps separate from the fold lines (moving off the $C_3=0$ plane), as is first visible in the bottom corner of Fig.\ \ref{fig:HyperbolicUmbilic}(c).

\subsection{\label{subsec:X9}The $X_9$ catastrophe}

In the absence of tilts ($\epsilon_i=0$), the BH trimer can be seen to have six independent control parameters: $\{U,K_R,K_X,\delta n_2,n_X,NK_Lt\}$ (or transformations thereof), thus the codimension-3 catastrophes that we have discussed are merely projections of a catastrophe embedded in a higher dimensional space. We shall argue in this and the following sections that the higher catastrophe that organizes the BH trimer dynamics is in fact the high order umbilic catastrophe known by its group-theoretic symbol, $X_9$. This complicated object has previously been the subject of detailed theoretical analysis by Borghi \cite{Borghi2012} and by Berry and Howls \cite{Berry2010}, and plays an important role in optical refraction through two-dimensional surfaces, such as water droplets \cite{Nye1986}, glass junctions \cite{BerryJunction}, gravitational lensing \cite{Nye1999}, and has also been discussed in the context of stochastic resonance in two dimensions \cite{Nicolis2012}. $X_9$ acts as an organizing centre for a multitude of lower catastrophes and we refer the reader to Fig.\ \ref{fig:abutement} for a ``bordering'' diagram showing its relationship to these subcatastrophes.

\begin{figure}\centering
	\includegraphics[width=0.9\columnwidth]{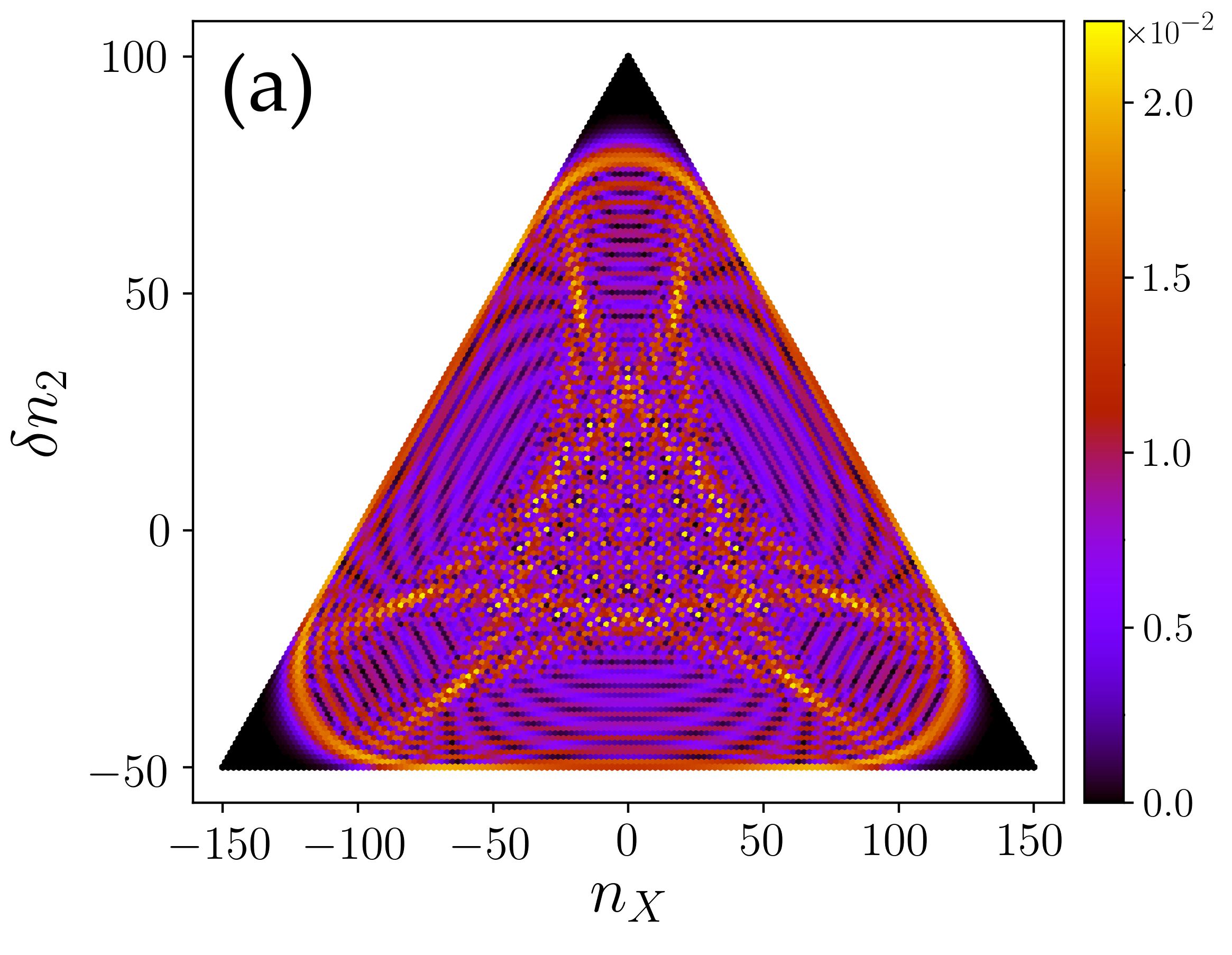}
	\includegraphics[width=0.9\columnwidth]{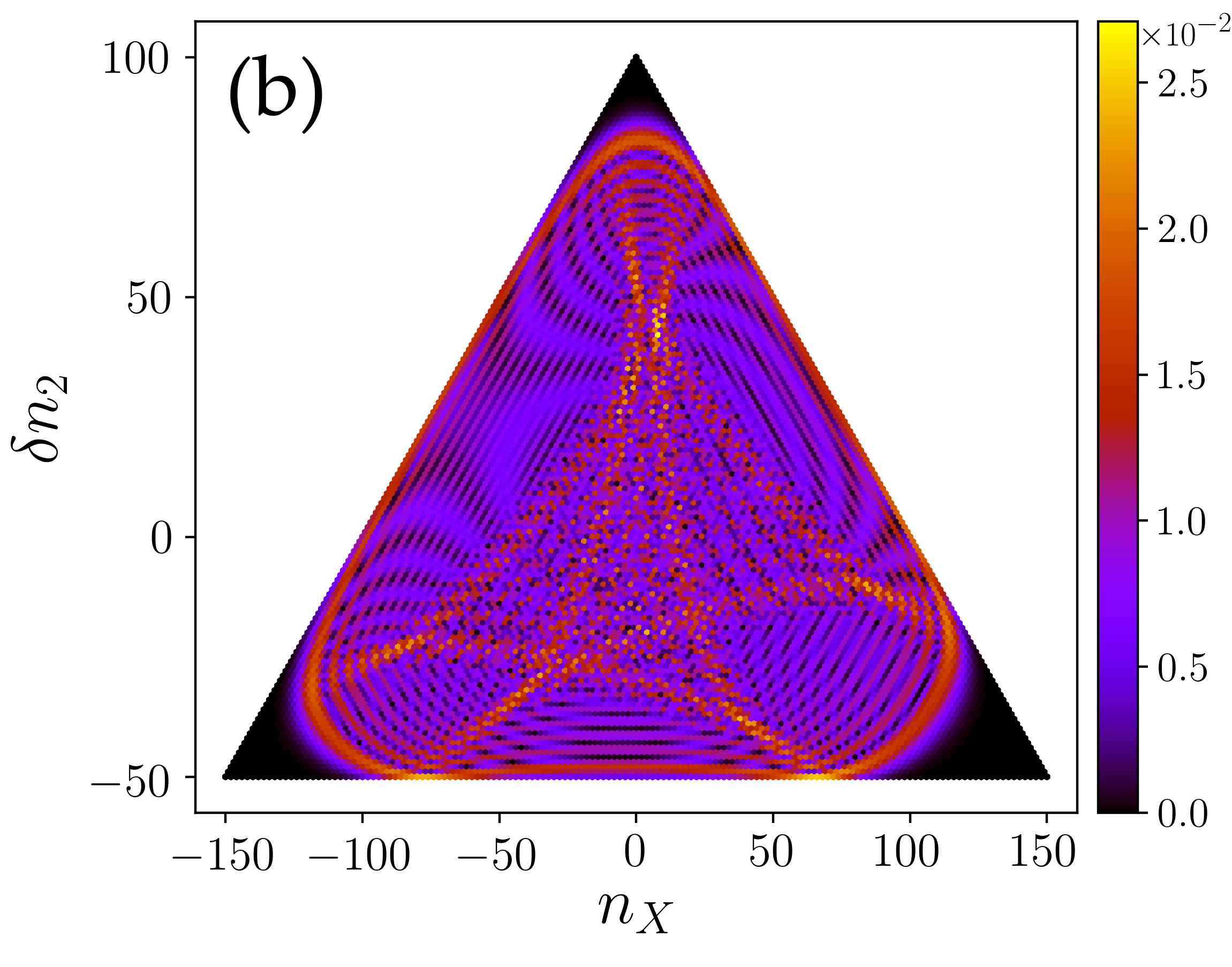}
	\caption{\label{fig:Gen2nCusp} $6$-cusped diffraction pattern surrounded by a fold line, typical of systems with 3-fold symmetry also organized by $X_9$. \textbf{Panel (a):} Fock space amplitudes after a quench starting from all particles evenly distributed in each well, $\ket{\delta n_2,n_X}=\ket{0,0}$. Here, $N=150$, $K_L=K_R=K_X\equiv J$, $U/J=0.04$ at $Jt/\hbar = 0.475$. \textbf{Panel (b):} As panel (a), now with $K_R=J$, $K_L=1.2J$, and $K_X=0.8J$, showing how an absence of symmetry does not destroy the caustics.}
\end{figure}

The $X_{9}$ catastrophe is of corank 2 and its complete 8-dimensional unfolding can be written as \cite{Nye1999},
\begin{align}
\Phi_{X_9^\pm}=&\;s_2^4+Ks_1^2s_2^2\pm s_1^4+C_7s_2^2s_1+C_6s_2s_1^2+C_5(s_2^2+s_1^2)\nonumber\\
&+C_4(s_2^2-s_1^2)+C_3s_2s_1+C_2s_2+C_1s_1 \ . \label{eq:PhiX9}
\end{align}
Although $X_9$ appears 8-dimensional, its control space has only seven parameters, while the eighth, $K$, is known as the \textit{modulus}. Catastrophes beyond codimension 5 (or above corank 3) can contain moduli which are different from regular control parameters in that they can only alter the caustic pattern geometrically  \cite{Gilmore1981}  (rather than topologically by changing the number of critical points \cite{Nye1986}), and cannot be removed via scaling arguments. This modulus has two excluded values: $K\neq \pm 2$, where the singularities achieve infinite codimension, meaning there are an infinite number of ways to unfold the singularity, and perturbations can lead to any number of coalescing critical points. 
As we shall see in the following sections, the $X_{9}$ germ $\Phi_{X_9^{\pm}}[\textbf{C}=\bm{0]}$ arises naturally as the base singularity in all the cases we study.

The high dimension of $X_{9}$ makes it hard to visualize and we shall therefore concentrate on  particular projections.
An example from BH trimer dynamics is shown in Fig.\ \ref{fig:Gen2nCusp}(a) where all the hopping amplitudes are  equal and leads to two superimposed elliptic umbilics slightly rotated from each other.  More precisely, the characteristic features  are a three-fold symmetric caustic featuring a six-cusped figure encapsulated by Airy-like fringes. This type of caustic is commonly encountered when light passes through liquid drops, where a 2$n$-cusped pattern is formed by drops with $n$-fold symmetry, all surrounded by an `oval' fold line \cite{Nye1986}. Perturbing the symmetry will not destroy this caustic structure, as shown in Fig.\ \ref{fig:Gen2nCusp}(b) where the symmetry of the hopping terms is broken. It can only be altered dramatically by changing the symmetry in a fundamental way, such as setting a hopping term to 0. 

It may appear that the control space of $X_9$ contains one too many parameters for the BH trimer.  This is because the physical constraint of number conservation restricts us from exploring the full mathematical control space. We shall see [Fig.\ \ref{fig:K6Hyperbolic}] that loosening the restriction $n_1+n_2+n_3=N$ allows us to access a complete section of the $X_9$ catastrophe since a point in Fock space is then specified by three coordinates rather than two, bringing the total number of control parameters to seven.

The case $K=+2$ in Eq.\ (\ref{eq:PhiX9}) does not give a proper structurally stable catastrophe but is conceptually important because it arises in models with circular symmetry, wherein a perfectly circular `spun cusp' is realized, punctured by an unstable axial caustic line. This is in fact the case we shall find within the QPM and for a perfectly triangular trimer (all hopping amplitudes equal) to be discussed in Section \ref{sec:triangulardimer}. 
The value $K=-2$ is also not a proper catastrophe but is important because it separates $X_9$ into two distinct sub-families $\prescript{0}{}{X_9}$ and  $\prescript{4}{}{X_9}$ which will come up in Section \ref{sec:Stability} when we go beyond the QPM.  The catastrophe germ for $K=0$, which for $C_7=C_6=C_3=0$ is equal to $\Phi_2(s_1)+\Phi_{2}(s_2)$, has led to the $X_9$ family being known as the `double cusp', sometimes even for $K\neq0$ \cite{Callahan1978,Callahan1981,Upstill1982}.

\section{\label{sec:kicked}Kicked Dynamics}

The patterns and shapes exhibited in Figs.\ \ref{fig:ThreeModeSlices}--\ref{fig:Gen2nCusp} are exact numerical solutions for the BH trimer model which is in general analytically intractable. They clearly resemble the caustics that catastrophe theory predicts, and also have the expected properties. However, it would be reassuring to have an analytical demonstration that in some tractable limit we really can map the dynamics to catastrophes. This is what we now do using a simplified kicked Hamiltonian. This not only allows us to analytically realize various versions of the $X_{9}$ catastrophe but also suggests an experimentally viable method for engineering precisely defined caustics.

Following an optical analogy where caustics are formed after light passes through `bad' lenses which deform the wavefront (an ideal lens will produce a perfect hemispherical wavefront that results in a point focus), the role of $\delta$-kicking is to produce a wavefront in Fock space that has distortions that upon further propagation will generate caustics like those we have seen for the full Hamiltonian. Since catastrophes are stable to perturbations the caustics we find in the kicked case will survive under more generic conditions. 

In the examples that follow, the interaction term proportional to $U$ will be flashed on instantaneously at $t=0$ and the system will afterwards evolve purely under the hopping terms. In ultracold atom systems time-dependent manipulation of interactions can be achieved using a Feshbach resonance \cite{Abeelen99}.  In general, when presented with Hamiltonians of the form,
\begin{equation}
	\hat{H}_\delta(t)=\hat{H}_0+\hat{H}_1\delta(t)\;,
\end{equation}
time-evolution can be achieved via the Floquet operator,
\begin{equation}
	\hat{\mathcal{F}}\equiv \mathrm{e}^{-\frac{\mathrm{i}}{\hbar}\hat{H}_0t}\mathrm{e}^{-\frac{\mathrm{i}}{\hbar}\hat{H}_1}\;.
\end{equation}
In previous work \cite{Mumford2017}, we demonstrated the presence of catastrophes of codimension 2 in (1+1)D kicked systems. Here, we employ and generalize this framework to higher-dimensional catastrophes in the triple-well system, in particular to the different available unfoldings of $X_9$. The $\delta$-kick is sometimes, but not always, necessary to see caustics in each of the cases we study in this section, since Fock space trajectories can still be focused by the hopping terms, a fact which can be demonstrated by ignoring $\hat{H}_1$. We include the kick as part of the calculations since it is more general and it allows the $\hat{H}_1$ term to mimic a tuneable `lens' as mentioned above.

We begin by considering the mean-field Hamiltonian,
\begin{align}\label{eq:dQPM}
	H_{\delta\text{QPM}}=&-\frac{2JN}{3}\cos\left(\phi_X-\phi_C\right)-\frac{2JN}{3}\cos\left(\phi_X+\phi_C\right) \nonumber \\ &-\frac{2K_XN}{3}\cos\left(2\phi_X\right)
	+\delta(t)\frac{\tilde{U}}{4}\left[3\delta n_2^2+n_X^2\right] \;,
\end{align}
which describes a kicked trimer within the quantum phase model and with a tuneable $K_X$ hopping term.  Note that the interaction strength in the kicked model has the units of $\hbar$ and is therefore given the symbol $\tilde{U}$ in Eq.\ (\ref{eq:dQPM}) in order to distinguish it from the original interaction energy $U$ in Eq.\ \eqref{eq:QuantumThreeMode}. We recall that the QPM is a valid approximation to the BH model when the mode occupation numbers are large, so that the square root factors can be neglected from $H_{\text{MF}}$ in Eq.\ \eqref{eq:HMF}.
In fact, we shall see that the QPM gives circularly symmetric caustics and will consider small corrections to the QPM later on in Section \ref{sec:Stability} in order to unfold the $X_{9}$ catastrophe that is orchestrating the dynamics from the shadows. To quote Nye \cite{Nye1986}: `...generic unfoldings may be best understood as perturbations of symmetrical ones'.

 The integrability of the kicked model is evident from the fact that analytic solutions for the classical trajectories can be found and are given in Appendix \ref{Appdx:DeltaKick}. To obtain quantum dynamics under the Floquet operator with $H_{\delta\text{QPM}}$   we re-promote the observables in Eq. \eqref{eq:dQPM} to operators such that they obey the Dirac number-phase commutators $[\hat{\phi},\hat{n}]=\mathrm{i}$.

\subsection{\label{subsec:LinearTrimer}Linear spatial configuration of trimer}
We will first consider the dynamics of the linear spatial configuration of the BH trimer which means we set $K_X=0$ in Eq.\ (\ref{eq:dQPM}), and further specialize to starting from an equal superposition of Fock states 
\begin{equation}\label{eq:FockSuper}
	\ket{\Psi(0)}=\sum_{\delta n_{2}',n_{X}'}\ket{\delta n_{2}',n_{X}'}\; .
\end{equation}
This state is a plane wave in the number difference basis but corresponds to a single phase state, $\ket{\phi_{X}=0,\phi_{C}=0}$. Being more concrete than we have in the earlier parts of this paper, we define the phase states as eigenstates of the phase operators and are the three-mode generalization of phase difference Bargmann states studied in Ref.\ \cite{Anglin2001}. These phase states are overcomplete and thus not strictly orthogonal for finite $N$ \cite{Mossmann2006}, but in what follows we operate on the assumption that $N\gg 1$ is large enough to approximate a complete set of states. Furthermore, in such a semiclassical regime we will take sums to be continuous integrals when convenient. 

The time-dependent state after undergoing Floquet evolution is given by 
\begin{equation}\label{eq:FockSuper2}
	\ket{\Psi(t)}=\sum_{\delta n_{2}',n_{X}'} \mathrm{e}^{-\frac{\mathrm{i}}{\hbar}\hat{\Phi}t}  \mathrm{e}^{-\frac{\mathrm{i}}{\hbar}\frac{\tilde{U}}{4}(3\delta n_{2}'^{2}+n_{X}'^{2})}  \ket{\delta n_{2}',n_{X}'}\;,
\end{equation}
where $\hat{\Phi}=\Phi(\hat{\phi}_{X},\hat{\phi}_{C})$ is the part of the Hamiltonian containing all the phase operators.  
To obtain the wavefunction in Fock space we project $\ket{\Psi(t)}$ onto the Fock basis by applying $\bra{\delta n_{2}, \delta n_{X}}$ and insert a resolution of identity $\scriptstyle\mathds{1}=\sum_{\phi_{X},\phi_{C}} \ket{\phi_{X}, \phi_{C}} \bra{\phi_{X}, \phi_{C}}$ between the exponential term and the ket $ \ket{\delta n_{2}',n_{X}'}$. Making use of relations such as $\langle \phi_{X}, \phi_{C}  \vert \delta n_{2}',n_{X}' \rangle = \exp [-\mathrm{i} (n_{X}' \phi_{X} + \delta n_{2}' \phi_{C}) ]$ and evaluating Gaussian integrals over the variables $n_{X}'$ and $\delta n_{2}'$ which appear at most quadratically,  we arrive at Eq.\ \eqref{eq:WavefunctionSum}.
Next, since the phase variables $\{\phi_{X},\phi_C\}$ are localized around zero, at least for short times, we expand the cosine terms to fourth order. Finally, under a change of variables, $\phi_C\to 18^{1/4}\phi_C$ and $\phi_X\to 18^{1/4}\phi_X$, we obtain Eq.\ \eqref{eq:KickedX9} which is in a form recognizable as the diffraction integral of $X_9$, although the symmetry of the QPM Hamiltonian and the initial state we have chosen restricts the unfolding so that the terms $C_3$, $C_6$, and $C_7$ do not yet appear (the remaining symmetries will be broken in Section \ref{sec:Stability}).  Note that both Eqns.\ (\ref{eq:WavefunctionSum}) and (\ref{eq:KickedX9}) have been written in a slightly more general form than necessary for the linear spatial configuration of the trimer by including the $K_{X}$ term so that they also apply to the triangular case discussed in Section \ref{sec:triangulardimer}.

\begin{widetext}
	\begin{align}\label{eq:WavefunctionSum}
		\Psi(\delta n_2,n_X,t)= \frac{4 \pi \hbar}{ \mathrm{i} \sqrt{3}  \tilde{U}}\sum_{\phi_X,\phi_C} \exp\bigg[\mathrm{i}\frac{2NJt}{3\hbar} & \left(\cos\left(\phi_X-\phi_C\right)+\cos\left(\phi_X+\phi_C\right)+\frac{K_X}{J}\cos\left(2\phi_X\right)\right) \nonumber \\ & +n_X\phi_X+\delta n_2\phi_C+\frac{\hbar}{3\tilde{U}}(\phi_C^2+3\phi_X^2) \bigg]
	\end{align}
	\begin{equation}\label{eq:KickedX9}
		\psi(\delta n_2,n_X,t)=A(t)\int\int\mathrm{d}\phi_X\mathrm{d}\phi_C\;\exp\Biggl[\mathrm{i}\frac{NJt}{\hbar}\biggl(\phi_X^4+K\phi_X^2\phi_C^2+\phi_C^4+\alpha\phi_X^2+\beta\phi_C^2+\zeta\phi_X+\eta\phi_C\biggr)\Biggr]
	\end{equation}
\end{widetext}

\begin{figure*}[t]\centering
        \includegraphics[width=0.46\columnwidth]{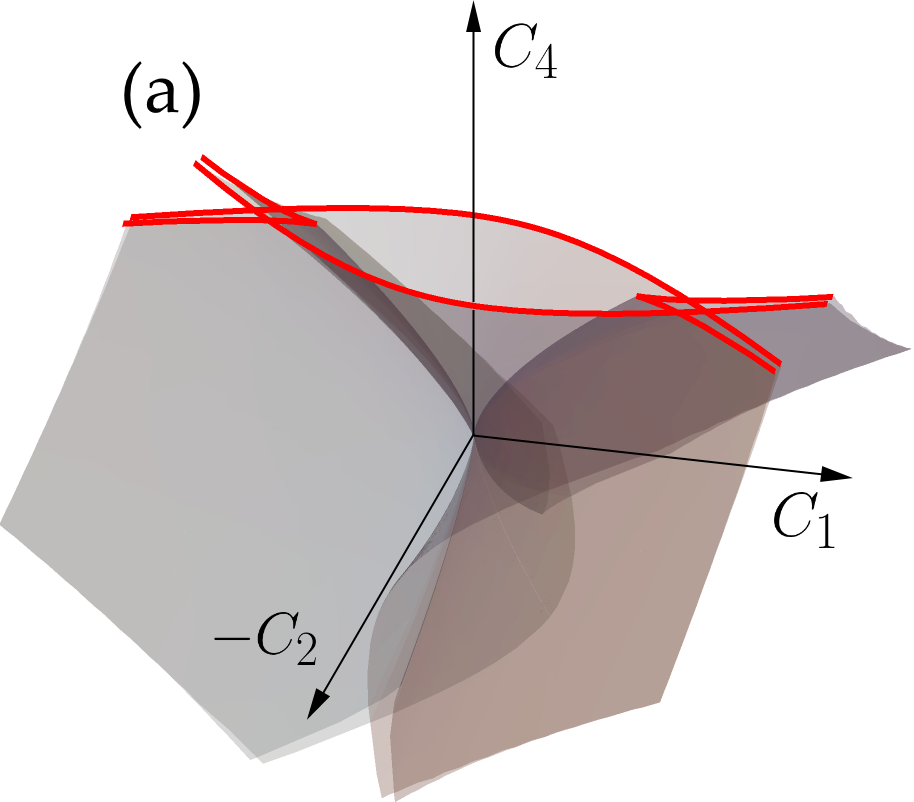}
	\includegraphics[width=0.54\columnwidth]{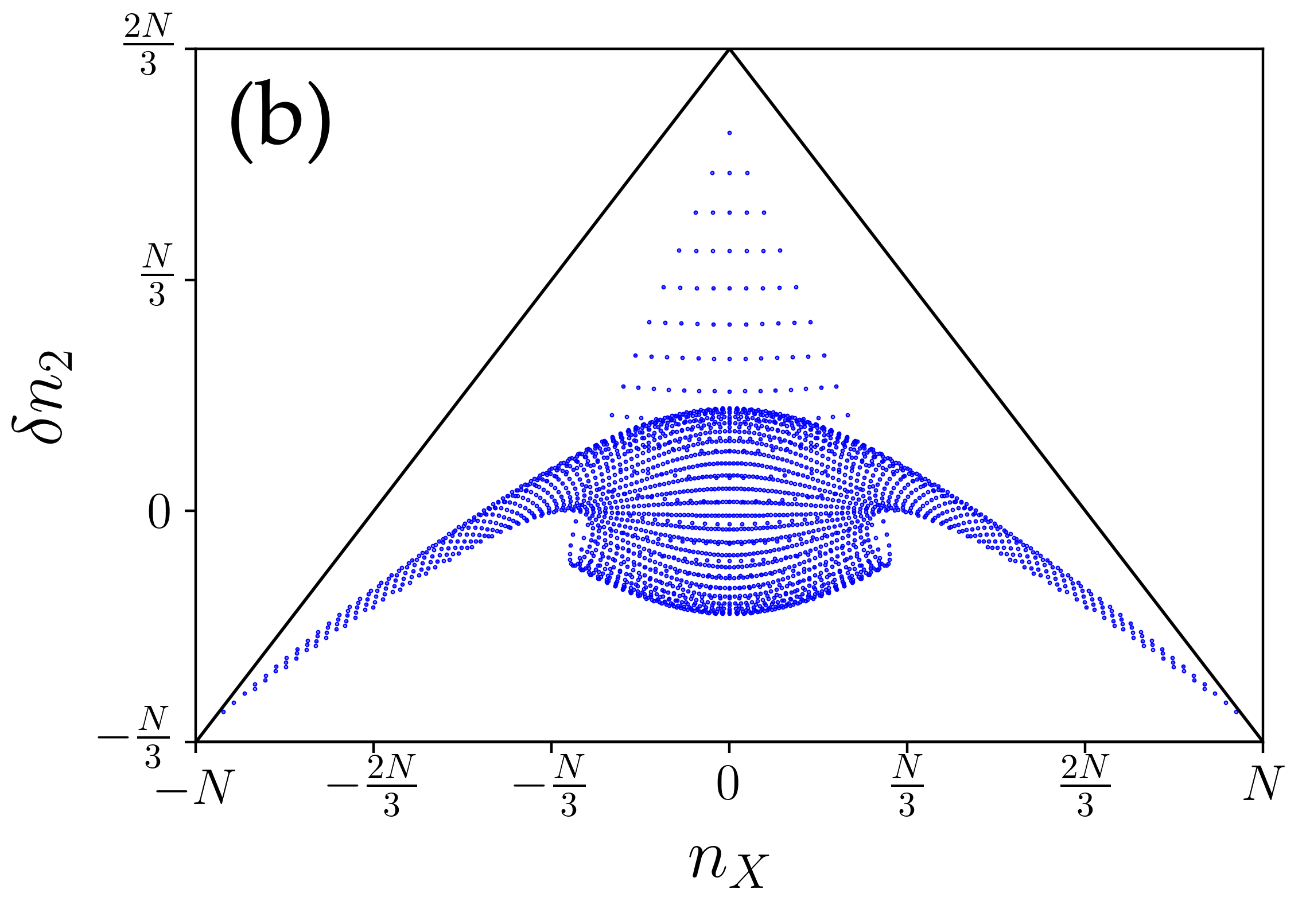}
	\includegraphics[width=0.54\columnwidth]{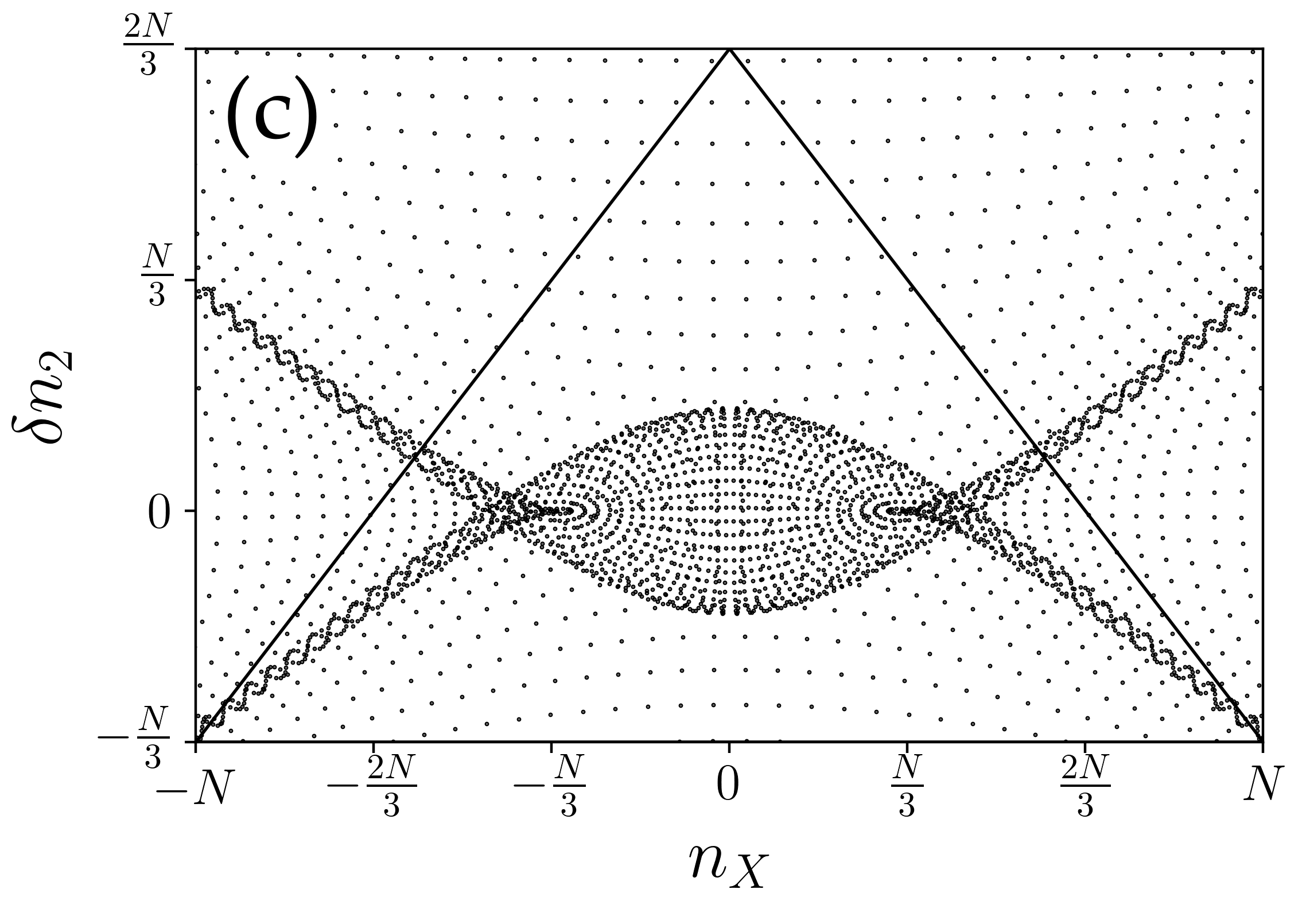}
	\includegraphics[width=0.49\columnwidth]{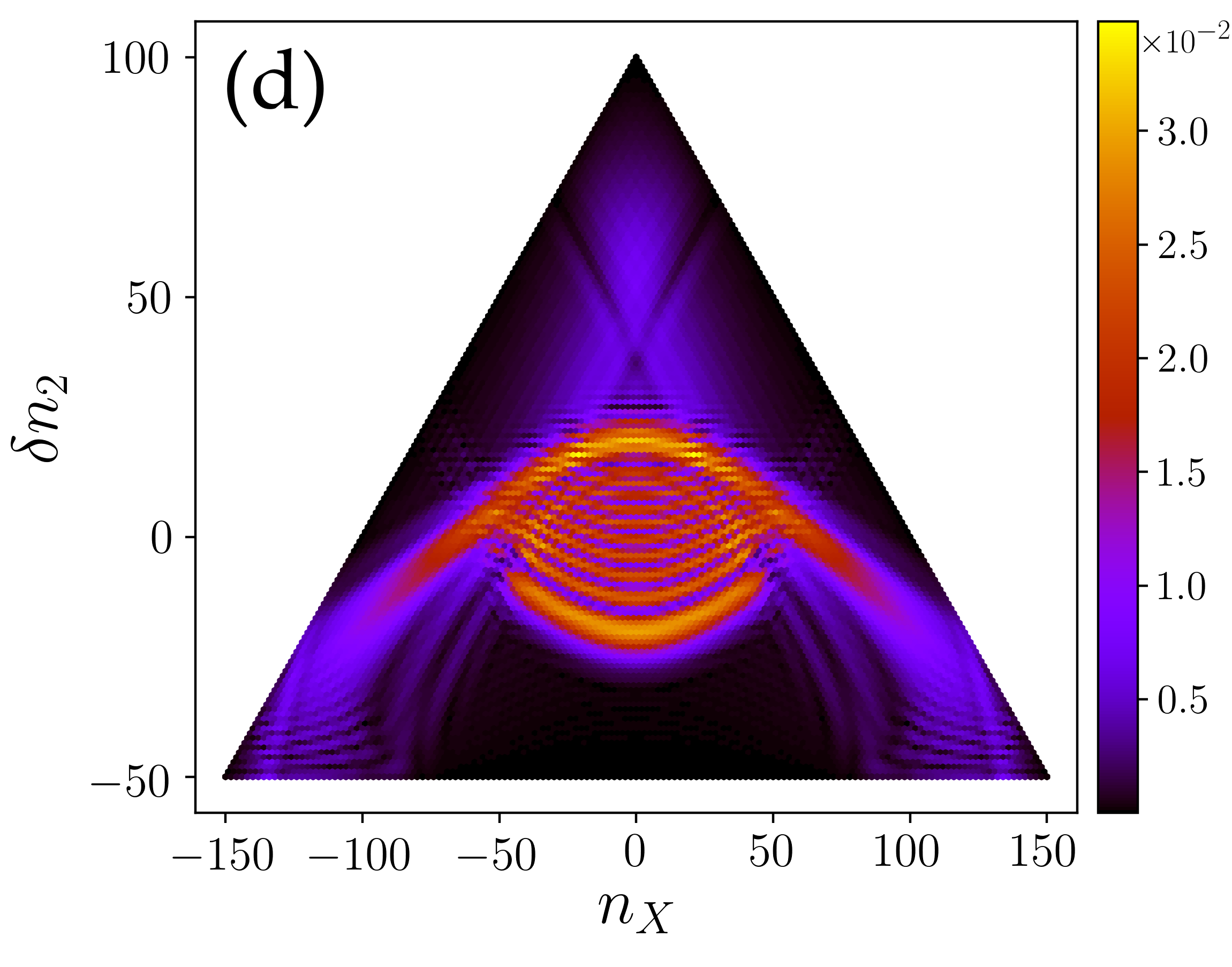}
	\caption{Hyperbolic sections of $X_{9}$ in the classical and quantum BH trimer.  \textbf{Panel (a):} Caustic surface for $\Phi_{X_9}^H$, as given in Eq.\ \eqref{eq:X9Hyperbolic}. The red highlights mark the intersection with the plane $C_4=\sqrt{2}$.  \textbf{Panel (b):} Distribution of points arising from classical trajectories of the $\delta$QPM Hamiltonian Eq.\ \eqref{eq:dQPM} (see Appendix \ref{Appdx:DeltaKick} for solutions of Hamilton's equations) for the linear trimer configuration $(K_X=0)$ at $t=\hbar^2/(JN\tilde{U})$ and $\alpha=C_{4}=\sqrt{2}$, starting in an equal spread of number differences, and with phase differences $\phi_{C}(0)=\phi_{X}(0)=0$. The resulting caustic is a partial section through the hyperbolic unfolding of $X_9$, restricted by the triangular Fock space (black solid lines) of physical classical paths. \textbf{Panel (c):} Same snapshot as in panel (b), now including some unphysical paths (such as those which have started or ended with $n_X>N$), to demonstrate how the caustic is restricted by the shape of Fock space (number conservation). \textbf{Panel (d):} Wavefunction amplitude at the same moment as panels (b) and (c), starting from an equal superposition of Fock states \eqref{eq:FockSuper} and evolved using the RN equations corresponding to the $\delta$QPM model (square root factors set to unity). Here, $N=150$, and $\tilde{U}/\hbar=0.02$. \label{fig:K6Hyperbolic}}
\end{figure*}

For the linear spatial configuration trimer, the modulus in Eq.\ (\ref{eq:KickedX9}) comes out to be $K=6$, and the control parameters are given by
\begin{align}
	\alpha=&\;\frac{3\sqrt{2}\hbar^2}{NJt\tilde{U}}-2\sqrt{2} \label{eq:alphadefinition}  \\%\overset{NJtU\gg \hbar^2}{\to}-2\sqrt{2}\\
	\beta=&\;\frac{\sqrt{2}\hbar^2}{NJt\tilde{U}}-2\sqrt{2}   \label{eq:betadefinition}   \\
	\zeta=&\;\frac{2^{1/4}\sqrt{3}\hbar }{NJt} n_X\\
	\eta=&\;\frac{2^{1/4} \sqrt{3}\hbar }{NJ t}\delta n_2
\end{align}
with,
\begin{equation}		
A(t)=\frac{4\sqrt{6}\pi\hbar}{U}\mathrm{e}^{-\mathrm{i}\frac{\pi}{2}}\mathrm{e}^{\mathrm{i}\frac{4NJt}{3 \hbar}}\;.
\end{equation}
The wave catastrophe described by Eq.\ (\ref{eq:KickedX9}) might appear four dimensional, with coordinates $\{\alpha,\beta,\zeta,\eta  \}$, but examination of Eqns.\ (\ref{eq:alphadefinition}) and (\ref{eq:betadefinition}) reveals that $\alpha$ and $\beta$ are not independent: $\alpha=3 \beta+4 \sqrt{2}$, and therefore it is really three dimensional. Let us consider the particular case $\alpha=-\beta$ which occurs naturally at the time $t=\hbar^2/(JN\tilde{U})$.  This gives a three-dimensional section of $X_{9}$ that is hyperbolic and described by the generating function
\begin{equation}\label{eq:X9Hyperbolic}
	\Phi_{X_9}^H=s_2^4+6s_1^2s_2^2+s_1^4+C_4(s_2^2-s_1^2)+C_2s_2+C_1s_1\;,
\end{equation}
which has previously been studied by Berry and Howls and is relevant to liquid-droplet lenses  \cite{Berry2010}. The corresponding theoretical caustic surface is plotted in panel (a) of Fig.\ \ref{fig:K6Hyperbolic} where a two-dimensional section at $\alpha=C_4=\sqrt{2}$ is highlighted in red. If we move slightly away from $t=\hbar^2/(JN\tilde{U})$, then $\alpha\neq -\beta$ leading to the introduction of a $C_5$ term.

The actual caustic formed by the classical dynamics [with Hamiltonian Eq.\ \eqref{eq:dQPM}] is plotted in Fig.\ \ref{fig:K6Hyperbolic}(b). This clearly resembles the analytical prediction but only half of it is present. This is because of the restriction of physical paths to always lie within the triangular Fock space  $-\frac{N}{3}\leq \delta n_2\leq \frac{2N}{3}$ and $-N\leq n_X\leq N$. If unphysical paths are allowed (i.e.\ non-number conserving paths such as those with $|n_X|>N$ or $n_2>N$), the full caustic is captured, as shown in Fig.\ \ref{fig:K6Hyperbolic}(c). Panel (d) shows the resulting quantum amplitudes in Fock space after time evolution under the corresponding RN equations. The quantum-classical correspondence is clear in this semiclassical regime  and we can identify the hyperbolic $X_9$ boundary, which is now dressed with characteristic interference fringes across the fold lines.

\subsection{Triangular spatial configuration of trimer}
\label{sec:triangulardimer}

\begin{figure}[t]\centering
	\includegraphics[width=0.49\columnwidth]{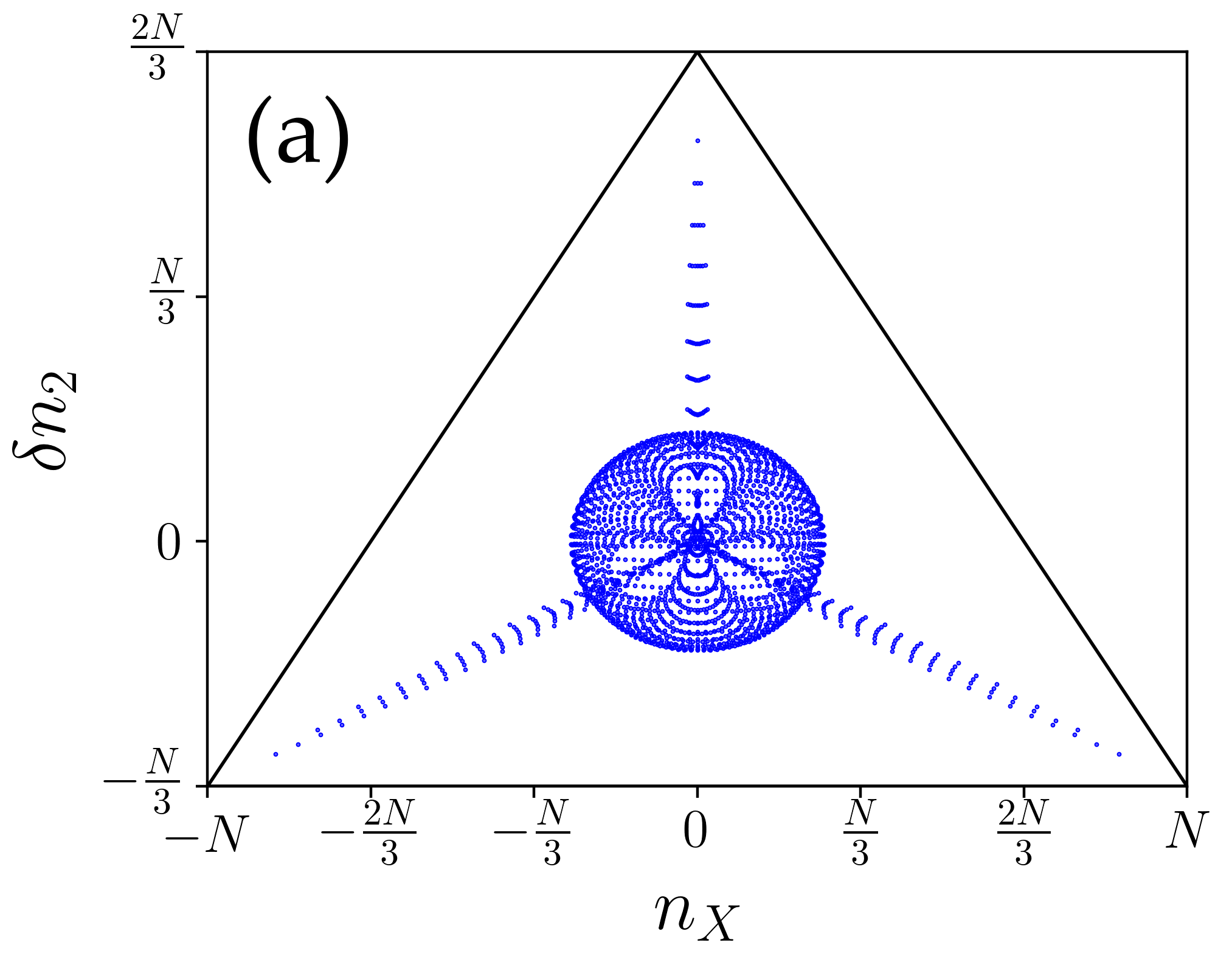}
	\includegraphics[width=0.49\columnwidth]{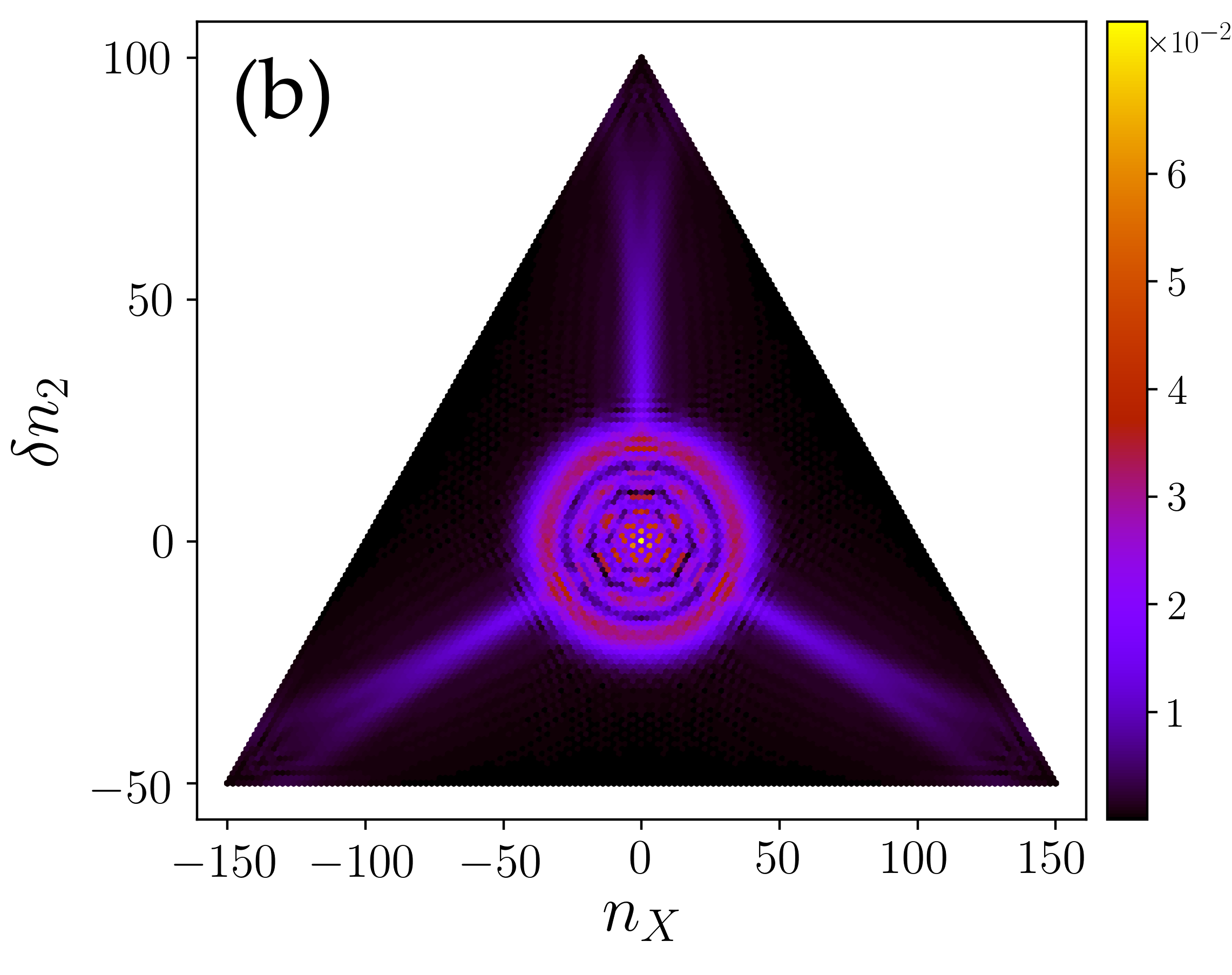}
	\includegraphics[width=0.4\columnwidth]{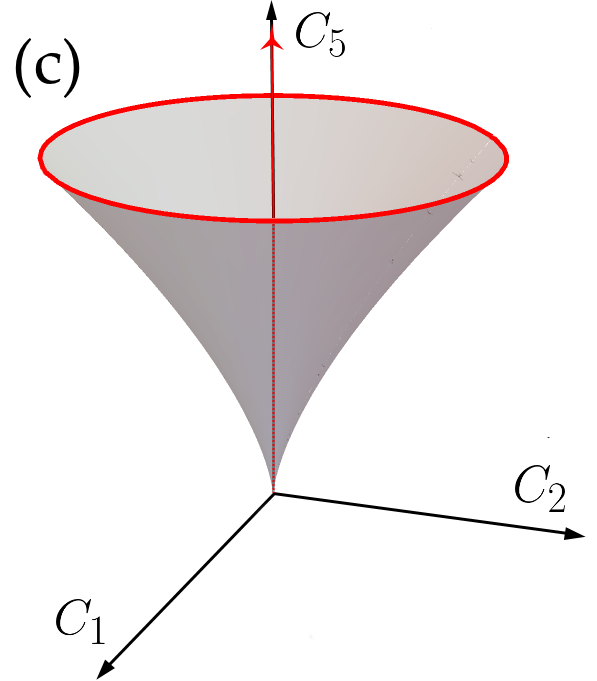}
	\caption{The spun cusp caustic generated by the triangular BH trimer $(K_X=J)$ within the $\delta$QPM.  \textbf{Panel (a):} Distribution of points arising from classical trajectories of the $\delta$QPM Hamiltonian Eq.\ \eqref{eq:dQPM} (see Appendix \ref{Appdx:DeltaKick} for solutions of Hamilton's equations) at time $t=\hbar^2/(JN\tilde{U})$ (twice $t_{\text{cusp}}$), starting from an equal spread of number differences, and with phase differences $\phi_{C}(0)=\phi_{X}(0)=0$. The circular caustic from the $K=2$ section of $X_9$ is clearly visible. \textbf{Panel (b):} Quantum wavefunction amplitude at the same time as panel (a), starting from an equal superposition of Fock states \eqref{eq:FockSuper} and evolved using linearized RN equations. $N=150$, and $\tilde{U}/\hbar=0.02$.  \textbf{Panel (c):} Caustic surface for the spun cusp, for negative $C_5$. The 2D circular caustic and the axial caustic are highlighted in red.\label{fig:K2SpunCusp}}
\end{figure}

Let us now consider the $\delta$QPM Hamiltonian Eq.\ \eqref{eq:dQPM} with $K_X=J$, corresponding to a system of three sites arranged in an equilateral triangle such that all three hopping amplitudes are equal. We follow the same procedure as the linear configuration case, except that in order to 
bring the wavefunction to the canonical form given in Eq.\ \eqref{eq:KickedX9} the two phase variables must be scaled differently:  $\phi_C\to 2^{1/4}\phi_C$ and $\phi_X\to 18^{1/4}\phi_X$.  We thereby obtain a modulus $K=2$ and find the following mapping between physical quantities and abstract control parameters,
\begin{align}
	\alpha=\beta=&\;\frac{\sqrt{2}\hbar^2}{\tilde{U}JNt}-2\sqrt{2}\label{eq:X9K2alphabeta}\\
	\zeta=&\;\frac{2^{1/4}\hbar}{NJt}n_X\label{eq:X9K2zeta}\\
	\eta=&\; \frac{2^{1/4}\sqrt{3}\hbar}{NJt}\delta n_2\label{eq:X9K2eta}
\end{align}
and,
\begin{equation}	A(t)=\frac{4\sqrt{2} \pi\hbar}{\tilde{U}}\mathrm{e}^{-\mathrm{i}\frac{\pi}{2}}\mathrm{e}^{\mathrm{i}\frac{2NJt}{\hbar}}\;.
\end{equation}
Once again we find that $\alpha$ and $\beta$ are not independent which this time leads to a spun cusp surrounding an axial caustic line described by the generating function
\begin{equation}
	\Phi_{X_9}^{\text{circ}}=s_2^4+2s_1^2s_2^2+s_1^4+C_5(s_2^2+s_1^2)+C_2s_2+C_1s_1 
\end{equation}
and pictured in Fig.\ \ref{fig:K2SpunCusp}. 
The axial caustic line is not generic and is in fact unstable. It arises from the rotational symmetry we have assumed in the $\delta$QPM with $K_X=J$. Physically speaking, the combination of isotropic hopping along with the simplified quantum phase Hamiltonian causes the system to not `feel' the triangular symmetry, and results in a perfectly circularly symmetric structure in Fock space. From the parameters \eqref{eq:X9K2alphabeta}, we can read off that the time at which the cusp point occurs is $t_{\text{cusp}}=\hbar^2/(2\tilde{U}JN)$. 

Fig.\ \ref{fig:K2SpunCusp} shows the dynamics of the wavefunction \eqref{eq:KickedX9} with control parameters given in Eqs.\ \eqref{eq:X9K2alphabeta}-\eqref{eq:X9K2eta} at $t=\hbar^2/(\tilde{U}JN)$, twice the cusp time. Panel (a) shows the circular caustic formed by the classical trajectories by mimicking the initial state Eq.\ \eqref{eq:FockSuper} with an even spread of initial points in Fock space. Panel (b) shows the wavefunction amplitude under the same conditions as (a), with clearly visible interference effects and a bright central Fock state amplitude $|\braket{0,0|\psi(t)}|$ corresponding to the axial caustic. The canonical spun cusp caustic surface is shown in panel (c), with a two-dimensional circular caustic outlined in red.  In Appendix \ref{Appdx:KXDependence} we give the control parameters for the  $X_9$ wavefunction [Eq.\ (\ref{eq:KickedX9})] for any intermediate value of $K_X$ between the linear and equilateral triangle configurations.

\subsection{Gaussian Initial states}
\label{sec:gaussian}

\begin{figure}[t]\centering
	\includegraphics[width=0.49\columnwidth]{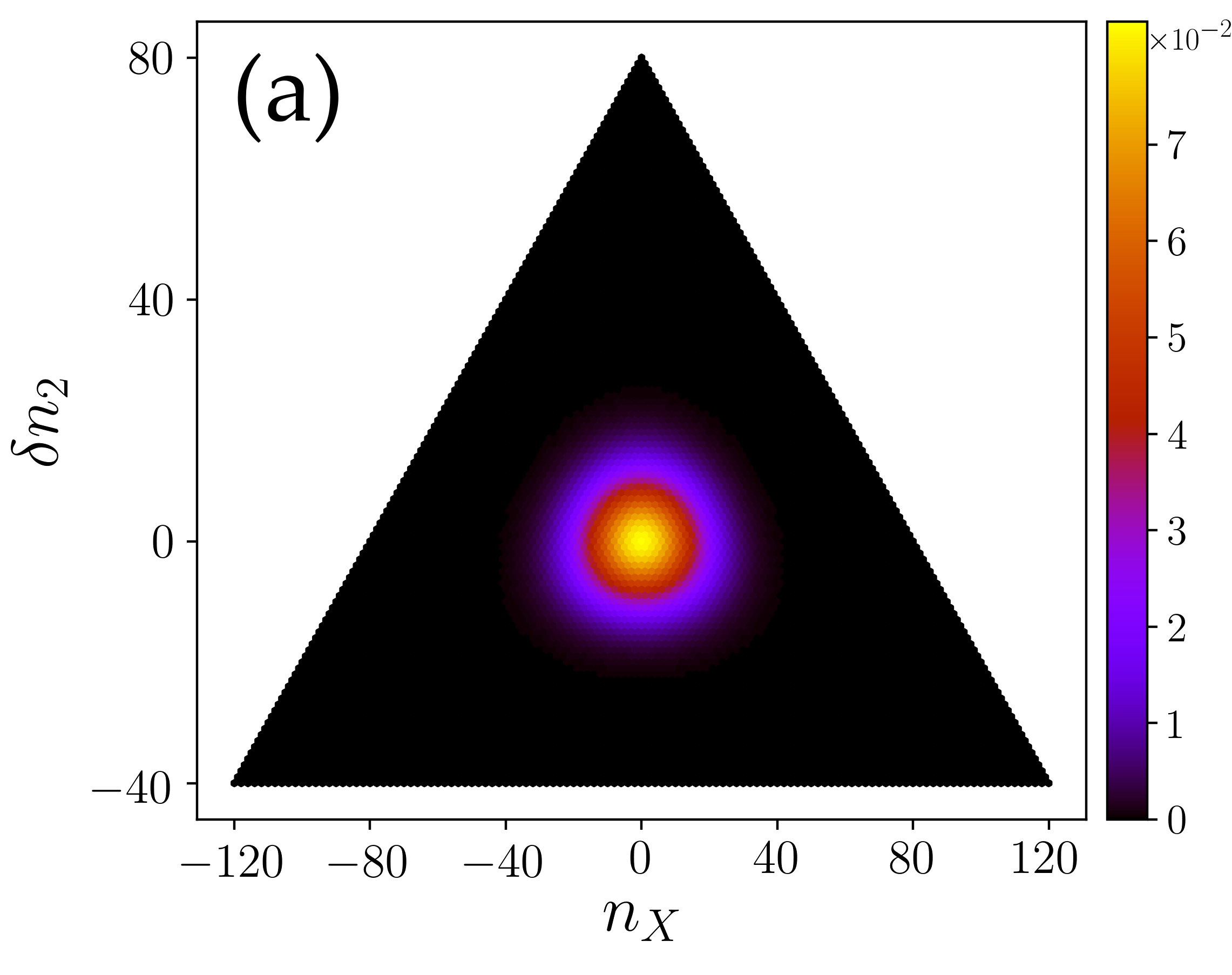}
	\includegraphics[width=0.49\columnwidth]{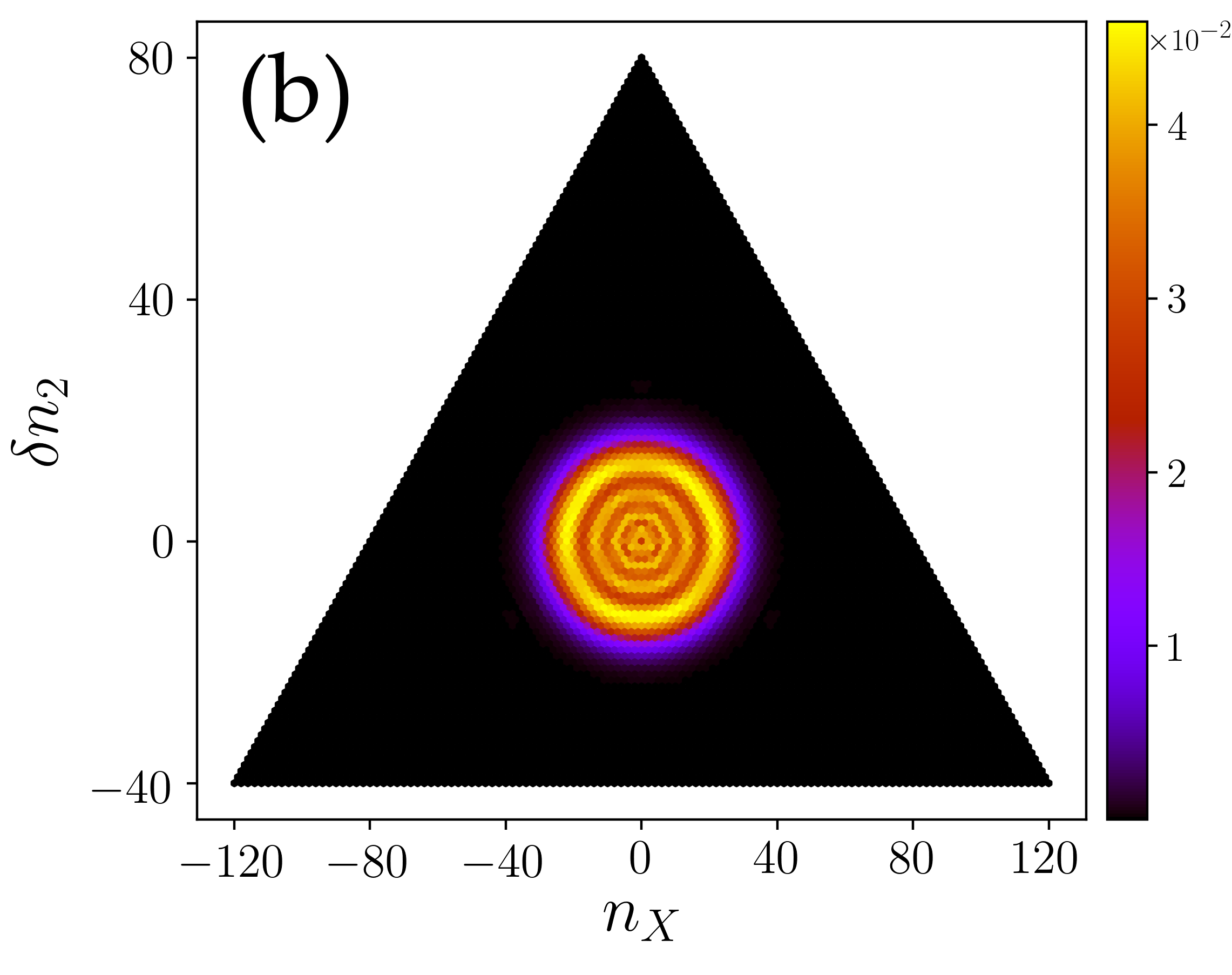}
	\caption{\label{fig:KickedGS} A Gaussian initial state gives the same caustic as a `plane wave'. \textbf{Panel (a):} The amplitude of the ground state of the noninteracting trimer model for $K_L=K_X=K_R\equiv J$ and $N=120$ has a Gaussian form. \textbf{Panel (b):} Time evolution of the ground state using the kicked Hamiltonian with $\tilde{U}/\hbar=0.08$, at time $t=2\hbar^2/(JN\tilde{U})$. Comparing with Fig.\ \ref{fig:K2SpunCusp} we see a strong resemblance indicating that initial states which differ significantly but have the same general form (flat near the center of Fock space) will give rise to qualitatively similar caustics. }
\end{figure}

Using the same system parameters, let us revisit our earlier choice of initial state as an equal superposition of Fock states [Eq.\ \eqref{eq:FockSuper}] which is the Fock space analogue of a plane wave. The primary goal of this choice is to demonstrate the natural focusing effect of the BH dynamics. We claim that any sufficiently wide spread of initial Fock states will yield qualitatively similar results with only slightly altered coefficients. 
As an example, consider the ground state of the trimer model \eqref{eq:QuantumThreeMode} for $U=0$,
\begin{equation}
\ket{\psi_G}=\sum_{n_1+n_2+n_3=N}\sqrt{\frac{N!}{3^Nn_1!n_2!n_3!}}\ket{n_1,n_2,n_3}
\end{equation}
where the sum is over all Fock space occupation numbers such that $n_1+n_2+n_3=N$. For more detail on the diagonalization of this model, see, e.g., Ref.\ \cite{Kolovsky2015} and references therein. In the limit of large $N$, this coherent state closely resembles a Gaussian centered at $\delta n_2=n_X=0$. After changing variables,
\begin{equation}
	\ket{\psi_G}\approx\frac{3^\frac{3}{4}}{\sqrt{2\pi N}}\sum_{\delta n_2,\;n_X}\mathrm{e}^{-\frac{3\ln 3}{8N}(3\delta n_2^2+n_X^2)}\ket{\delta n_2,n_X} \ .
\end{equation}
Note that the coefficients of $\delta n_2^2$ and $n_X^2$ are not the same since $n_X$ ranges from $-N$ to $N$ and $\delta n_2$ ranges from $-N/3$ to $2N/3$.
The assumption of localized phase-state contributions around $\bm{\phi}=0$ still applies, since,
\begin{equation}
	\braket{\phi_X,\phi_C,\Theta|\psi_G}\approx\frac{ 4\sqrt{2\pi N}}{3^{\frac{3}{4}}\ln 3}\mathrm{e}^{\mathrm{i}\Theta N/3}\mathrm{e}^{-\frac{2N}{9\ln 3}(\phi_C^2+3\phi_X^2)}\;,
\end{equation} 
which for large $N$ becomes narrowly peaked around $\phi_{X}=\phi_{C}=0$, while widely spread in the number difference values.  Starting from this initial state, we use the Floquet operator to propagate and the results are shown in Fig.\ \ref{fig:KickedGS} where we see that despite the fact that the Gaussian differs significantly from a plane wave we still obtain qualitatively the same caustic as in Fig.\ \ref{fig:K2SpunCusp}.  Note that the result of applying the Floquet operator is to take $U\to U-\mathrm{i}\frac{3\ln 3}{2N}$, implying that for finite $N$ there is no longer a well-defined cusp point since the solution $\alpha=\beta=0$ cannot occur for a real value of $t$. The outer circular caustic surface does, however, remain and we see a diffraction pattern reminiscent of the $K=2$ spun cusp. Furthermore, in the semiclassical regime $N \gg 1$ we can closely approximate a cusp point.

\begin{figure*}[t]\centering
	\begin{tabular}{cccc}
		$H_{\delta\mathrm{QPM}}$& $H_{\delta\mathrm{MF}}$ & $H_{\triangle}$ & Quantum wavefunction\\
		\includegraphics[width=0.49\columnwidth]{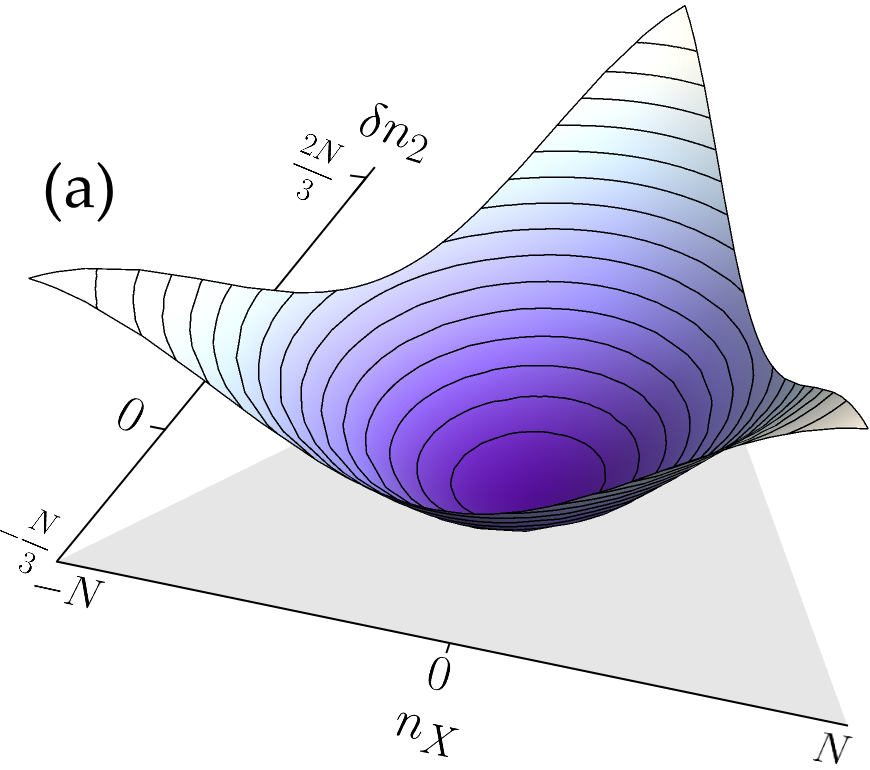}&
		\includegraphics[width=0.49\columnwidth]{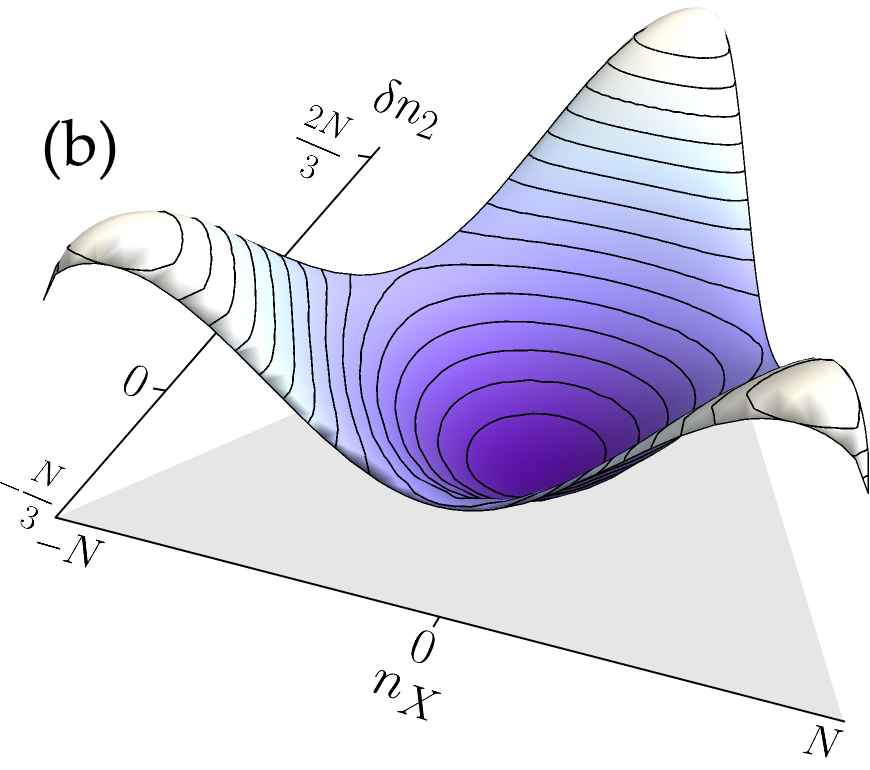}&
		\includegraphics[width=0.49\columnwidth]{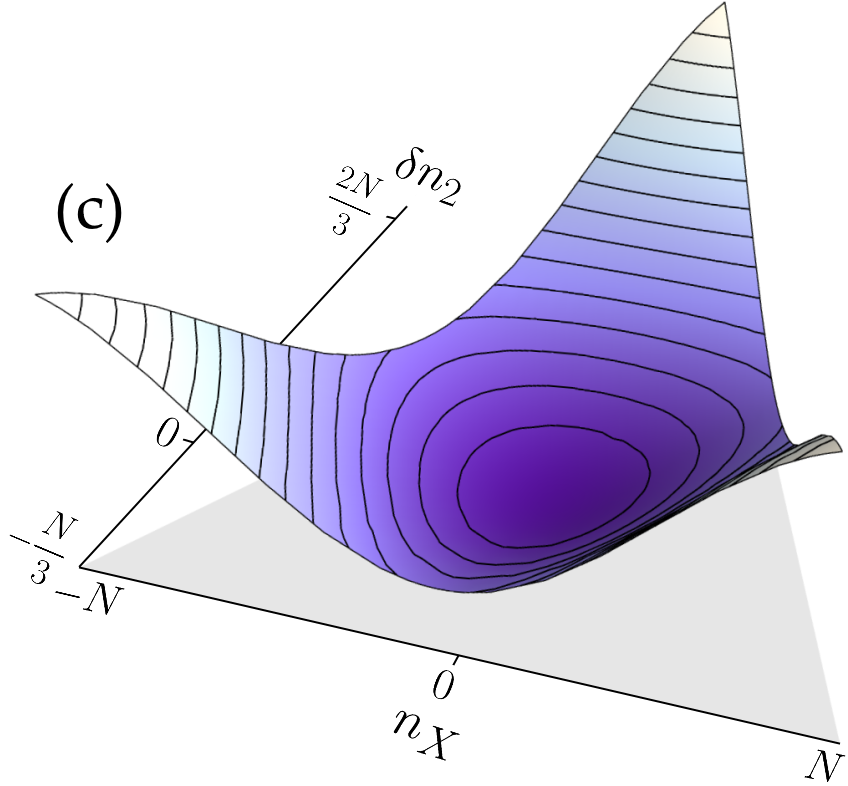}&
		\includegraphics[width=0.49\columnwidth]{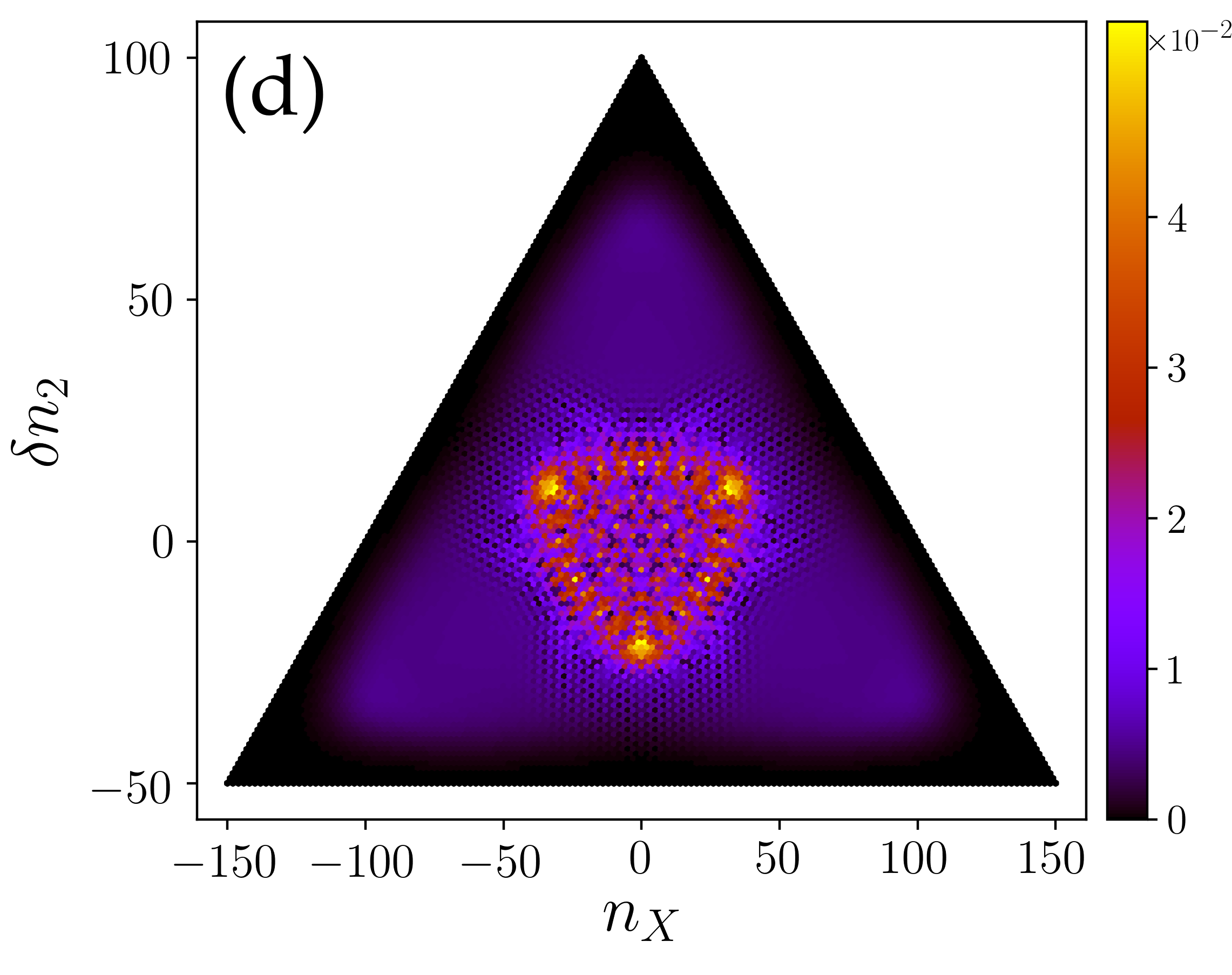}
	\end{tabular}
	\caption{\label{fig:DeformedCusp} Breaking circular symmetry by going beyond the QPM. \textbf{Panel (a):} The energy surface determining classical trajectories as provided by $H_{\delta \mathrm{QPM}}$ in Eq. \eqref{eq:dQPM} immediately following the $\delta$-kick at $t=0$. The initial values of the dynamical variables that generate this surface are  $\phi_{X}=\phi_{C}=0$ and an equal superposition of all number differences; the phase trajectories are bent by the $\delta$-kick as explained in Appendix \ref{Appdx:DeltaKick}.  The overall scale is arbitrary and lighter colours represent higher energy. Note the circular symmetry as seen from the contour lines.   \textbf{Panel (b):}  Same as panel (a) but now with the $\delta$-kicked mean-field Hamiltonian $H_{\delta \mathrm{MF}}$ given in Eq. \eqref{eq:dMF} with all hopping amplitudes equal. The circular symmetry is broken by the square root factors and the resulting three-fold symmetric `valleys' will act to pinch the caustic away from the corners of Fock space. \textbf{Panel (c):} Same as panel (a) but now with $H_{\triangle}$ as given in Eq.\ (\ref{eq:KickedFirstOrder}). In this Hamiltonian the square root factors have been expanded to first order and this is enough to break the circular symmetry and replace it with a three-fold symmetric energy surface. \textbf{Panel (d):} Quantum wavefunction obtained using the full RN equations with $\delta$-kicked interactions (quantum equivalent of $H_{\delta \mathrm{MF}}$) and the same parameters as used in Fig. \ref{fig:K2SpunCusp} (b). The chosen time is $t=\hbar^2/(JN\tilde{U})$ which is twice that of where the cusp point appears. }
\end{figure*}

\section{\label{sec:Stability} Beyond the quantum phase model}
In the QPM the effects of depletion of the modes [accounted for by the square root factors in the RN equations Eq.\ \eqref{eq:RamanNath}] are ignored. This is often a good approximation in experimentally realizable superfluids, including arrays of Josephson junctions \cite{Otterlo1995,Fazio01} and also some regimes of atomic BECs in optical lattices \cite{Cataliotti2001,Orzel2001,Hadzibabic2004,Schori2004,Xu2006,Schweikhard2007}. However, in terms of caustics it can lead to some special situations such as the appearance of a $K=2$ spun-cusp caustic for the triangular trimer with $K_{X}=J$, as discussed in the previous section. This has circular symmetry which is of course non-generic and, indeed, the modulus $K=2$ is excluded from the $X_9$ catastrophe. The symmetry is broken by any perturbation which is noncircular, such as choosing $K_X \neq J$.  Counter-intuitively, another way to break the symmetry is to go beyond the QPM by including the effects of the square root factors. For our purposes of illustrating generic many-body caustics, it suffices to expand the square roots and keep just the first order corrections because this is enough to generate the $X_9$ catastrophe (even when all the hopping amplitudes are equal), thereby providing an instructive example of structural stability, or lack thereof, for caustics that do not correspond to catastrophes.

\subsection{Triangular deformations: a path integral formulation }

The mean-field Hamiltonian that retains the square root factors, has  $\delta$-kicked interactions and all-equal hopping is given by 
	\begin{align}
		 H_{\delta\text{MF}} & =-J\sqrt{2(\delta n_2+\tfrac{N}{3})(n_X-\delta n_2+\tfrac{2N}{3})}\cos(\phi_X-\phi_C) \nonumber \\  -J & \sqrt{2(\delta n_2+\tfrac{N}{3})(-n_X-\delta n_2+\tfrac{2N}{3})}\cos(\phi_X+\phi_C) \nonumber \\ -J & \sqrt{(\delta n_2-\tfrac{2N}{3})^2-n_X^2}\cos(2\phi_X)+\delta(t)\frac{\tilde{U}}{4}\left[3\delta n_2^2+n_X^2\right]  \label{eq:dMF} .
		 \end{align}
	
In Fig.\ \ref{fig:DeformedCusp}(a) and (b) we compare the energy surfaces produced by  $H_{\delta\text{QPM}}$ and $H_{\delta\text{MF}}$. It is evident from the contour lines that  $H_{\delta\text{MF}}$ breaks the circular symmetry and replaces it with a triangular one. However, as explained above, our interest is more in generic many-body caustics rather than specific  models, so all we really need to do is perturb away from the QPM and hence we expand the square roots in $H_{\delta\text{MF}}$ and keep only the first order corrections:
\begin{align}
	H_{\triangle}=&\;-\frac{J}{2}\left(n_X+\delta n_2+\frac{2N}{3}\right)\cos\left(\phi_X-\phi_C\right)\nonumber\\
	&-\frac{J}{2}\left(-n_X+\delta n_2+\frac{2N}{3}\right)\cos\left(\phi_X+\phi_C\right) \nonumber \\
	&-J\left(\frac{2N}{3}-\delta n_2\right)\cos\left(2\phi_X\right)
	+\delta(t)\frac{\tilde{U}}{4}\left[3\delta n_2^2+n_X^2\right]\nonumber \\
	& \equiv \; NJ\Phi_{\triangle}+\delta(t)\frac{\tilde{U}}{4}\left[3\delta n_2^2+n_X^2\right] \ . \label{eq:KickedFirstOrder}
\end{align}
As can be seen from the energy surface in Fig.\ \ref{fig:DeformedCusp}(c),  this `triangular' Hamiltonian retains the basic triangular symmetry possessed by the full mean-field model. The cone shape of the energy surface present in all three models has an overall focusing effect on trajectories, but the triangular shape of the latter two models channels the trajectories away from the three corners of Fock space and effectively pinches the outer part of the caustic in three spots. This can be seen in the quantum version shown in panel (d) created using the RN equations under no approximations at $t=2t_{\text{cusp}}=\hbar^2/(JN\tilde{U})$. The caustic evolves at roughly the same speed in both the classical and quantum cases, and three bright spots appear both before and after the cusp point and correspond to the three valleys on the triangular energy surfaces. Another crucial feature is that there is no bright patch at the center of Fock space, due to the destruction of the axial caustic which is a feature peculiar to circular symmetry.

\begin{figure}[t]
	\centering
	\includegraphics[width=0.49\columnwidth]{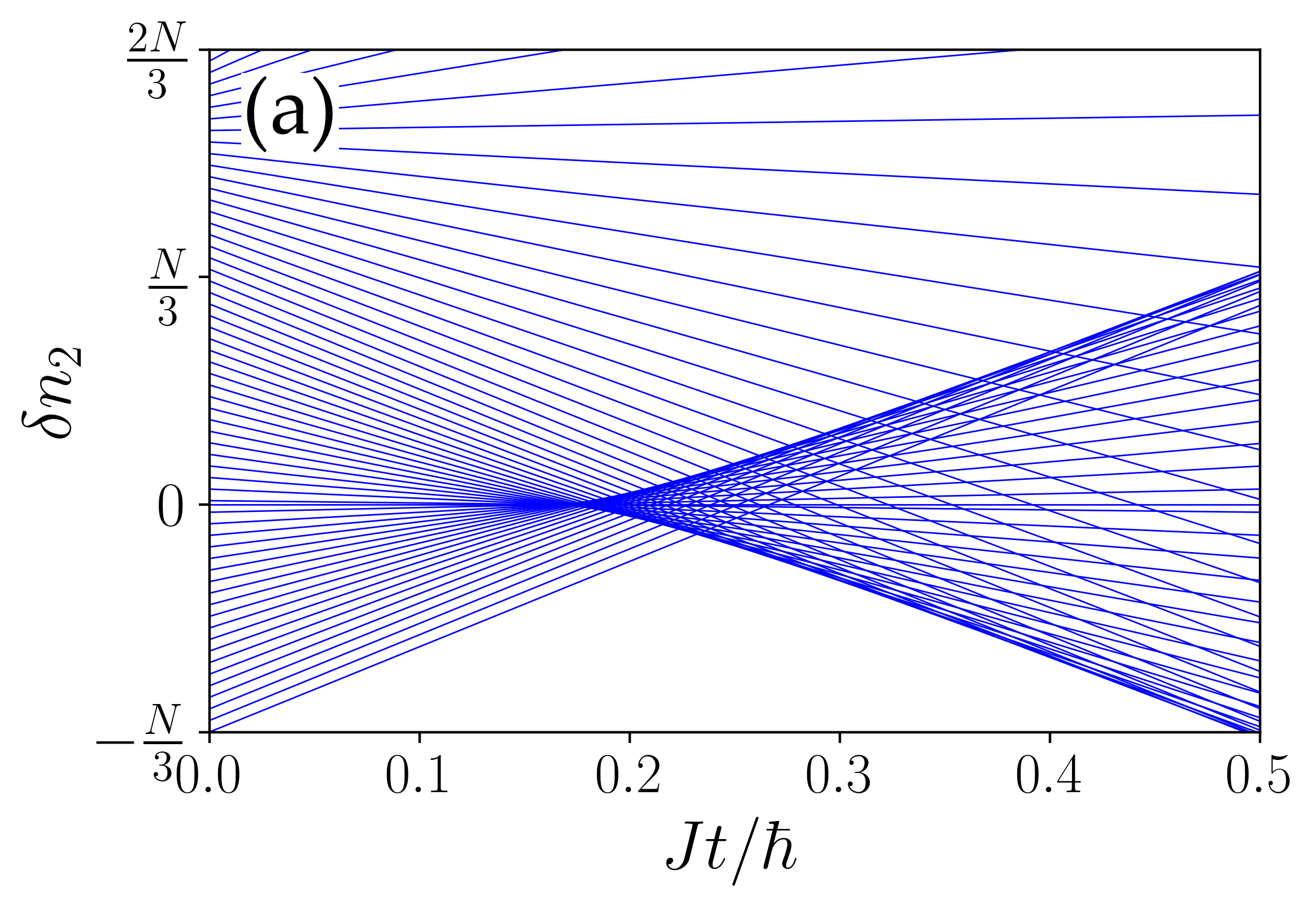}
	\includegraphics[width=0.49\columnwidth]{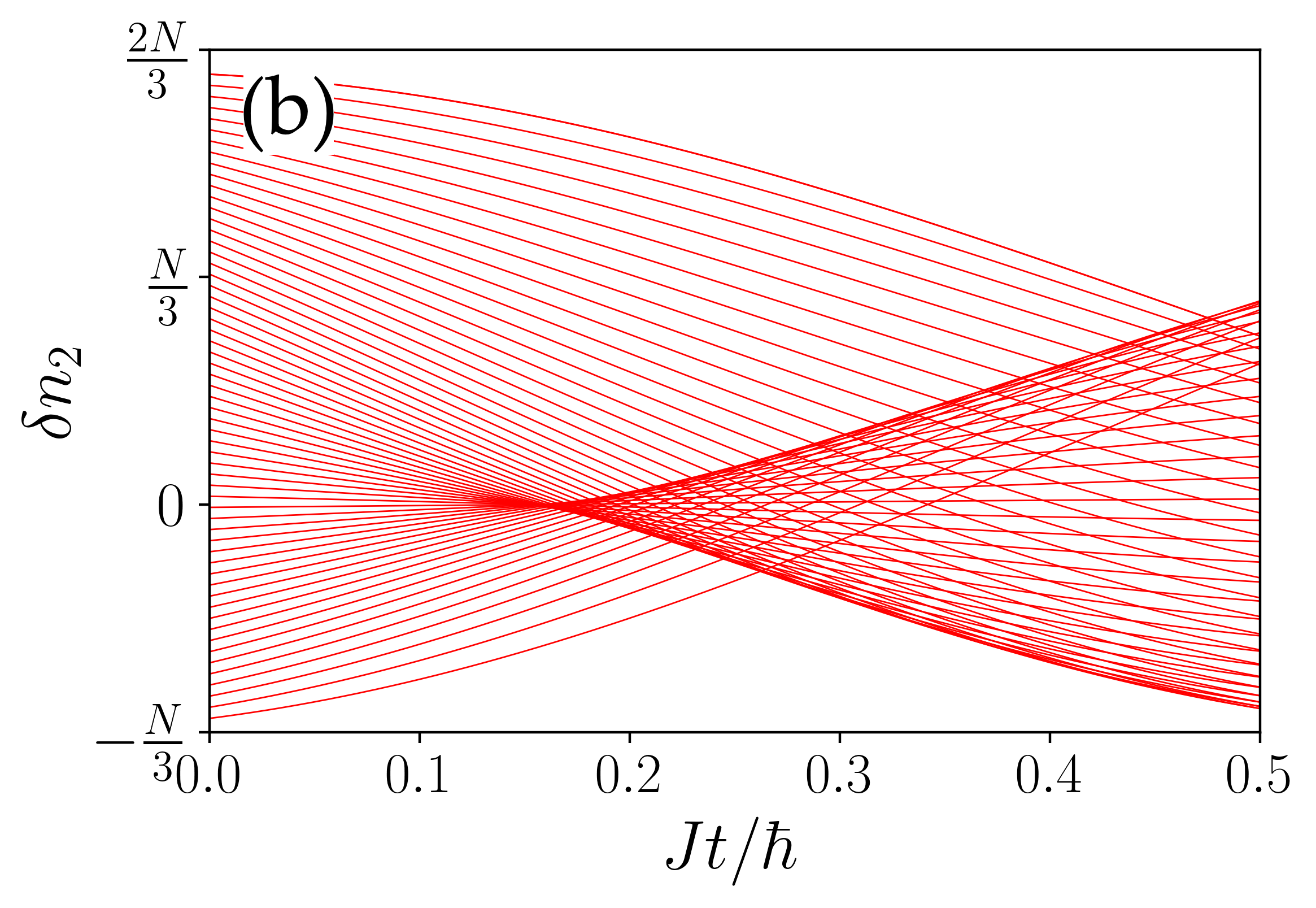}
	\caption{\label{fig:KickedCusp} Stability of cusp formation in the $n_{X}=0$ plane against inclusion of beyond QPM effects. \textbf{Panel (a):} Set of classical trajectories for the $\delta$QPM when $n_X=0$ starting from a range of $\delta n_2$ values and kicked with $\tilde{U}/\hbar=0.02$. The cusp point here is $Jt/\hbar\approx0.167$. \textbf{Panel (b):} Exact trajectories (without approximation to the square root factors) for the triangular trimer with a kicked interaction term as given by $H_{\delta \mathrm{MF}}$. The cusp structure remains intact and is only curved near the edge of Fock space, showing that the $\delta$QPM is a good approximation for this sub-catastrophe and that caustics have structural stability against perturbations.}
\end{figure}

 In order to treat this problem analytically we need to reconsider our approach, since calculating an analytic form for the wavefunction as a single diffraction integral in the same manner as we did for the $\delta$QPM is not possible for the full Hamiltonian. 
More precisely, in the $\delta$QPM there is a separation between the phase and number difference variables, like in a standard $H=p^2/2m + V(x)$ type Hamiltonian, allowing us to integrate out the number (momentum) variable using gaussian integrals leaving a diffraction integral purely in terms of the phase (position) variables. The presence of the square root factors effectively gives us a potential $V(x,p)$ which depends on both position and momentum variables. This difficulty is not reduced by expanding out the square root factors and we instead resort to a phase space path integral-style formulation for the wavefunction: 
\begin{align}\label{eq:PathIntegralWaveFn}
	\psi(n_X,\delta n_2,t)=&\;\braket{n_X,\delta n_2|\hat{\mathcal{F}}|\psi(0)}\\
	\propto&\;\int\mathcal{D}\bm{\phi}\mathcal{D}\bm{n}\;\mathrm{e}^{\mathrm{i}\mathcal{S}[\bm{n}(\tau),\bm{\phi}(\tau)]} 
\end{align}
where we have introduced the bold vector notation $\bm{\phi}=(\Theta,\phi_X,\phi_C)$ and $\bm{n}=(\frac{N}{3},n_X,\delta n_2)$. In the semiclassical regime the dominant contributions to the path integral come from the evaluation of the action $\mathcal{S}$ along  \textit{classical paths} which we denote by $\{\bar{\bm{n}},\bar{\bm{\phi}}\}$. However, the standard WKB approximation blows up precisely at caustics because these are places where saddles of $\mathcal{S}$ coalesce. Following Schulman \cite{Schulman} and Dangelmayr and Veit \cite{Dangelmayr1979}, a proper treatment of the problem shows that the leading contribution to the path integral close to caustics can be factorized into a part involving the phase along the classical path and a part which is a diffraction integral
\begin{equation}
	\psi(\bm{n},t)\propto \mathrm{e}^{\mathrm{i}\mathcal{S}(\bar{\bm{n}},\bar{\bm{\phi}},t)}\int\int\mathrm{d} s_1\mathrm{d}s_2\;\mathrm{e}^{\mathrm{i}\lambda\Phi_Q(\textbf{s};\textbf{C})}\;,
\end{equation}
(in the above cited papers the diffraction integral is sometimes referred to as a `generalized Airy function').
For the BH trimer model, $\bm{s}$ is two-dimensional and dependent on $\bm{\phi}$, and the control parameters $\bm{C}$ depend on the remaining parameters of the system, including $\bm{n}$.

The action that appears in Eq.\ \ref{eq:PathIntegralWaveFn} can be derived by breaking up the time evolution operator into infinitesimal steps, i.e.\   applying the Trotter prescription to the operator $\hat{\mathcal{F}}$ with $H_{\triangle}$. The details of this calculation are presented in Appendix \ref{Appdx:PathIntegral} where we find 
\begin{widetext}
	\begin{align}
		\mathcal{S}[  \bm{n}(\tau),\bm{\phi}(\tau)] & =  \;\int_0^{\tfrac{NJt}{\hbar}}\mathrm{d}\tau  \left[ \bm{n}\cdot\dot{\bm{\phi}} - \Phi_{\triangle} \right] +\frac{\hbar}{3 \tilde{U}}\left[\phi_C^2(0)+3\phi_X^2(0)\right] \label{eq:FirstOrderAction} \\
	=	\int_0^{NJt/\hbar}\mathrm{d}\tau\Biggl[&\left(\frac{1}{2}-\frac{5}{8N}\delta n_2\right)\phi_X^4+\left(\frac{1}{18}+\frac{1}{24N}\delta n_2\right)\phi_C^4+\left(\frac{1}{3}+\frac{1}{4N}\delta n_2\right)\phi_X^2\phi_C^2-\frac{1}{6N}n_X\left(\phi_X^3\phi_C+\phi_X\phi_C^3\right) \nonumber  \\
		&\;-\left(2-\frac{3}{2N}\delta n_2\right)\phi_X^2-\left(\frac{2}{3}+\frac{1}{2N}\delta n_2\right)\phi_C^2+\frac{n_X}{N}\phi_X\phi_C+\bm{n}\cdot\dot{\bm{\phi}}\Biggr]\nonumber+\frac{\hbar}{3\tilde{U}}\left[\phi_C^2(0)+3\phi_X^2(0)\right] 	\ .
	\end{align}
\end{widetext}
Comparing to the standard relation $L=p \dot{x}-H$, we see that the generating function $\Phi$ plays the role of a Lagrangian.

The action given in Eq.\ \eqref{eq:FirstOrderAction} is not in a canonical form for any unfolding of $X_9$, since the coefficients of the fourth order phase variables have not been appropriately scaled away yet, and the terms $\phi_X^3\phi_C$ and $\phi_X\phi_C^3$ are still present. It is possible to remove these terms by an appropriate change of variables, resulting in the introduction of the cubic unfolding terms $\phi_X^3$ and $\phi_X\phi_C^2$, which lead to triangular symmetry and a stable caustic. However we shall not attempt this here since it involves the simultaneous solution of five equations of quartic and cubic order. We shall instead proceed by restricting ourselves to the projection $n_X=0$. In Fig.\ \ref{fig:KickedCusp}  we plot the classical trajectories in the $n_X=0$ plane and compare between the $H_{\delta\mathrm{QPM}}$ and the $H_{\delta\mathrm{MF}}$ cases. As expected, in (1+1)D the stable caustics are cusps and indeed the cusp point occurs nearly simultaneously for $H_{\delta\mathrm{QPM}}$ and the $H_{\delta\mathrm{MF}}$. This is because the square roots multiplying the cosines do not drastically affect the shape of the focusing surface near the centre of Fock space. Rather, the effects of the square roots only become significant near the edges of Fock space, a region in which the QPM approximation becomes inaccurate. Indeed, we see from Fig.\ \ref{fig:KickedCusp} that the effect of the square root factors is to make fold lines of the full model, shown in panel (b), become curved near the boundaries. Thus, although the rotational symmetry of the caustic is removed by beyond QPM effects, other features of the caustic are robust against such changes, a result that follows from structural stability.

 \begin{figure}[t]
	\centering
	\includegraphics[width=1.0\columnwidth]{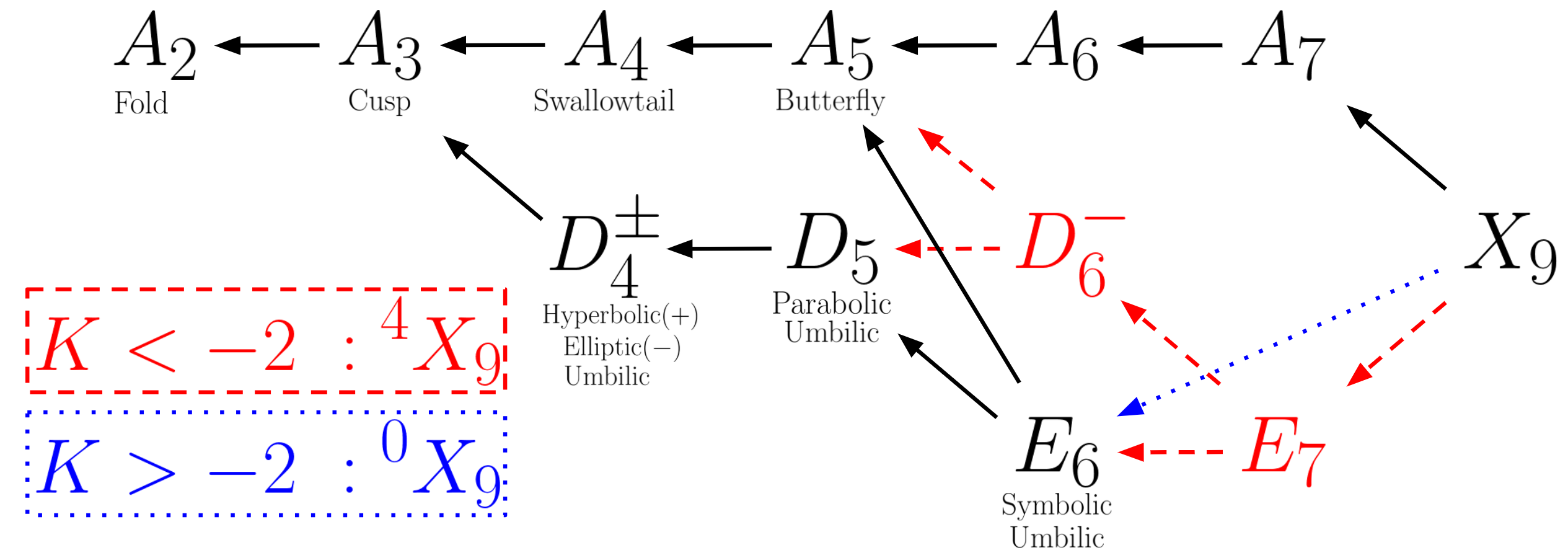}
	\caption{\label{fig:abutement} Bordering (or abutment) diagram for the catastrophe $X_9$, adapted from Nye \cite{Nye1999} and obtained using theorems of catastrophe projection. Each catastrophe of higher order contains (many, not necessarily all) catastrophes of lower order, shown by the direction of arrows. The subfamilies of $X_9$ lead to two different sub-catastrophe sets. ${}^4X_9$ contains the singularities $E_7$ and $D_6$, shown by the red dashed arrows. ${}^0X_9$ does not contain $E_7$ and will only contain $E_6$, shown by the blue dotted arrow.}
\end{figure}

\begin{figure}[t]
	\centering
	\includegraphics[width=0.49\columnwidth]{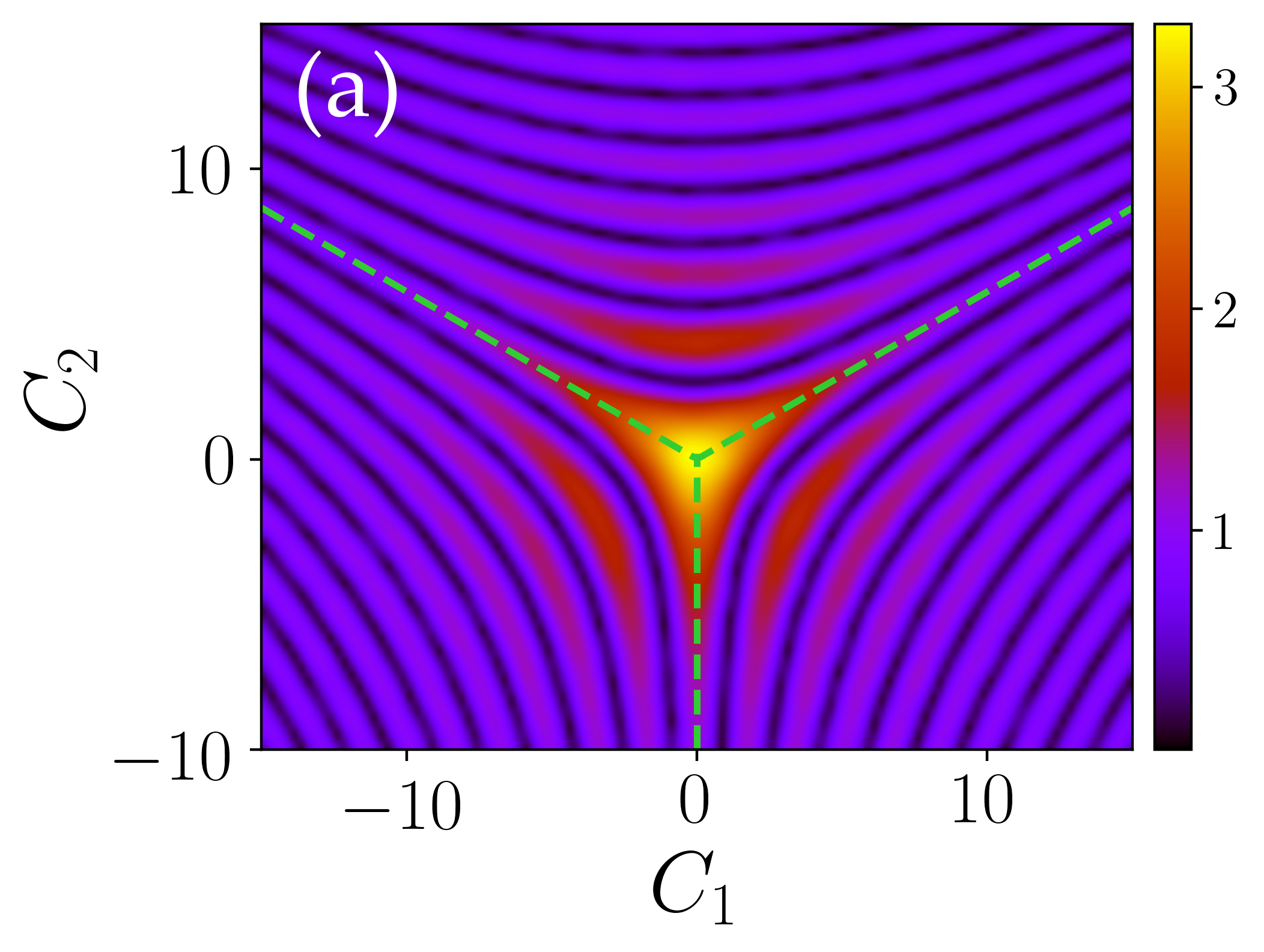}
	\includegraphics[width=0.49\columnwidth]{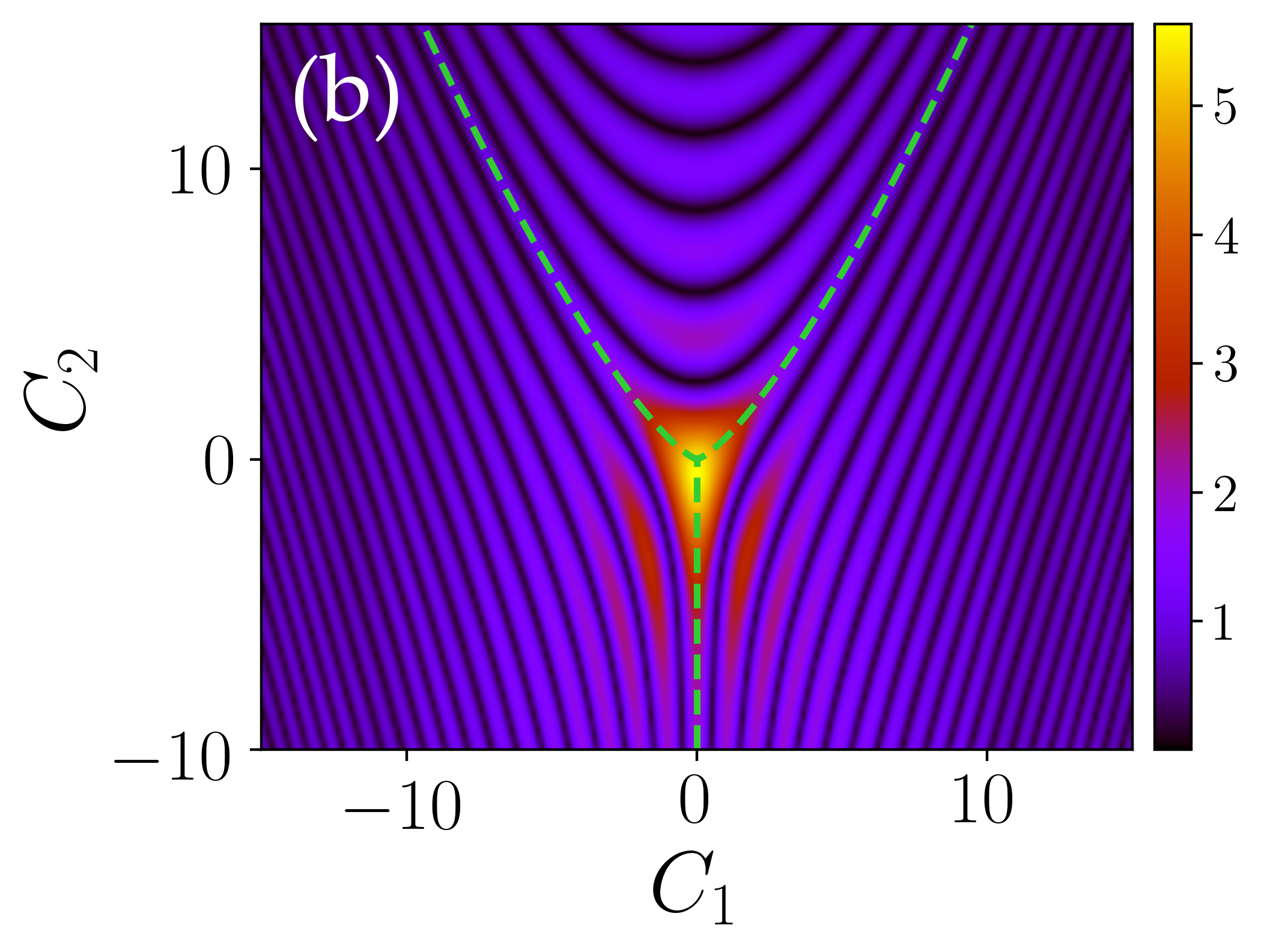}
	\caption{\label{fig:focaldetail} Most singular sections of the diffraction patterns for $D_4^-$ and $D_6^-$ showing how the foci of the two sub-families of $X_{9}$ differ. \textbf{Panel (a)} The focal plane of $D_4^-$ revealing an elliptic umbilic focus. \textbf{Panel (b)} The focal plane of $D_6^-$  revealing another, but subtly different type of elliptic umbilic-like structure. In the case of $D_4^-$ the caustic in its canonical form is threefold symmetric with the brightest ribs (traced with green dashed lines) meeting in straight lines at angles of $2\pi/3$. For $D_6^-$ the bright central ribs meet along \textit{curved} lines obeying $27C_2^4=64C_1^3$. The presence of a $D_6^-$ focus can in principle be a diagnostic tool for determining the presence of ${}^4X_9$.}
\end{figure}

Nevertheless, the $n_X=0$ projection does allow us to account for some beyond-QPM effects and even obtain qualitative features such as the subfamily of $X_9$ that results from the breaking of the circular symmetry in $H_{\triangle}$ and $H_{\delta \mathrm{MF}}$.  In particular, after scaling phase variables to put $X_9$ in its canonical form, the modulus for the restricted problem for $H_{\triangle}$ becomes,
\begin{align}
	K_{n_X=0}(\delta n_2)=&\;(8 N+6\delta n_2) \sqrt{\frac{1}{3 \delta n_2+4 N}} \sqrt{\frac{1}{4 N-5 \delta n_2}}\;. \label{eq:multiK}
\end{align}
This result warrants some explanation because the wavefunction contains a range of the Fock space variables $(\delta n_{2},n_{X})$, and hence there is no single well defined value of the modulus. However, because all physical paths must lie in the interval $-\frac{N}{3}\leq\delta n_2\leq \frac{2N}{3}$, we have to first order, $\frac{6}{\sqrt{17}}\leq K\leq 6$ with $K=2$ occurring only at $\delta n_2=0$. This range of values for $K$ indicates that the system symmetry selects the $\prescript{0}{}{X_9}$ subfamily of $X_{9}$, and specifically its $\prescript{0}{}{X_9^+}$ variant, called \textit{compact} by Callahan \cite{Callahan1981}. A schematic plot, known as an `abutment' or `bordering' diagram is given in Fig.\ \ref{fig:abutement} which summarizes the relationships between the sub-catastrophes which can appear within $X_{9}$  \cite{Nye1999}. We see that the $\prescript{0}{}{X_9}$ subfamily does not contain the umbilic catastrophes $E_7$ or $D_6$ given in Table \ref{tab:catastrophetable2}, both of which are instead members of the $\prescript{4}{}{X_9}$ subfamily.  Diagnosing the family of $X_9$ experimentally via the sub-catastrophes is in principle possible via a careful analysis of elliptic umbilic foci [Fig.\ \ref{fig:causticgallery}(c), the last time slice in Fig.\ \ref{fig:3DPlot}(b), and Figs.\ \ref{fig:EllipticUmbilic}(a) and (b)  all show elliptic umbilic foci]. As illustrated in Fig.\  \ref{fig:focaldetail}, elliptic umbilic foci are three-fold symmetric about straight axes of symmetry meeting at the origin, while for $D_6^-$ two of these lines become curved and the pattern is only two-fold symmetric (see Nye \cite{Nye1987} for more details).

There is an important lesson to be learned from Eq.\ (\ref{eq:multiK}) and the fact there is no single value of the modulus (except locally, at the origin of Fock space). Catastrophes have their origin in topology, and as such there is considerable flexibility as to their precise shape. In fact each catastrophe forms an equivalence class, where different specific realizations within each class are related by smooth transformations (diffeomorphisms) of state variables and control parameters (there is no smooth mapping between different classes). The dynamics of a nonlinear system such as the  BH model results in a caustic of the $X_9$ class which slowly varies in space-time but which is not destroyed by nonlinearities.

\begin{figure}[t]
	\centering
	\includegraphics[width=0.49\columnwidth]{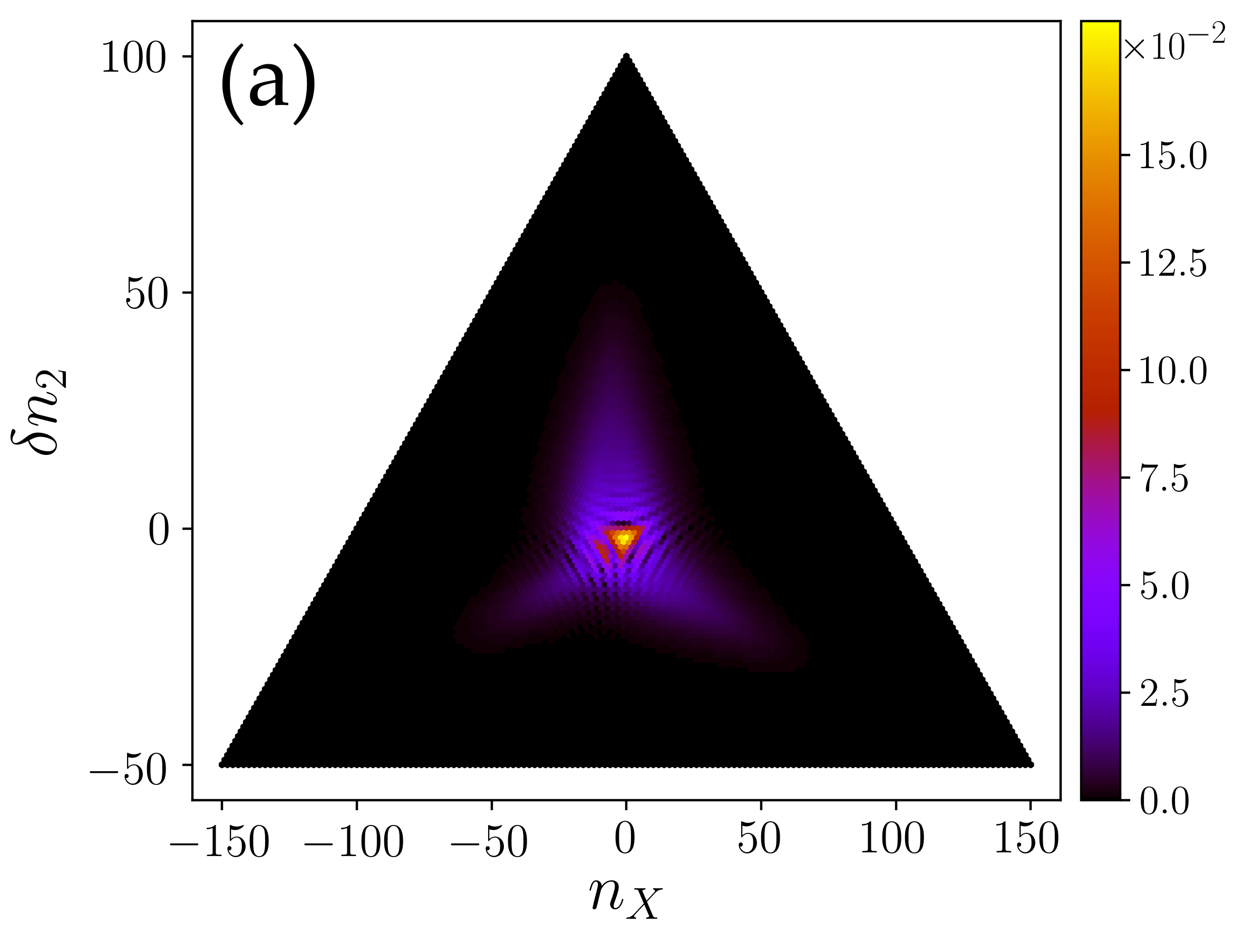}
	\includegraphics[width=0.47\columnwidth]{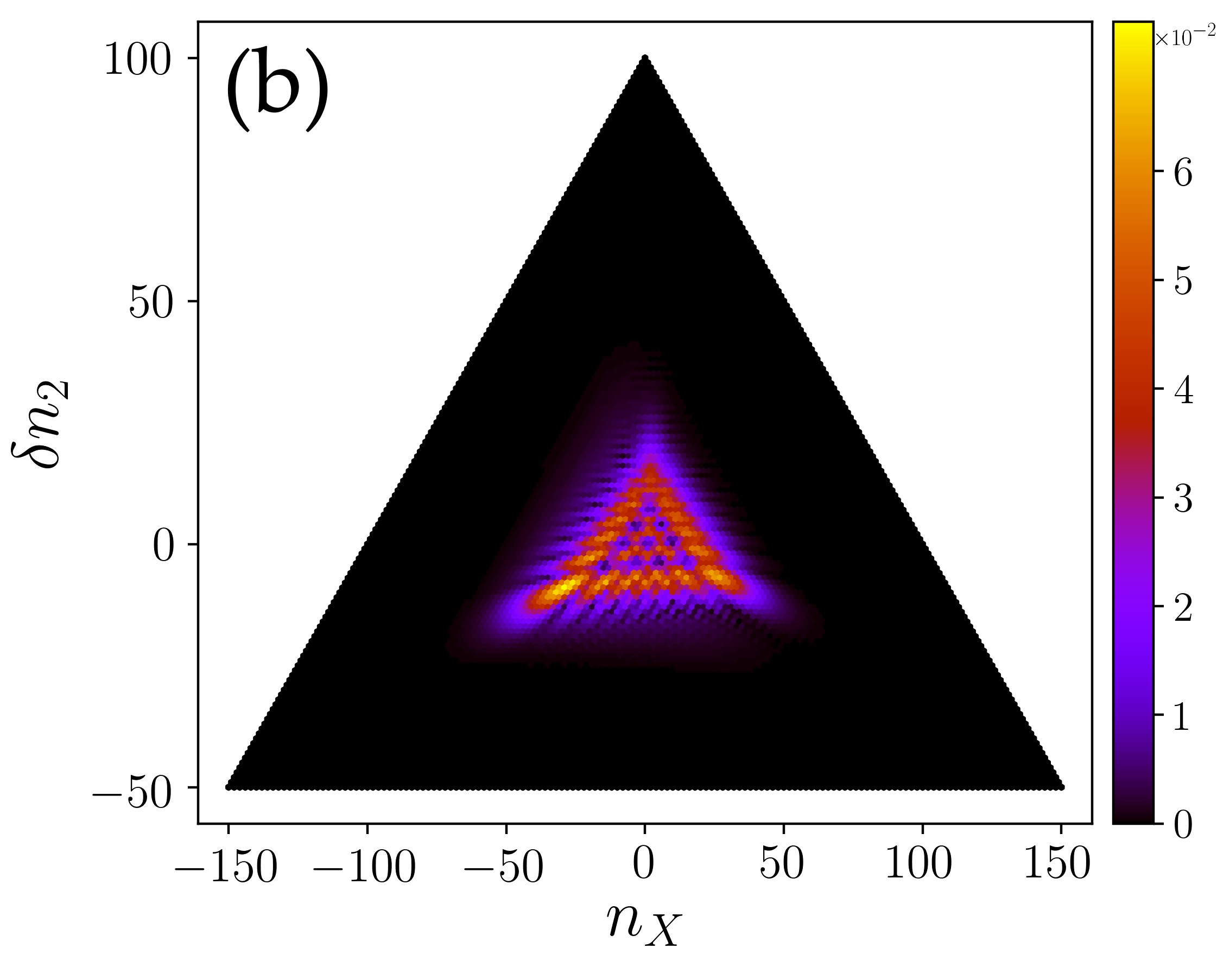}
	\caption{\label{fig:DifferentHoppings} Distorted elliptic umbilic due to unequal hopping strengths using the full trimer Hamiltonian. $K_R=1.1K_L$, $K_X=1.2K_L$, $U/K_L=0.01$ and $N=150$, starting from an equal superposition of Fock states. \textbf{Panel (a):} Distorted focus at $K_Lt/\hbar = 0.431$. \textbf{Panel (b):} Unfolded at $K_Lt/\hbar = 0.513$}
\end{figure}

\subsection{Unequal hopping amplitudes}
So far our consideration of beyond-QPM dynamics has centered on the effect of mode depletion accounted for by the square root factors. However, one can also consider the situation where we have already included the effects of the mode depletion but then additionally break the symmetry further by making the hopping amplitudes unequal (a similar effect could be had by including a bias, $\epsilon_i \neq 0$). In Fig. \ref{fig:DifferentHoppings} we see the effects of changing the hoppings so that they are slightly detuned from one another $(K_R \neq K_L\neq K_X)$. The elliptic umbilic caustic becomes distorted and asymmetric, but due to the structural stability of the underlying catastrophe it remains intact and recognizable. In particular, panel (a) shows the effect on the focus, which can be shifted in space and time but clearly retains its form. Likewise for panel (b) which shows a time slice somewhat after the focus.

\section{Role of interactions}
\label{sec:interactions}

\begin{figure}[t]
	\centering
	\includegraphics[width=0.49\columnwidth]{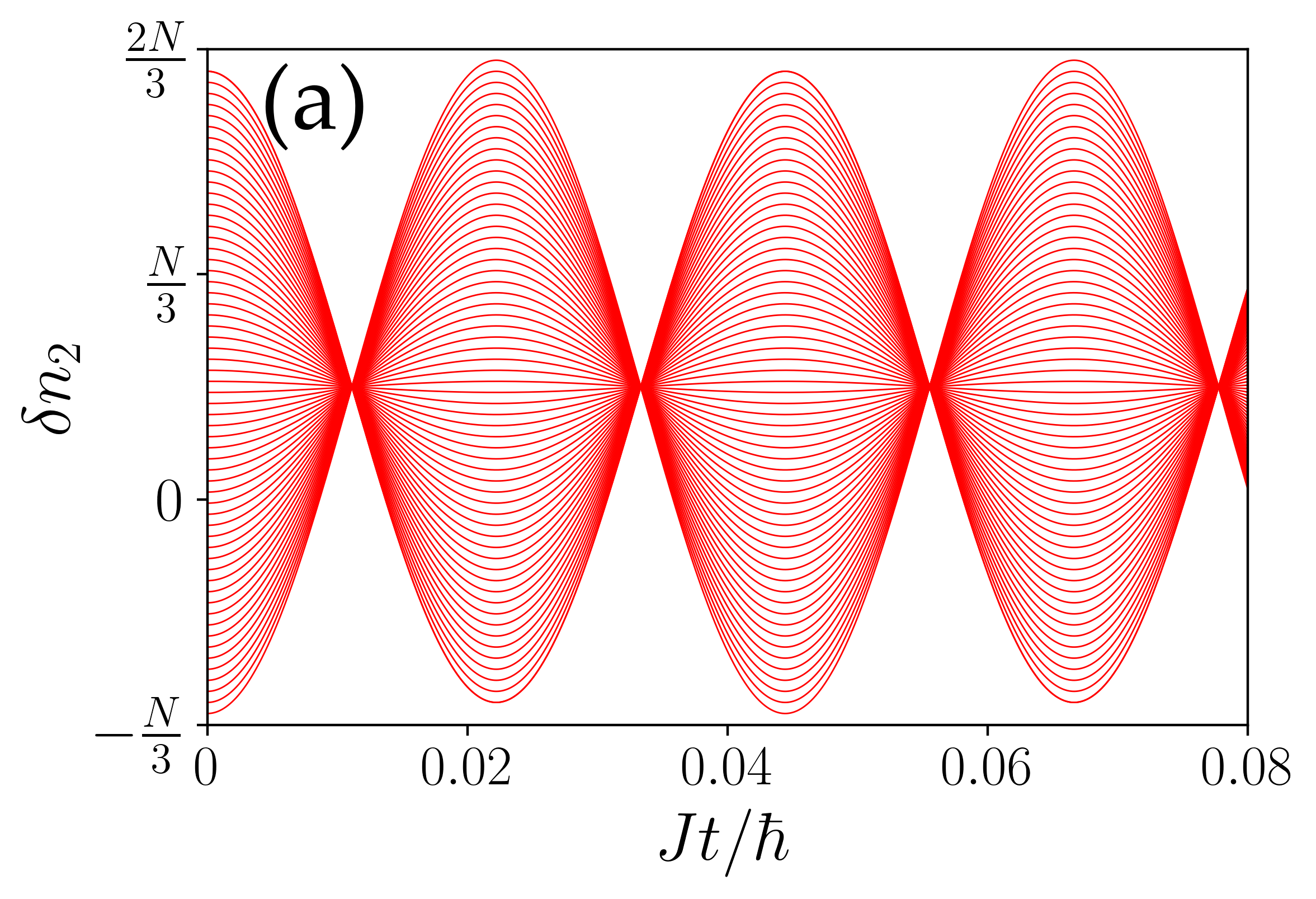}
	\includegraphics[width=0.49\columnwidth]{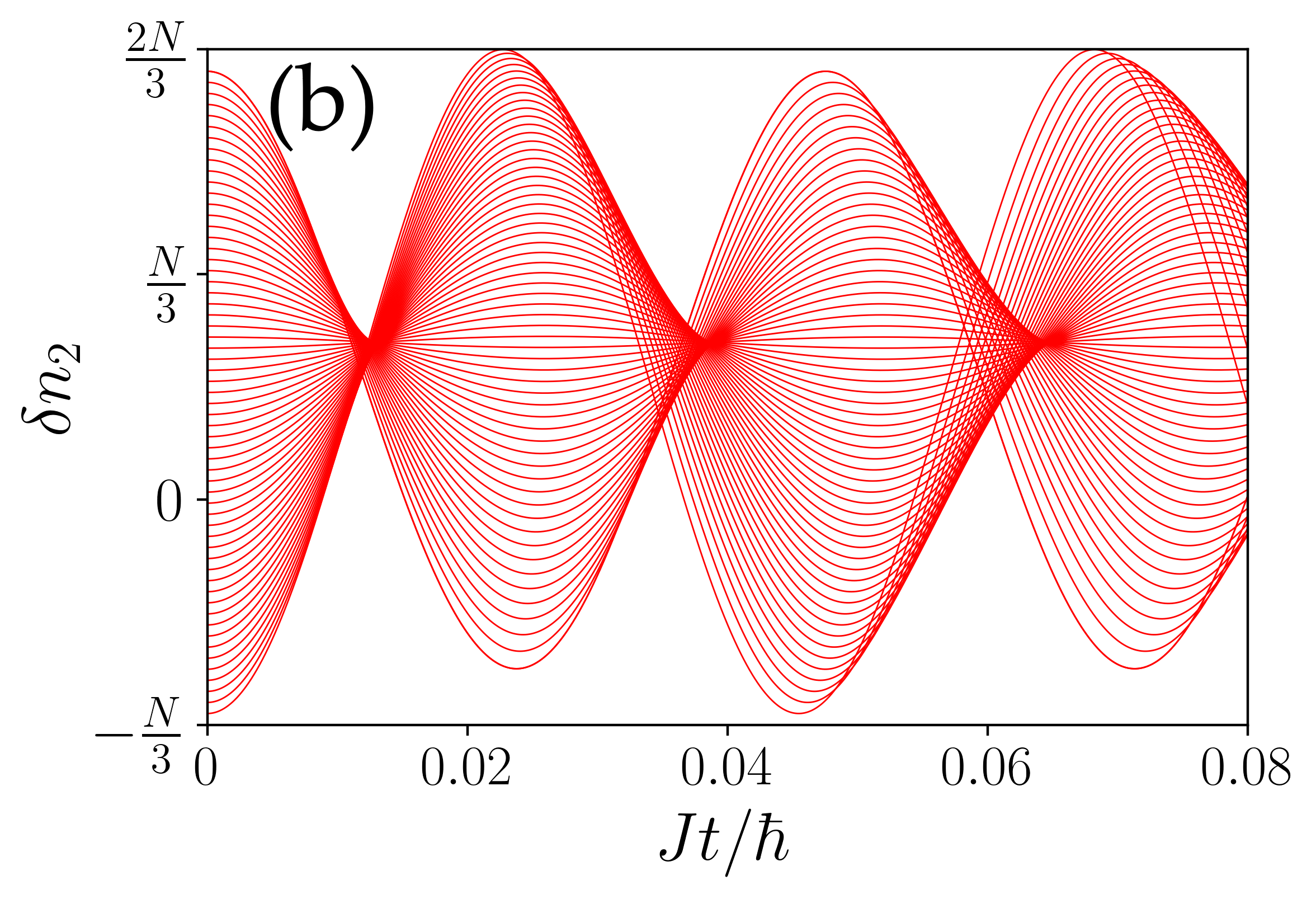}
	\caption{\label{fig:UnstableFocusing} Instability of perfect focusing events to the introduction of interactions. \textbf{Panel (a):} Set of classical trajectories starting from an equal spread of Fock states, with $K_L=K_R$, $U=K_X=0$. Focusing events are isolated, meaning that all trajectories meet at a point.  \textbf{Panel (b):} Same initial state as panel (a), but now with weakly attractive interactions, $K_L=K_R\equiv J$ and $U/J=-0.03$. A similar result was found for repulsive interactions in panel (a) of Fig.\ \ref{fig:ThreeModeSlices}, except that the cusps open in the opposite direction for attractive interactions.}
\end{figure}

In their original setting of natural optics (rainbows etc.), caustics occur in a linear and hence integrable system. Similarly, the BH dimer is an integrable, albeit nonlinear, system as long as energy is conserved. The trimer, by contrast, is not generally integrable. In our analytic calculations in this paper we used $\delta$-kicked interactions such that the time evolution is integrable. While this made calculations possible, it does raise the question of the stability of caustics in the presence of constant interactions. It is therefore worth emphasizing again that all the (numerical) examples of caustics shown in Section \ref{sec:BHtrimer} were obtained with the interactions switched on throughout the time evolution. Moreover, interactions can sometimes be \textit{necessary} for catastrophes to fully manifest in both the dimer and trimer. This is best understood with an example, as given in Fig.\ \ref{fig:UnstableFocusing}. In panel (a) $U=0$, $K_L=K_R=J$, and $K_X=0$, for which the quantum revival time is $Jt_{\text{rev}}/\hbar = \pi/\sqrt{2}$ (the recurrence time is twice this time). Starting from an even spread of Fock states, we observe a set of classical trajectories in the $n_X=0$ plane, similar to Fig.\ \ref{fig:ThreeModeSlices}(a) except that here the trajectories form isolated focal points. According to catastrophe theory these are unstable in two dimensions and indeed, in panel (b) where interactions are present the focal points are unfolded to cusps. For variety we have chosen attractive interactions $U<0$ here whereas Fig.\ \ref{fig:ThreeModeSlices} has repulsive interactions $U>0$. The difference is that we find forward-opening cusps for $U>0$ and backward-opening cusps for $U<0$. The effect is similar when starting from a highly focused state, such as a Fock state, where a point focus will recur infinitely unless interactions are introduced and structurally stable caustics form. The nonlinearity introduced by interactions is therefore crucial to fully unfold the caustics (essentially by introducing different periods for different amplitudes of excitation).

\begin{figure}[b]\centering
	\includegraphics[width=0.49\columnwidth]{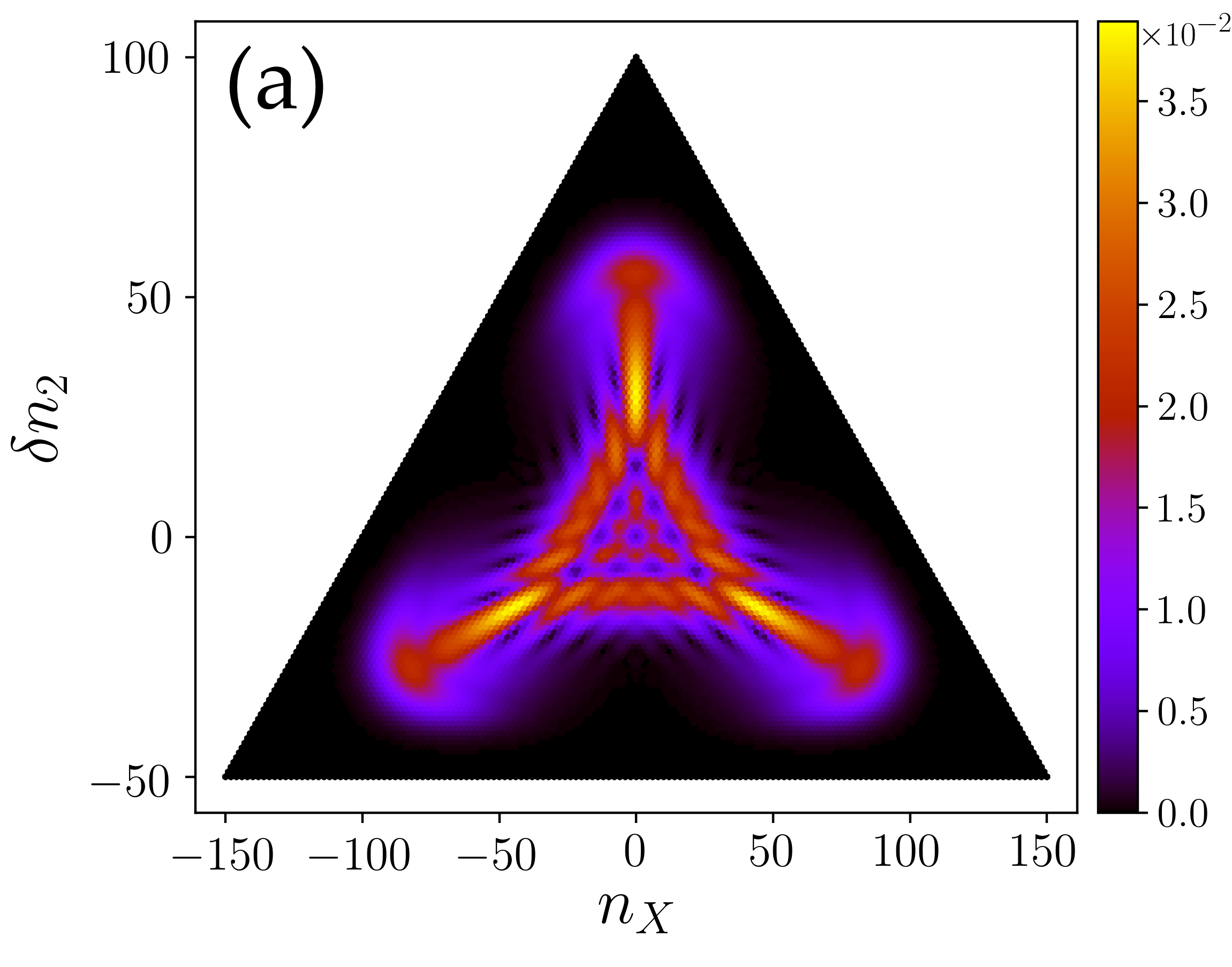}
	\includegraphics[width=0.49\columnwidth]{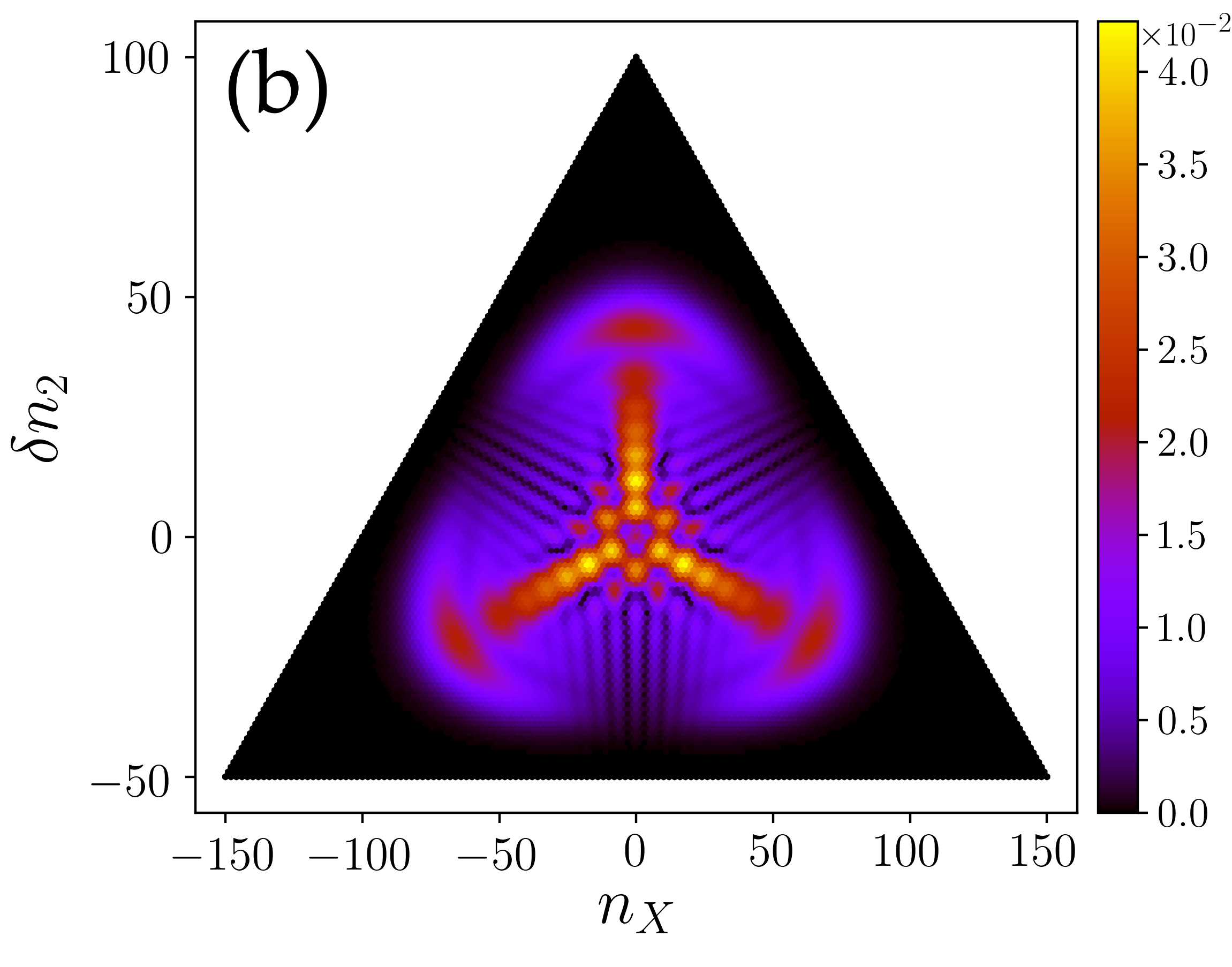}
	\vspace{-0.1cm}\includegraphics[width=0.49\columnwidth]{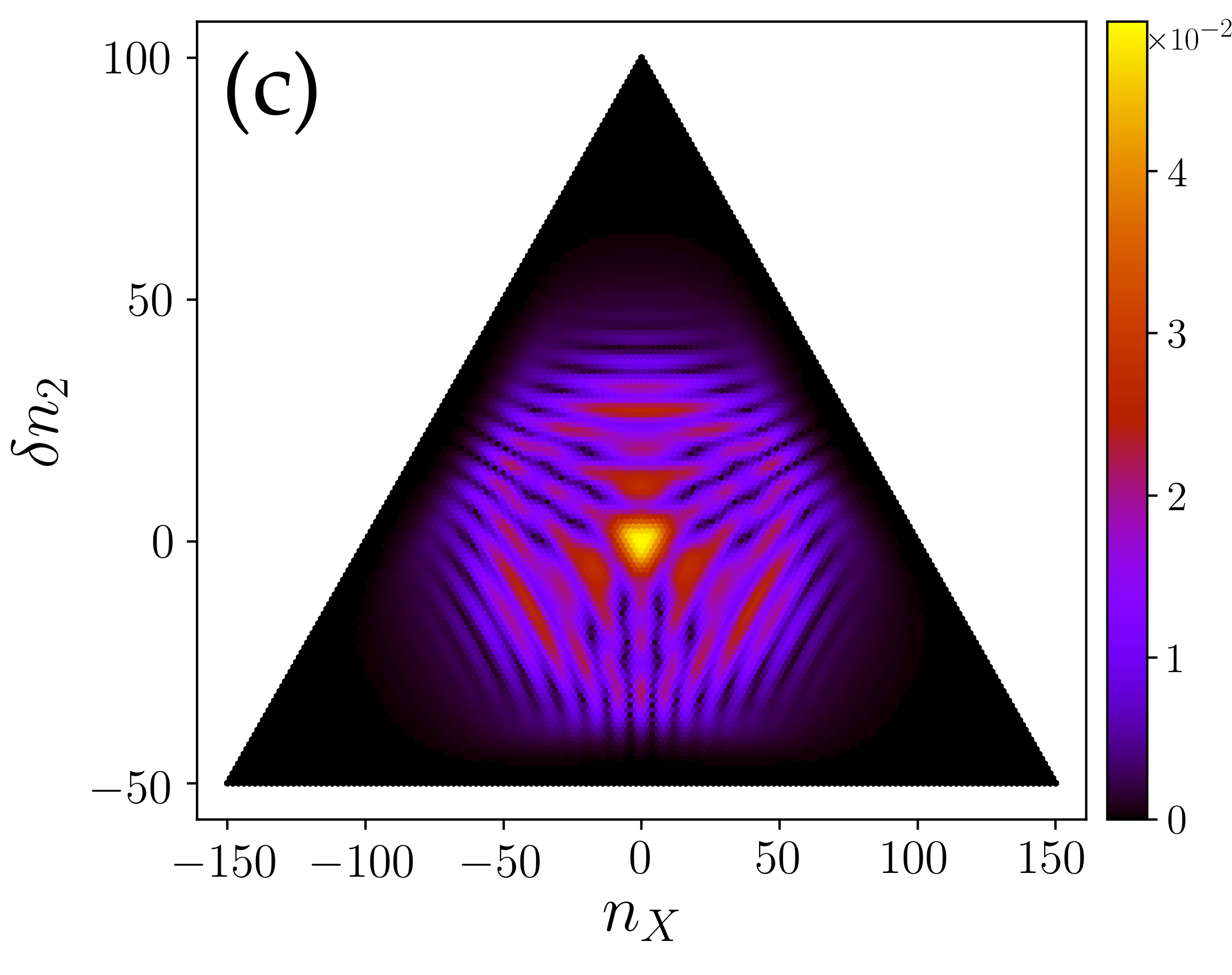}
	\raisebox{0.1\height}{\includegraphics[width=0.44\columnwidth]{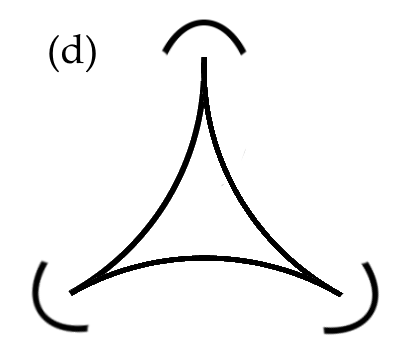}}
	\caption{\label{fig:NegativeU} Formation of fold lines around cusps due to attractive interactions. \textbf{Panel (a):} Same as panel (c) of Fig.\ \ref{fig:EllipticUmbilic}, now with $U/J=-0.01$. The elliptic umbilic diffraction pattern is now surrounded by bright fringes, which consist of small fold lines.  \textbf{Panels (b)-(c):} Same as panel (a) but at $Jt/\hbar=0.641$ and $Jt/\hbar=0.818$. As the elliptic umbilic focus is approached, the fold lines around each cusp extend and form a ring. \textbf{Panel (d):} Schematic of a section of the elliptic umbilic caustic, with `lips' surrounding each cusp point. }
\end{figure}

Another example of dynamics with attractive interactions is shown in Fig.\ \ref{fig:NegativeU}. The effect of negative $U$  is to pull the caustic outwards towards the corners of Fock space in comparison to the case of the repulsive $U$ shown in Fig.\ \ref{fig:HyperbolicUmbilic}. This means that the elliptic umbilic emerges more clearly before the central focusing event, but also results in fold `lips' around the edges of each cusp point. The formation of lips around an elliptic umbilic caustic appears to be a threefold symmetric version of the fourfold unfolding of $X_9$ with a negative modulus studied in \cite{Nye1986} (see also Fig.\ \ref{fig:Gen2nCusp}).  These lips continue to extend into long fold lines as time progresses and ultimately intersect one another after the elliptic umbilic focus, reminiscent of the triple glass junction studied by Berry \cite{BerryJunction} and later elaborated on by Nye \cite{Nye1999}.

\begin{figure}[t]\centering
	\includegraphics[width=0.85\columnwidth]{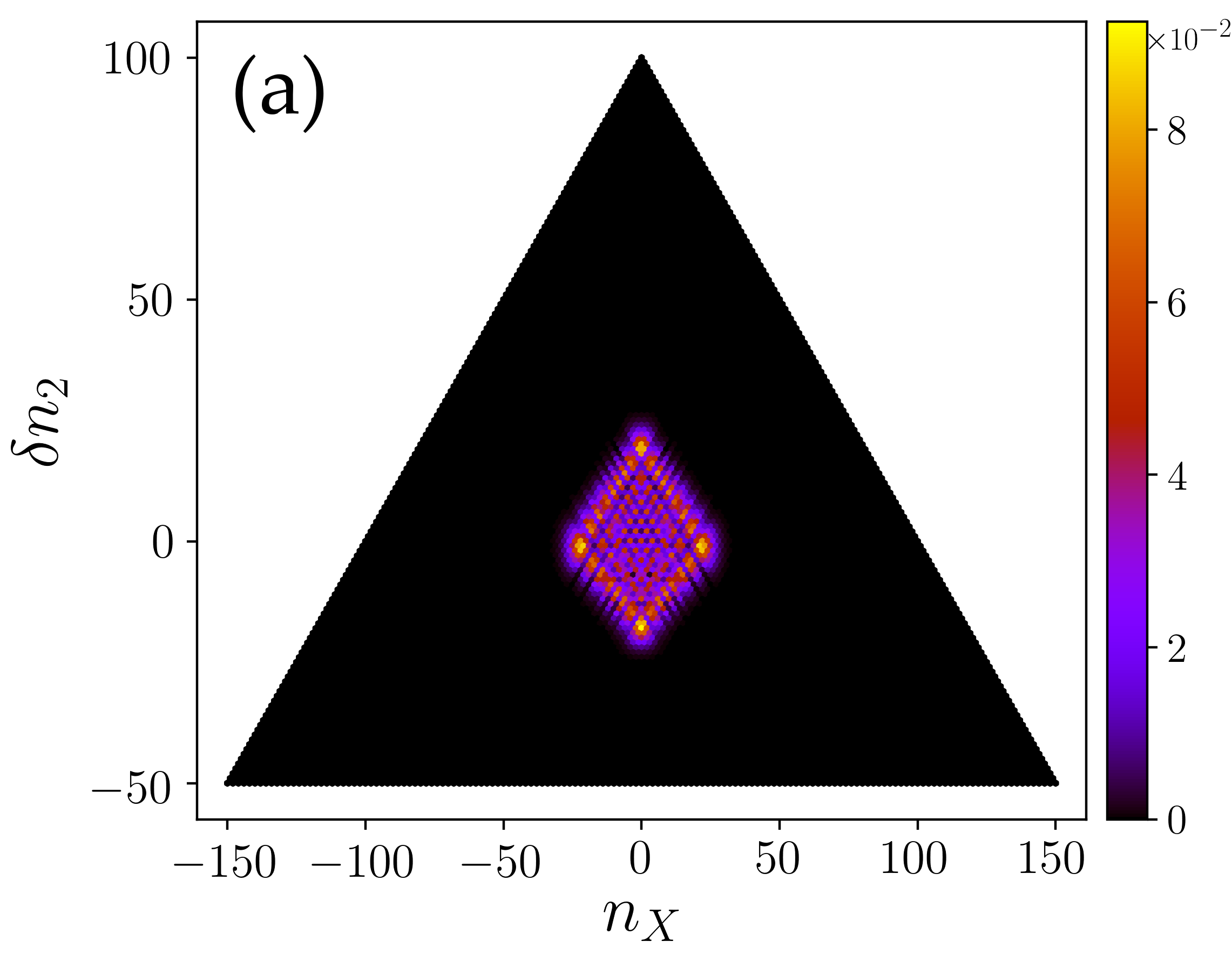}
	\includegraphics[width=0.85\columnwidth]{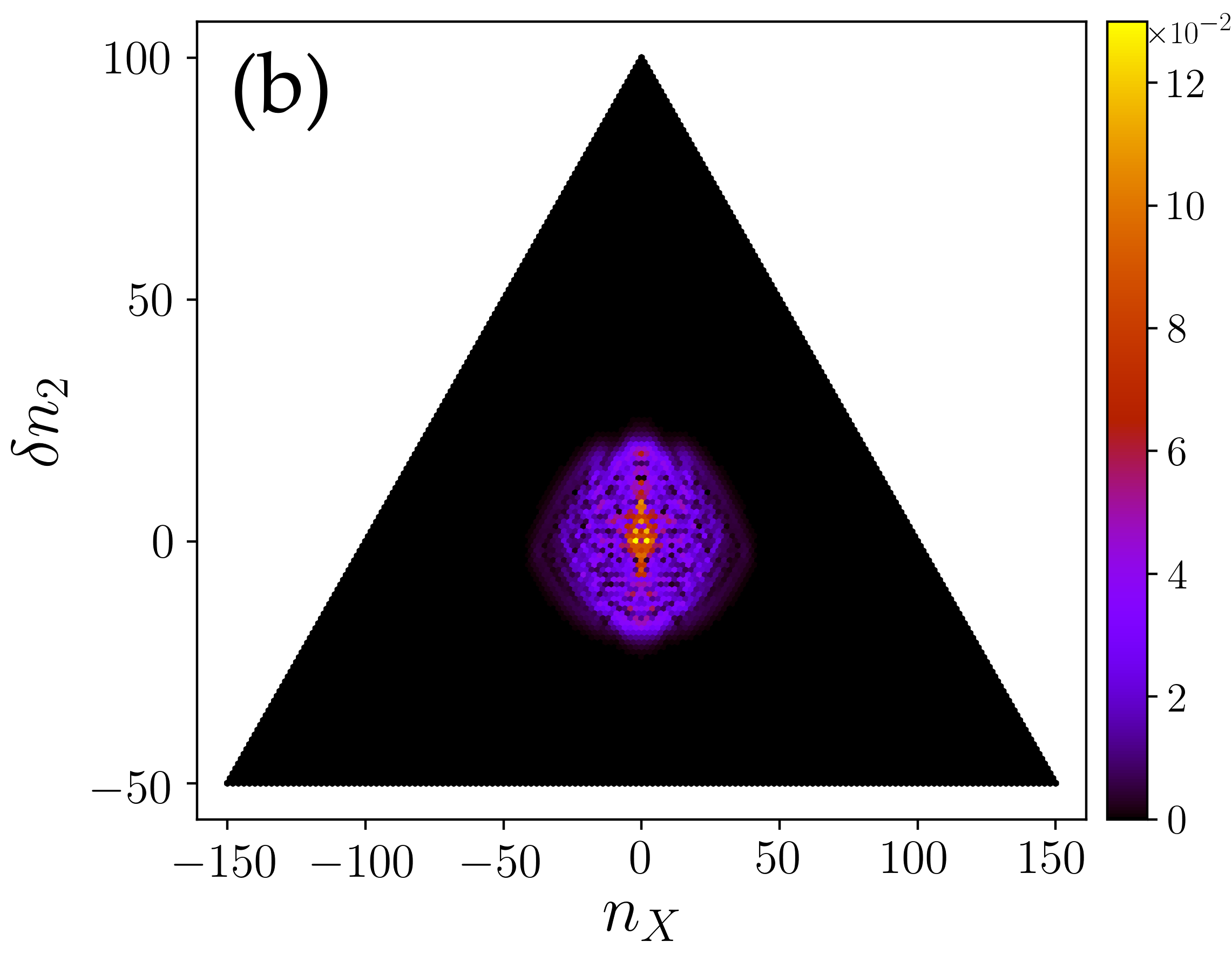}
	\caption{\label{fig:UEqJ} Catastrophe formation at high interaction strength. In this figure, we show exact (numerical) dynamics deep in the chaotic regime with $U=J$ in the linear configuration of the BH trimer where $K_{L}=K_{R}\equiv J$, and $K_X=0$. Caustic formation follows similarly to previously discussed cases with weaker interactions, however stronger interactions compress the dynamics to a smaller region in Fock space. \textbf{Panel (a):} Formation of a fourfold diffraction pattern at $Jt/\hbar=0.13$, similar to the one seen in Fig.\ \ref{fig:HyperbolicUmbilic}(a), but its largest extent is smaller than the weaker interaction case. \textbf{Panel (b):} At $Jt/\hbar=0.36$, the wavefront has become highly distorted, although strong focusing (fluctuations) remains.}
\end{figure}

Finally, we note that our choice of parameters in this paper has been guided empirically so as to make the catastrophes as visible as possible. For example, the elliptic and hyperbolic umbilic unfoldings shown in Section \ref{sec:BHtrimer} were made using relatively weak interactions in the range $U/J= 0.01J$ to $U/J=0.25J$. This regime is called the strong-tunneling (or Josephson \cite{Wilsmann2018}) regime (see Refs.\ \cite{Lee2006,Arwas2014,Kolovsky2007,Kolovsky2020,Bradly2012,Gallemi2015}), and exhibits moderate quantum revivals, allowing the catastrophes to be clearly identified because a single caustic stretches across a good fraction of Fock space.  Stronger interactions result in more powerful effective focusing potentials which, in the case of repulsive interactions, compress the wavefunction into a small region around the center of Fock space.  In Fig.\ \ref{fig:UEqJ}(a) we show that the canonical hyperbolic umbilic catastrophe is still visible even when $U=J$, although Fig.\ \ref{fig:UEqJ}(b) indicates that at longer times this wavefunction evolves into a highly intricate structure (which will eventually reveal the discreteness of Fock space as the wavefunction fringes reach small scales). The highly distorted wavefronts that arise in this situation would probably be best described using the statistical version of catastrophe theory developed in the context of light passing through a turbulent atmosphere  \cite{berry77}, where caustics manifest themselves as extreme amplitude events that occur more frequently than expected from random Gaussian fluctuations. This is the freak/rogue wave paradigm recently explored in microwave \cite{hohmann10} and optical \cite{Solli2007,Arecchi2011,Akhmediev2013,Marsal2014,Mathis2015,Pierangeli15,Mattheakis2016,Safari17,Zannottibook} experiments.

\section{\label{sec:Discussion} Conclusions and Experimental perspectives}

Using a combination of exact numerical computations and analytic calculations (in the $\delta$-kicked case), we have demonstrated the existence of simple and higher caustics in integrable and nonintegrable quantum many-body dynamics. More precisely, using a variety of initial states (broad and narrow in Fock space) we have shown that caustics occur generically following a quench in the BH dimer and trimer models even when the latter is in the chaotic regime (e.g.\ when $U \sim J$).  In the semiclassical regime caustics dominate the many-body wavefunction which takes on characteristic patterns that are stable against perturbations to both the Hamiltonian and initial conditions.  Catastrophe theory provides a mathematically rigorous tool for predicting and analyzing these universal wavefunctions. The universality of wave catastrophes is underlined by the fact that we find the same basic caustic structures  in the optics of liquid droplet lenses whose principal curvatures have been modified by a triangular perimeter, as shown in  Fig.\ \ref{fig:causticgallery}, as we do in the triangular Fock space of the BH trimer.

One may ask, if quantum many-body caustics are generic and striking why have they not been seen already? The answer is that they have been seen following quenches in experiments on ultracold atoms in optical lattices and trapped ions in the form of so-called light cones  \cite{Cheneau2012,Fukuhara2013,Langen2013,Richerme14,Jurcevic14}. As shown in our previous paper on spin chains  \cite{Kirkby2019}, and as we explain in Section \ref{sec:Cats} in this paper, light cones are caustics closely analogous to ship wakes. The ones seen experimentally so far are fold catastrophes (which have an Airy function profile \cite{Cheneau2012}) and are the simplest in the hierarchy we discuss in Sec.\ \ref{sec:Cats}. Turning this around, the higher caustics we predict can be viewed as \textit{generalized light cones}.

Caustics will manifest themselves in experiments as \textit{singularity dominated fluctuations}  that are stronger than the random gaussian fluctuations one expects during generic chaotic dynamics \cite{Berry1977}. For cold atoms in an optical lattice this means strong  fluctuations (which form universal patterns) in the probability distribution for populations of different sites, see Figs.\ \ref{fig:3DPlot},\ref{fig:TwoMode},\ref{fig:ThreeModeSlices},\ref{fig:EllipticUmbilic},\ref{fig:HyperbolicUmbilic},\ref{fig:Gen2nCusp},\ref{fig:K6Hyperbolic},\ref{fig:K2SpunCusp},\ref{fig:KickedGS},\ref{fig:DeformedCusp},\ref{fig:KickedCusp},\ref{fig:DifferentHoppings},\ref{fig:UnstableFocusing},\ref{fig:NegativeU}, and \ref{fig:UEqJ}. In fact, thanks to advances in imaging such as the quantum gas microscope \cite{Bakr2009,Sherson2010}, it is now possible to monitor the population of single sites \textit{in situ} in optical lattices at the single atom level and thereby directly measure the Fock-space probability distribution, just as was done in the observation of light cones \cite{Cheneau2012} and many-body localization \cite{Schreiber2015,Choi2016}.  Alternatively, miscible spin-1 gases (as can be realized using $^{23}$Na) where the atoms occupy the same spatial mode, also offer a highly controllable environment for exploring integrable and nonintegrable three-mode many-body dynamics where 
the populations of the three Zeeman sublevels can be obtained using Stern-Gerlach type measurements \cite{Evrard2021}. 

The above mentioned experimental examples suggest that it may soon be possible to map out wave catastrophe patterns in many-body wavefunctions in some detail. However, it should be borne in mind that there are fundamental differences between classical and quantum waves and this impacts the information we can extract in a single projective measurement.
In classical waves, the wave catastrophe patterns can be captured in a single shot like in Fig.\ \ref{fig:causticgallery}, whereas in the quantum case a measurement in the Fock basis will randomly collapse the wavefunction to give us a single Fock state. Thus, a single absorption image of the BH trimer case will yield just two numbers: the two relative population differences between the three wells (assuming total number conservation). The experiment must be repeated many times under the same conditions in order to build up a probability distribution for the population differences and thereby reveal the caustics in the probability distribution. The whole scheme should then be repeated for different evolution times if we wish to map out the full three dimensional dynamical caustic. This is a challenging task, and there is no getting around the fact that quantum many-body wavefunctions are high dimensional objects containing a huge amount of information, but it is important to note that measuring the probability distribution is a simpler task than full quantum tomography (see \cite{Mumford2019} for a discussion). 
It should also be pointed out that caustics equally occur in the phase-difference variables \cite{Mumford2017} conjugate to the number-difference variables and hence caustics can alternatively be seen by releasing the atoms from the wells and imaging after some time of flight to allow the atom clouds from each well to overlap and interfere \cite{andrews97}. 

In traditional measurements on condensed matter systems it is not the full wavefunction that is usually measured directly but one- and two-point correlation functions that are obtained. Although we have chosen to focus on the wavefunction in this paper, it is the key object needed to calculate correlation functions and we have seen here how it takes on universal forms. Furthermore, in our previous work on light-cones in integrable systems \cite{Kirkby2019} we showed that due to the hierarchy of catastrophes and the projection identities they obey (such that the higher ones contain the lower ones), correlation functions also contain caustics. For example, the equal-time two-site correlation function on a spin chain can be expressed as the product of two single quasiparticle wavefunctions evaluated at different points (see Section VII of \cite{Kirkby2019}) and gives codimension 3 catastrophes such as the hyperbolic umbilic. Based on the results of the present paper, we expect that following a quench correlation functions for nonintegrable systems will also display caustics. 

Rather than a finished theory, the results presented in this paper are merely one step on the road to understanding caustics in quantum many-body systems. In going to systems with a larger number of wells we encounter higher dimensional Fock spaces and hence higher dimensional catastrophes. In these cases it is easier to proceed by going over to the statistical version of catastrophe theory mentioned in Section \ref{sec:interactions} and developed in the context of random focusing of light passing through a turbulent atmosphere  \cite{berry77}, that has also been applied to freak waves in hydrodynamics and optics \cite{Solli2007,Arecchi2011,Akhmediev2013,Marsal2014,Mathis2015,Pierangeli15,Mattheakis2016,Safari17,Zannottibook}. The statistical theory aims to predict the `twinkling exponents' of fluctuations. Perhaps surprisingly, it is not the higher order (more singular) catastrophes that necessarily dominate these fluctuations due their relative rarity. Moreover, the existing theory for classical waves will need to be revised since in the quantum case there is a new scale provided by the discretization in Fock space that can alter the finest details and potentially change the exponents.

\acknowledgments We are grateful to the Natural Sciences and Engineering Research Council of Canada (NSERC) for funding, and thank Professor Sir Michael Berry for comments on the manuscript.

\appendix

\section{\label{Appdx:DeltaKick} Classical Trajectories of the $\delta$QPM}

Using Hamilton's equations of motion,
\begin{align}
	\dot{n}_2=&\;-\frac{1}{\hbar}\frac{\partial H}{\partial \phi_C}\\
	\dot{n}_X=&\;-\frac{1}{\hbar}\frac{\partial H}{\partial \phi_X}\\
	\dot{\phi}_C=&\;\frac{1}{\hbar}\frac{\partial H}{\partial n_2}\\
	\dot{\phi}_X=&\;\frac{1}{\hbar}\frac{\partial H}{\partial n_X}
\end{align}
The $\delta$-kick allows classical trajectories to be calculated directly via integration, 
\begin{align}
	\phi_C(t)=&\;\phi_C(0)+\frac{3U}{2\hbar}\delta n_2(0)\Theta(t)\\
	\phi_X(t)=&\;\phi_X(0)+\frac{U}{2\hbar}n_X(0)\Theta(t)\;,
\end{align}
where we make use of the Heaviside function,
\begin{equation}
\Theta(t)=\begin{cases}
1&t>0\\
0&t\leq 0
\end{cases}
\end{equation}

First, for the triangular configuration of the trimer, assuming a classical analogue of an equal superposition of Fock states corresponds to an ensemble of classical trajectories, each with different $\{\delta n_2(0),n_X(0)\}$, but all $\phi_{C}(0)=\phi_{X}(0)=0$, then for $t>0$,
\begin{align}
	\delta n_2(t)=&\;\delta n_2(0)-\frac{4J N t}{3\hbar} \cos\left(\frac{U}{2\hbar}n_X(0)\right)\sin\left(\frac{3U}{2\hbar}\delta n_2(0)\right)\\
	n_X(t)=&\;n_X(0)-\frac{4 J Nt}{3\hbar}\Biggl[\cos\left(\frac{3U}{2\hbar}\delta n_2(0)\right)\sin\left(\frac{U}{2\hbar}n_X(0)\right)\\
	&+2\cos\left(\frac{U}{2\hbar}n_X(0)\right)\sin\left(\frac{U}{2\hbar}n_X(0)\right)\Biggr]\nonumber
\end{align}
Note that the trajectories $n_X(0)=\delta n_2(0)=0$ have no time dependence and correspond exactly to the unstable axial caustic in the $K=2$ excluded family of $X_9$.

In the case of the linear spatial configuration ($K_X=0$), then $\delta n(t)$ remains the same, while,
\begin{equation}
	n_X(t)=\;n_X(0)-\frac{4 J Nt}{3\hbar}\cos\left(\frac{3U}{2\hbar}\delta n_2(0)\right)\sin\left(\frac{U}{2\hbar}n_X(0)\right)\;.
\end{equation}

\section{\label{Appdx:KXDependence} Dependence of $\delta$QPM wavefunction on $K_X$}
For general $K_X$, the Hamiltonian \eqref{eq:dQPM} yields wavefunctions of the same form as in Eq. \eqref{eq:KickedX9}, but now with modulus,
\begin{equation}
	K=\frac{6}{\sqrt{\frac{8K_X}{J}+1}}
\end{equation}
and control parameters,
\begin{align}
	\alpha=&\;\frac{\sqrt{2}\left(3\hbar^2-2NJt U-4 N t U K_X\right)}{NJtU \sqrt{\frac{8K_X}{J}+1}}\\%\overset{Nt\gg 1}{\to}-\frac{2 \sqrt{2} \left(1+2 \bar{K}_X\right)}{\sqrt{8 \bar{K}_X+1}}\\
	\beta=&\;\frac{\sqrt{2} \left(\hbar ^2-2 J N t U\right)}{J N t U}\\
	\zeta=&\;\frac{2^{1/4}\sqrt{3}\hbar }{NJt\left(\frac{8K_X}{J}+1\right)^{1/4}} n_X\\%\overset{Nt\gg 1}{\to}0\\
	\eta=&\;\frac{2^{1/4} \sqrt{3}\hbar }{NJ t}\delta n_2 %\overset{Nt\gg 1}{\to}0
\end{align}
and
\begin{equation}
	A(t)=\left(\frac{324}{8K_X/J+1}\right)^{1/4}\frac{4\pi\hbar}{\sqrt{3}U}\mathrm{e}^{-\mathrm{i}\frac{\pi}{2}}\mathrm{e}^{\mathrm{i}\frac{2NJt}{\hbar}} \ .
\end{equation}

\section{\label{Appdx:PathIntegral}Derivation of the Path Integral}
Starting with the Hamiltonian given in Eq.\ \eqref{eq:KickedFirstOrder} which goes beyond the QPM by including the square root factors to first order and hence includes effects due to the triangular boundaries of Fock space, we expand the cosine terms to fourth order in the phase difference coordinates to give
\begin{widetext}
	\begin{align} \label{eq:hdelta_appendix}
		H_\triangle\approx&\;-NJ\Biggl[\left(\frac{1}{2}-\frac{5}{8N}\delta n_2\right)\phi_X^4+\left(\frac{1}{18}+\frac{1}{24N}\delta n_2\right)\phi_C^4+\left(\frac{1}{3}+\frac{1}{4N}\delta n_2\right)\phi_X^2\phi_C^2-\frac{1}{6N}n_X\left(\phi_X^3\phi_C+\phi_X\phi_C^3\right)\\
		&\;-\left(2-\frac{3}{2N}\delta n_2\right)\phi_X^2-\left(\frac{2}{3}+\frac{1}{2N}\delta n_2\right)\phi_C^2+\frac{n_X}{N}\phi_X\phi_C\Biggr]+\delta(t)\frac{\tilde{U}}{4}\left[3\delta n_2^2+n_X^2\right]\nonumber\\
		\equiv&\; NJ\Phi_{\triangle}+\delta(t)\frac{\tilde{U}}{4}\left[3\delta n_2^2+n_X^2\right] \ .
	\end{align}
Notice that relative to the expansion of $H_{\delta\mathrm{QPM}}$, all of the circularly symmetric terms get a perturbation proportional to $\delta n_2$ and we also pick up some non-circularly symmetric terms proportional to $n_X$. Applying the Floquet operator incorporating this Hamiltonian to an initial state comprising of an equal superposition of all Fock states, and projecting onto the Fock basis, we obtain the wavefunction
\begin{align}
	\psi(n_X,\delta n_2,t)=&\;\sum_{n_X',\delta n_2'}\mathrm{e}^{-\mathrm{i}\frac{U}{4}\left[3{\delta n_2'}^2+{n_X'}^2\right]}\bra{n_X,\delta n_2}\mathrm{e}^{\mathrm{i}\frac{NJt}{\hbar}\Phi_\triangle(\hat{n}_X,\delta \hat{n}_2,\hat{\phi}_C,\hat{\phi}_X)}\mathclap{\mathop{}\limits_{\overbrace{\scriptstyle\mathds{1}\approx\sum_{\Theta,\phi_X,\phi_C}\ket{\Theta,\phi_X,\phi_C}\bra{\Theta,\phi_X,\phi_C}}}}\ket{n_X',\delta n_2'}\\
	=  \int & \mathrm{d}\Theta\mathrm{d}\phi_X\mathrm{d}\phi_C   \bra{n_X,\delta n_2}\mathrm{e}^{\mathrm{i}\frac{NJt}{\hbar}\Phi_\triangle(\hat{n}_X,\delta \hat{n}_2,\hat{\phi}_C,\hat{\phi}_X)t}\ket{\Theta,\phi_X,\phi_C}\mathrm{e}^{-\mathrm{i}\frac{\Theta N}{3}}\underbrace{\sum_{n_X',\delta n_2'}\mathrm{e}^{-\mathrm{i}\frac{U}{4\hbar}\left[3{\delta n_2'}^2+{n_X'}^2\right]}\mathrm{e}^{-\mathrm{i}\left[n_X'\phi_X+\delta n_2'\phi_C\right]}}_{\approx \frac{4\pi\hbar }{\mathrm{i}\sqrt{3}\tilde{U}}\exp\left[\frac{\mathrm{i}\hbar}{3\tilde{U}}\left(\phi_C^2+3\phi_X^2\right)\right]}
\end{align}
where in the first line we have indicated where a resolution of the identity in terms of phase states should be inserted, and in second line we have turned the resulting double sums over phase variables into integrals as well as indicating that the double sum over the primed number variables can be approximated by gaussian integrals. Let us now focus on the matrix elements of $\mathrm{e}^{\mathrm{i}\frac{NJt}{\hbar}\hat{\Phi}_\triangle}$,  
\begin{align}
	\bra{n_X,\delta n_2}\mathrm{e}^{\mathrm{i}\frac{NJt}{\hbar}\Phi_\triangle(\hat{n}_X,\delta \hat{n}_2,\hat{\phi}_C,\hat{\phi}_X)} & \ket{\Theta,\phi_X,\phi_C}\approx \;    \bra{n_X,\delta n_2}	\left(\mathds{1}+\mathrm{i}\frac{NJt}{M\hbar}\hat{\Phi}_\triangle \right)^{M}\ket{\Theta,\phi_X,\phi_C}\\
	=&\;\bra{n_X,\delta n_2}\underbrace{\left(\mathds{1}+\mathrm{i}\frac{NJt}{M\hbar}\hat{\Phi}_\triangle\right)\left(\mathds{1}+\mathrm{i}\frac{NJt}{M\hbar}\hat{\Phi}_\triangle\right)...\left(\mathds{1}+\mathrm{i}\frac{NJt}{M\hbar}\hat{\Phi}_\triangle\right)}_{M \mathrm{ times}}\ket{\Theta,\phi_X,\phi_C}\;,
\end{align}
where $M$ is an integer giving the number of infinitesimal time steps $t/M$ into which the propagation is decomposed. We shall assume that $M \gg N \gg 1$.
Switching to the bold vector notation $\bm{\phi}=(\Theta,\phi_X,\phi_C)$ and $\bm{n}=(\frac{N}{3},n_X,\delta n_2)$ for brevity, we insert resolutions of the identity
\begin{equation}
	\mathds{1}=\int\mathrm{d}\bm{\phi}\ket{\bm{\phi}}\bra{\bm{\phi}}
\end{equation}
 between each set of parentheses
  \begin{align}
	 \bra{n_X,\delta n_2}  \mathrm{e}^{\mathrm{i}\frac{NJt}{\hbar}\Phi_\triangle(\hat{n}_X,\delta \hat{n}_2,\hat{\phi}_C,\hat{\phi}_X)} & \ket{\Theta,\phi_X,\phi_C}\approx \;  \int\mathrm{d}\bm{\phi}_{M-1} \ldots \mathrm{d}\bm{\phi}_{1} \ \bigg\{ \bra{n_X,\delta n_2}  	\left(\mathds{1}+\mathrm{i}\frac{NJt}{M\hbar}\hat{\Phi}_\triangle \right) \ket{\bm{\phi}_{M-1}} \nonumber \\
	\times & \;\bra{\bm{\phi}_{M-1}} \left(\mathds{1}+\mathrm{i}\frac{NJt}{M\hbar}\hat{\Phi}_\triangle\right) \ket{\bm{\phi}_{M-2}} \times \ldots \times \bra{\bm{\phi}_{1}}  \left(\mathds{1}+\mathrm{i}\frac{NJt}{M\hbar}\hat{\Phi}_\triangle\right)  \ket{\bm{\phi}_{0} } \bigg\} \;, \label{eq:matrixelement_appendix}
\end{align}
 where we have used $\ket{\bm{\phi}_{0} }$ to denote $ \ket{\Theta,\phi_X,\phi_C} $. In order to evaluate each matrix element in this product, we note that $\hat{\Phi}_\triangle$ has the form
 \begin{equation}
	\hat{\Phi}_\triangle=\left(\frac{1}{2}-\frac{5}{8N}\hat{\delta n}_2\right)\hat{\phi}_X^4+\left(\frac{1}{18}+\frac{1}{24N}\hat{\delta n}_2\right)\hat{\phi}_C^4+\left(\frac{1}{3}+\frac{1}{4N}\hat{\delta n}_2\right)\hat{\phi}_X^2\hat{\phi}_C^2-\frac{1}{6N}\hat{n}_X\left(\hat{\phi}_X^3\hat{\phi}_C+\hat{\phi}_X\hat{\phi}_C^3\right)+...
\end{equation}
where the operator ordering is assumed to be the same as the classical expression given in Eq.\ (\ref{eq:hdelta_appendix}) such that all the number operators lie to the left of the phase operators. We therefore insert resolutions of the identity over the number states $\mathds{1}=\int\mathrm{d}\bm{n} \ket{\bm{n}}\bra{\bm{n}}$
inside the matrix elements so that the number operators can act to the left and the phase operators to the right
\begin{align} \bra{\bm{\phi}_{j+1}} \left(\mathds{1}+\mathrm{i}\frac{NJt}{M\hbar}\hat{\Phi}_\triangle\right)  \ket{\bm{\phi}_{j}} \longrightarrow &  \bra{\bm{\phi}_{j+1}}  \int\mathrm{d}\bm{n}_{j} \ket{\bm{n}_{j}} \bra{\bm{n}_{j}} \ \left(\mathds{1}+\mathrm{i}\frac{NJt}{M\hbar}\hat{\Phi}_\triangle\right)\ket{\bm{\phi}_{j}} \\   &
	\approx \int \mathrm{d}\bm{n}_{j}\;\exp \left[ \mathrm{i}\bm{n}_{j}\cdot (\bm{\phi}_{j+1}-\bm{\phi}_{j}) \right] \ \exp \left[ \mathrm{i} \frac{NJt}{M\hbar}\Phi_\triangle(\bm{n}_{j},\bm{\phi}_{j}) \right]
\end{align}
where to obtain the second line we have used the relation $\langle \bm{\phi} \vert \bm{n} \rangle = \exp [\mathrm{i} \bm{n} \cdot \bm{\phi}] $ twice and $\Phi_\triangle (\bm{n}_{j},\bm{\phi}_{j})$ is now a function of ordinary variables rather than operators.
Thus, all the matrix elements apart from the most lefthand one  in Eq.\ (\ref{eq:matrixelement_appendix}) contribute a phase factor $\exp \left[ \mathrm{i} \frac{NJt}{M\hbar}\Phi_{\triangle}(\bm{n}_{j},\bm{\phi}_{j})+  \mathrm{i}\bm{n}_{j}\cdot(\bm{\phi}_{j+1}-\bm{\phi}_{j})\right] $, and the wavefunction becomes
\begin{align}
	\psi(n_X,\delta n_2,t)= \int \mathrm{d} \bm{\phi}_{M-1}  \int\prod_{j=0}^{M-2}\mathrm{d}\bm{\phi}_{j}\mathrm{d}\bm{n}_{j}  \  \underbrace{B(n_{X},\delta n_{2}, \bm{\phi}_{M-1}) }_{\text{boundary term}}   & \  \exp \left[\mathrm{i}\sum_{j=0}^{M-2}\left\{\frac{NJt}{M \hbar }\Phi_\triangle(\bm{n}_{j},\bm{\phi}_{j})+\bm{n}_{j}\cdot(\bm{\phi}_{j+1}-\bm{\phi}_{j})\right\} \right] \nonumber \\&\times  \left(\frac{4\pi\hbar}{\mathrm{i}\sqrt{3}\tilde{U}}  \exp \left[\frac{\mathrm{i}\hbar}{3 \tilde{U}}\left( \{\phi_C\}_{0}^{\ 2}+3 \{ \phi_X \}_{0}^{\ 2} \right)\right] \right)  \ .
\end{align}	
The boundary term comes from the most lefthand matrix element  in Eq.\ (\ref{eq:matrixelement_appendix}) and is also a pure phase factor 
\begin{equation}
B (n_{X},\delta n_{2}, \bm{\phi}_{M-1}) = \exp \left[ \mathrm{i} \frac{NJt}{M \hbar }\Phi_\triangle[n_{X},\delta n_{2},\bm{\phi}_{M-1}]-n_{X} \ \{ \phi_{X} \}_{M-1} - \delta n_{2} \ \{ \phi_{C} \}_{M-1} \right] \ .
\end{equation}
in which the notation $ \{ \phi_{X} \}_{M-1}$ and $ \{ \phi_{C} \}_{M-1}$ is used for the individual components of $\bm{\phi}_{M-1}$ and likewise $ \{ \phi_{X} \}_{0}$ and $ \{ \phi_{C} \}_{0}$ is used for the individual components of $\bm{\phi}_{0}$. 

A condition that the above expansion of the propagator into $M$ terms is an accurate approximation is that $NJt/(\hbar M) \equiv \varepsilon$ is a small quantity. Assuming this to be the case we can write
\begin{align}
B (n_{X},\delta n_{2}, \bm{\phi}_{M-1}) +	\varepsilon\sum_{j=0}^{M-2}\left\{\Phi_\triangle[\bm{n}_{j},\bm{\phi}_{j}]+\bm{n}_{j}\cdot\frac{\bm{\phi}_{j+1}-\bm{\phi}_{j}}{\varepsilon}\right\}\sim \int_0^{NJt/\hbar}\mathrm{d}\tau\left\{\Phi_\triangle[\bm{n}(\tau),\bm{\phi}(\tau)]+\bm{n}(\tau)\cdot\dot{\bm{\phi}}(\tau)\right\} \ .
\end{align}
This becomes exact when $\varepsilon \rightarrow 0$, or equivalently $M \rightarrow \infty$ so that we can express the wavefunction as a path integral
\begin{equation}
	\psi(n_X,\delta n_2,t)=\frac{4\pi\hbar}{\mathrm{i}\sqrt{3}\tilde{U}}\int\mathcal{D}\bm{\phi}\mathcal{D}\bm{n}\; \exp \left[\mathrm{i}\int_0^{NJt/\hbar}\mathrm{d}\tau\{\Phi_\triangle[\bm{n}(\tau),\bm{\phi}(\tau)]+\bm{n}(\tau)\cdot\dot{\bm{\phi}}(\tau)\} \right]  \exp \left[\frac{\mathrm{i}\hbar}{3\tilde{U}}\left\{\phi_C^2(0)+3\phi_X^2(0) \right\} \right]
\end{equation}
where
\begin{equation}
	\mathcal{D}\bm{\phi}\mathcal{D}\bm{n}= \lim\limits_{M\to\infty} \mathrm{d} \bm{\phi}_{M-1} \prod_{j=0}^{M-2}\mathrm{d}\bm{\phi}_{j}\mathrm{d}\bm{n}_{j}  = \lim\limits_{M\to\infty} \prod_{j=0}^{M-1}\mathrm{d}\bm{\phi}_{j}\mathrm{d}\bm{n}_{j} \ \delta[\bm{n}_{M-1}-(n_{X},\delta n_{2})]
\end{equation}

\end{widetext}

\end{document}